\documentclass[sigconf, nonacm]{acmart}
\usepackage{multirow}
\usepackage{graphicx}
\usepackage{array,booktabs}
\usepackage{caption}
\usepackage{subcaption}
\usepackage{tabularx}

\AtBeginDocument{%
  }

\acmConference[CCS]{The ACM SIGSAC Conference on Computer and
  Communications Security}{October 14--18,
  2024}{Salt Lake City, USA}

\begin{document}

\title{Impact of EIP-4844 on Ethereum: Consensus Security, Ethereum Usage, Rollup Transaction Dynamics, and Blob Gas Fee Markets}

\author{Seongwan Park}
\email{sucre87@snu.ac.kr}
\orcid{0000-0001-9026-5114}
\authornote{Both authors contributed equally to the paper}
\affiliation{%
  \institution{Seoul National University}
  \city{Seoul}
  \country{Republic of Korea}
  \postcode{08826}
}

\author{Bosul Mun}
\email{bsbs8645@snu.ac.kr}
\orcid{0009-0007-5619-9409}
\authornotemark[1]
\affiliation{%
  \institution{Seoul National University}
  \city{Seoul}
  \country{Republic of Korea}
}

\author{Seungyun Lee}
\affiliation{%
  \institution{Seoul National University}
  \city{Seoul}
  \country{Repulic of Korea}
}

\author{Woojin Jeong}
\affiliation{%
 \institution{Seoul National University}
  \city{Seoul}
  \country{Repulic of Korea}
  }

\author{Jaewook Lee}
\affiliation{%
 \institution{Seoul National University}
  \city{Seoul}
  \country{Repulic of Korea}
  }
\author{Hyeonsang Eom}
\affiliation{%
 \institution{Seoul National University}
  \city{Seoul}
  \country{Repulic of Korea}
}

\author{Huisu Jang}
\authornote{Corresponding author}
\affiliation{%
  \institution{Soongsil University}
  \city{Seoul}
  \country{Republic of Korea}}

\renewcommand{\shortauthors}{Seongwan Park et al.}

\begin{abstract}    
    On March 13, 2024, Ethereum implemented EIP-4844, designed to enhance its role as a data availability layer. While this upgrade reduces data posting costs for rollups, it also raises concerns about its impact on the consensus layer due to increased propagation sizes. Moreover, the broader effects on the overall Ethereum ecosystem remain largely unexplored.

    In this paper, we conduct an empirical analysis of EIP-4844’s impact on consensus security, Ethereum usage, rollup transaction dynamics, and the blob gas fee mechanism. We explore changes in synchronization times, provide quantitative assessments of rollup and user behaviors, and deepen the understanding of the blob gas fee mechanism, highlighting both enhancements and areas of concern post-upgrade.
    
\end{abstract}






\maketitle
\begin{sloppypar}

\section{Introduction}
Increasing scalability without compromising security and decentralization is regarded as a core challenge\cite{bez2019scalability} for most public blockchains, such as Bitcoin and Ethereum. To address the growing demand for transactions and pave the way for mass adoption, diverse approaches have been explored since 2017, including plasma chains\cite{poon2017plasma}, sidechains, and state channels\cite{dziembowski2018general}. More recently, strategies like increasing the block gas limit\cite{ethresearch_gaslimit} and parallelizing execution\cite{garamvolgyi2022utilizing,gelashvili2023block} have garnered attention.

Among these solutions, rollups have emerged as a primary point of research since 2018\cite{ethresearch_rollup,fernando2023poster,visscher2022poster}, becoming a critical part in the recent Ethereum roadmap. Distinct from earlier methods such as Plasma and sidechains, which struggled with data availability problem\cite{poon2017plasma} and centralization, rollups could benefit from the robust security of the Ethereum mainnet\cite{nazirkhanova2022information}. They process transactions off-chain and post summarized batches back to Layer 1 for final verification. This approach significantly reduces the computational burden on Ethereum, potentially lowering transaction fees while maintaining strong security.

Security advantages have propelled rollups to significant prominence, with the total economic value secured by rollup solutions surpassing 40 billion dollars as of April 2024, according to L2beat\cite{l2beat2024}. Leading platforms like Arbitrum, Optimism, and Base process approximately five million transactions daily, and dozens of new rollups are planning to launch.

Despite these gains in scalability, challenges remain due to the limited capacity of Ethereum mainnet\cite{neiheiser2023practical}, which serves as a data availability layer essential for validating rollup transactions. In response, Ethereum researchers are attempting to enhance Ethereum's function as a DA layer for rollups. 


A pivotal development in addressing these limitations is the introduction of EIP-4844, or Proto-Danksharding, which was implemented on March 13, 2024\cite{eip4844spec}. This protocol introduces blobs, a new data structure temporarily accessible for 18 days, unlike traditional calldata that is permanently stored. This adjustment aims to reduce the cost of data posted by rollups significantly. EIP-4844 also ushers in a new blob gas fee market, marking the first introduction of a multidimensional fee market in Ethereum. While this has successfully reduced rollup data posting costs\cite{l2fees}, a thorough examination of its broader impacts on Ethereum ecosystem is crucial.

In this paper, we aim to provide a comprehensive analysis of EIP-4844’s impact on consensus security, Ethereum usage, rollup transaction dynamics, and the new blob gas fee mechanism, offering insights that could help evaluate the protocol change.

\textbf{Motivation.}
The changes introduced by EIP-4844 could increase the time required for nodes to validate slots and reach consensus, potentially affecting Ethereum's consensus security. This increased data size, up to 768KiB per slot in extreme cases, may lead to more forked and missing slots, impacting chain stability. Additionally, this could create disparities among validators, as those with better resources may have an advantage in successfully proposing beacon blocks(slot)\cite{xiao2020modeling}


Moreover, understanding the dynamics of the new blob gas fee market is crucial. This market manages the fees associated with new blob structures. Deeply comprehending this market is essential for evaluating fee mechanisms, enhancing predictability, and developing optimized fee strategies for decentralized applications (DApps). Insights from studies like \cite{mamageishvili2023efficient}, which explore optimal batching strategies in single-dimensional gas fee markets, underscore the need for adaptations in this new multidimensional context.

Furthermore, an empirical analysis of how EIP-4844 affects Ethereum usage and rollup behavior is imperative. It is important to determine whether there has been a significant change in total user engagement or an increase in rollup fees. Such analysis will help evaluate the effectiveness of this protocol change and guide further improvements based on the real user behavior data. Previous studies like \cite{liu2022empirical}, which analyzed the impacts of EIP-1559 on user waiting times, orphan slot occurrence, and fee dynamics, have provided valuable insights into Ethereum's evolving fee markets. However, research specifically focused on EIP-4844 remains scarce, indicating a need for comprehensive studies that can inform future enhancements and ensure robustness in Ethereum's infrastructure.

\textbf{Challenges and Our Approach.}
Conducting a comprehensive analysis of EIP-4844 posed several significant challenges, each requiring effective solutions to ensure the robustness and accuracy of our findings:

\textit{Data Collection on Slot Synchronization Delays.} Unlike persistent on-chain data, information on slot synchronization delays is ephemeral and highly variable, influenced by factors such as hardware capabilities and geographic location. To capture a comprehensive range of real-time data on slot reception, processing, and synchronization times, we deployed three Ethereum full nodes on AWS instances with identical hardware specifications distributed across Paris, Singapore, and Virginia. This setup allowed us to observe slot sync times across different network conditions. However, it also posed challenges, such as occasional node downtime and the complex task of managing log files to accurately interpret each timing event. We ensured that nodes remain online to collect accurate data and meticulously extract relevant timing information from client source code. These efforts were crucial for ensuring the reliability of our findings.

\textit{Data Gathering from Multiple Rollup Networks.} The varied architectures and rapid block times of rollups present significant data collection challenges. These challenges are amplified by the unique transaction function names used by each rollup. To address these complexities, we first identified known rollup addresses through resources like Etherscan\cite{etherscan2024} and L2BEAT\cite{l2beat2024}. For ten major rollups\footnote{Arbitrum One, Optimism, Base, Blast, Starknet, zkSync Era, dYdX V3, Linea, Mode, Scroll}, we analyzed transaction functions to classify them according to their purpose—either for data availability or execution. Leveraging block explorers, our own archive nodes, and tools like Ethernow\cite{ethernow2024}, we meticulously collected and decoded transaction data. This allowed us to compile a detailed dataset, such as user delays and transaction volumes, which was crucial for our in-depth analysis.

\textit{Evaluating the Blob Gas Fee Market.} 
One major challenge was selecting an appropriate analysis period for the blob gas markets, given their volatility since inception. Initially, the blob gas base fee mostly stayed at the minimum of 1 wei, with occasional spikes, such as a peak at 654 Gwei, before reverting to 1 wei. To effectively capture the characteristics of the blob gas base fee, we focused on a specific period\footnote{From block number 19,518,097 to block number 19,587,588} where it exceeded 0.1 Gwei, allowing for a more meaningful analysis of its behavior under varied market conditions.

Another is that the absence of a direct priority fee mechanism for blobs added complexity to assessing market dynamics. We developed a metric that combines the gas priority fee with gas used and blob gas base fee. Validated through VAR modeling, this metric effectively quantifies user demand reflection in blob gas pricing, providing crucial insights into the market's functionality.


\textbf{Our contributions.}
Our main contributions in this paper are as follows:

\begin{itemize}
\item We detail the potential negative impacts of EIP-4844 on Ethereum's consensus security, specifically addressing the increase in fork rates and identifying their primary causes. This analysis helps clarify concerns within the community about the stability of the network post-update.

\item Through comprehensive visualizations and statistical analyses, we illustrate the effects of EIP-4844 on Ethereum usage and rollup transactions. Our findings confirm whether the upgrade successfully incentivized rollup activities through reduced fees and explore the changes in user delay, indicating an overall increase.

\item We introduce and justify the blob gas priority fee metric, demonstrating its utility in predicting blob gas base fees. It provides insights into the new blob gas fee market dynamics. We also discuss the evaluation of blob gas fee mechanism design utilizing the priority fee metric.

\item Our extensive data collection includes time series data on slot arrivals, processing times, and blob arrivals, as well as detailed rollup transactions and Ethereum usage metrics from various sources. We make this dataset fully available to the research community, providing a valuable resource for further investigation.

\end{itemize}

\section{Background}
\subsection{Rollup}
Ethereum operates as a decentralized state machine where transactions, bundled into blocks, transition the state representing account balances and smart contract values. Achieving consensus on state changes among all network participants is crucial for maintaining the integrity of the blockchain \cite{wood2014ethereum}.

Scalability remains a significant challenge for Ethereum, largely because every node must execute all transactions and reach consensus. In contrast to centralized systems like VISA, which handle over 2000 transactions per second\cite{klarman2018bloxroute}, Ethereum manages about 15 transactions per second. This limitation restricts wider blockchain adoption \cite{bez2019scalability}.

\begin{figure}[ht]
  \centering
  \includegraphics[width=\linewidth]{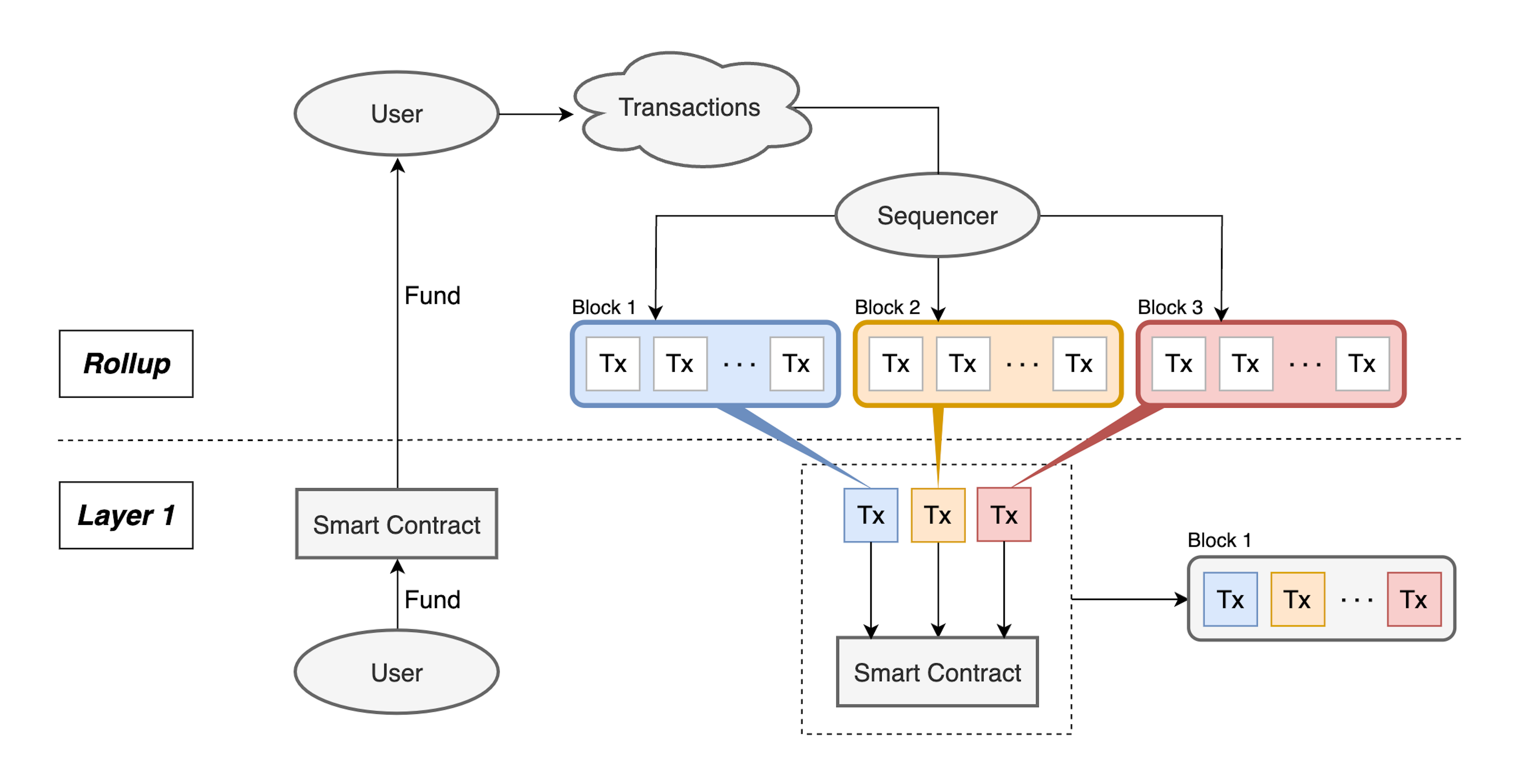}
  \caption{Interaction between rollup and Layer 1.}
  \label{fig:2_1_rollup_layer1}
  \Description{}
\end{figure}

Rollups address Ethereum’s scalability challenges by using a two-layer model: off-chain execution and on-chain settlement. In this model, transactions are processed off-chain and summarized back to a smart contract on the Ethereum mainnet. This method significantly reduces transaction fees and increases scalability by lessening the load on the base layer\cite{sguanci2021layer}. Additionally, because rollups post their results to the mainnet, they inherit the security properties of Ethereum, ensuring that off-chain computations can be verified. The transaction workflow in rollups includes:

\begin{enumerate}
    \item Submission of transactions to the rollup chain, either directly or via a Layer 1 bridging smart contract.
    \item Execution of transactions in batches by rollup sequencers.
    \item Aggregation and reporting of the transaction results and updated state summaries by sequencers to their corresponding smart contracts on the Ethereum mainnet.
    \item On-chain verification of the state updates' correctness on the Ethereum mainnet.
\end{enumerate}

Rollups are classified into two types based on their verification methods on Ethereum: Optimistic and Validity rollups. Validity rollups, such as zk-rollups, submit transaction data along with a new state root and a validity proof—a concise proof confirming the accuracy of the new state root derived from the executed transactions. In contrast, Optimistic rollups publish only the transaction data and state root, assuming correctness unless challenged during a dispute period initiated by verifiers if inconsistencies in the posted state changes are detected \cite{thibault2022blockchain}.

\subsection{EIP-4844}

\subsubsection{Data Availability in Rollups}
Data availability(DA) ensures that specific data can be accessed at a given point in time and verified as accessible at that same point in the future. Unlike permanent data storage, DA does not imply indefinite data retention but ensures that data is temporarily accessible for verification and auditing purposes.

In the context of Ethereum rollups, DA is crucial for ensuring user security\cite{tas2022accountable}, with numerous researches focusing on enhancing this aspect\cite{krol2023data,hall2023foundations,sheng2021aced}. It guarantees that users can access transaction data temporarily to verify state updates or reconstruct the rollup's current state. For instance, in optimistic rollups, users require DA extending beyond the challenge period (typically 7-12 days) to validate state changes confidently.

The assurance of DA is vital because, without it, users must rely solely on the trustworthiness of rollup operators. This reliance can expose users to risks if operators act maliciously or withhold data, compromising the integrity of the rollup\cite{huang2024data}.

Prior to the implementation of EIP-4844, Ethereum rollups typically used Ethereum’s calldata—a space within transactions designated for storing function call arguments—for DA. This method involves storing compressed transaction data in calldata, where the cost is 16 gas per non-zero byte and 4 gas per zero byte\cite{kotzer2024sok}. EIP-4844 aims to refine this model by introducing more efficient data handling mechanisms to reduce costs and improve scalability, addressing both the economic and performance limitations of previous approaches.

\subsubsection{Blob-carrying transaction}
The Ethereum network has a current block gas limit of 30M gas, which theoretically allows for a maximum block size of 1.8MB when filled solely with calldata. This scenario represents a potential worst-case scenario for network block size. Further reducing the cost of calldata to enhance scalability risks permitting unsustainable block sizes, posing significant network overloading risks \cite{buterin2024danksharding}.

To mitigate these concerns and improve DA without significantly impacting the network’s maximum block size, EIP-4844, also known as Proto-DankSharding, introduces the concept of blob-carrying transactions. These transactions incorporate a new data structure called a blob, which consists of 4096 field elements, each 32 bytes in size, amounting to a total of 128KiB per blob. Rather than storing blobs directly within transactions, they are represented by a versioned hash of the blob's KZG commitment hash and are temporarily maintained on the consensus layer for a duration of 18 days before deletion.

\begin{figure}[ht]
  \centering
  \includegraphics[width=\linewidth]{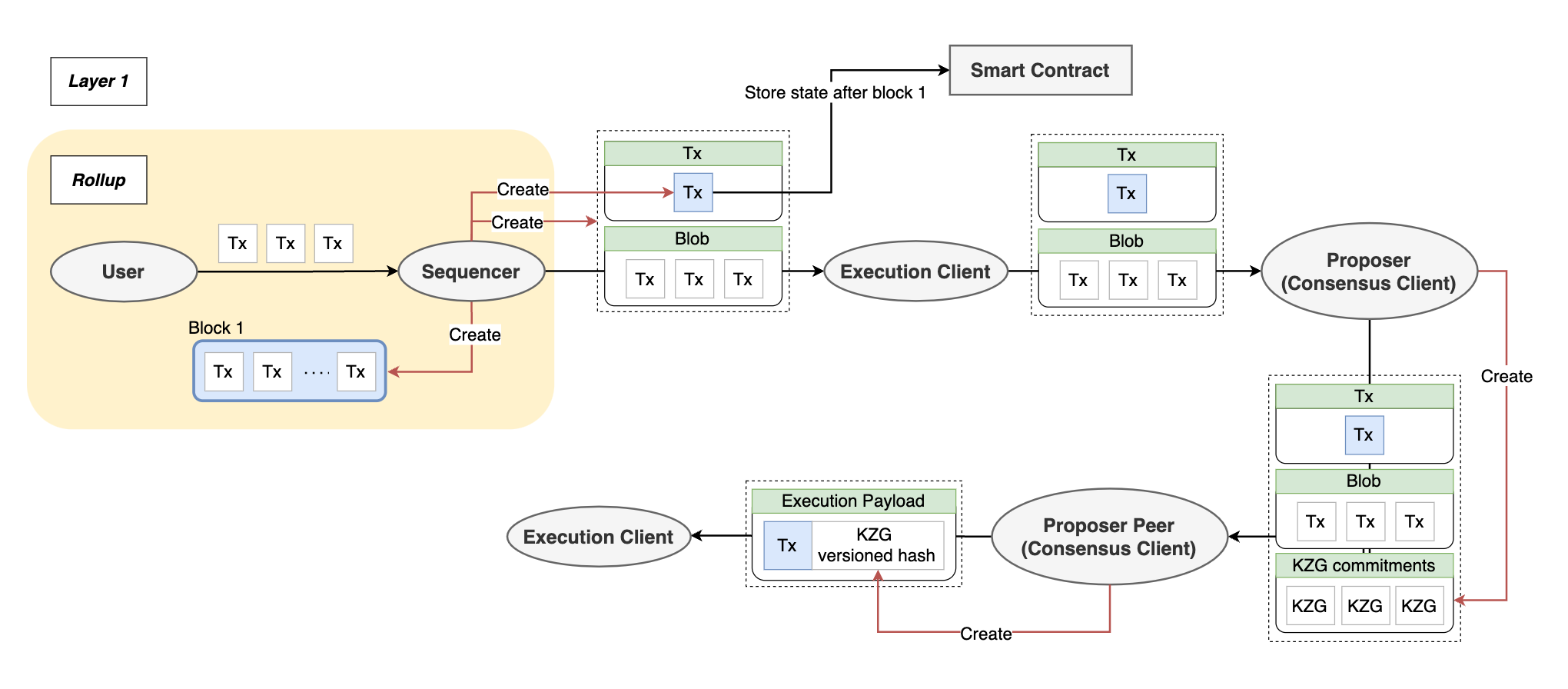}
  \caption{Lifecycle of a blob}
  \label{fig:2_2_blob_life_cycle}
  \Description{}
\end{figure}

The design choice to exclude blobs from permanent storage on the execution layer significantly enhances gas efficiency. This is because blobs, unlike traditional calldata, do not incur the high gas costs associated with permanent data storage. The lifecycle of a blob, from creation to expiration, is illustrated in Figure \ref{fig:2_2_blob_life_cycle}.

\subsubsection{Blob gas fee market}
The implementation of EIP-4844 introduced a distinct fee market for blob gas, incurred by transactions carrying blobs. Each blob consumes a constant \(131,072 = 2^{17}\) blob gas units. The base fee for one unit of blob gas is dynamically adjusted every block in response to network congestion, mirroring the mechanism established by EIP-1559\cite{roughgarden2021transaction} for gas.

The blob gas base fee adjustment follows a specific formula aimed at maintaining an optimal number of blobs per block. The target number of blobs is set at three per block. The base fee is adjusted based on the actual usage compared to this target:
\[
B_{\text{blob gas}, k+1} = B_{\text{blob gas}, k} \times \exp\left(\frac{u - t}{8t}\right)
\]
where:
\begin{itemize}
    \item \(B_{\text{blob gas}, k}\) represents the base fee for blob gas in block \(k\),
    \item \(u\) denotes the total blob gas used in block \(k\),
    \item \(t\) is the target blob gas usage, set at \(3 \times 131,072\) blob gas units per block.
\end{itemize}



\subsection{VAR(Vector Autoregression)}
VAR is a statistical model designed to analyze multivariate time series data\cite{hamilton2020time}. It extends the univariate autoregressive model by allowing multiple interdependent time series to be modeled simultaneously. In a VAR model, each variable is a linear function of past lags of itself and past lags of other variables in the system. A VAR model of order \( p \) (VAR(p)) is specified as follows:

\begin{equation}
    Y_t = c + \Phi_1 Y_{t-1} + \Phi_2 Y_{t-2} + \cdots + \Phi_p Y_{t-p} + \epsilon_t
\end{equation}

where \( Y_t \) is a vector of endogenous variables at \( t \), \( c \) is a constant vector, \( \Phi_1, \Phi_2, \ldots, \Phi_p \) are coefficient matrices, and \( \epsilon_t \) is an error term vector. Unlike simple linear regression which assesses static relationships, VAR models the dynamic interaction among the variables over time, capturing the internal mechanics of systems.





\section{Data}
\textbf{Data Availability.} All data and code utilized in this study are openly accessible to ensure the reproducibility of our analyses and to support further research. These resources can be found at \footnote{\url{https://github.com/etelpmoc/eip4844}}. 

\subsection{Consensus security data}
To mitigate geographical biases and better isolate the impact of network speed variations across different locations, we deployed three Ethereum full nodes with homogeneous hardware configurations and client versions. Each node was hosted on an AWS t3.xlarge instance equipped with Ubuntu 22.04, featuring 4 vCPUs and 16GB of memory, and ran identical software stacks—Geth 1.13.14\cite{go_ethereum} as the execution client and Prysm v5.0.0\cite{prysm2024} as the consensus client. These nodes were located in distinct geographic regions: Virginia, Paris, and Singapore. This setup ensured that any observed differences in data propagation and processing times could be predominantly attributed to network latency rather than variations in hardware performance or software configurations.

In this study, we analyze slots from 8,570,000 to 8,626,175 representing eight days prior to the implementation of EIP-4844, and slots 8,626,176 to 8,839,999 covering approximately four weeks following its introduction. However, as indicated in the subsequent analysis of the fork rate, there exists data anomalies between slots from 8,720,000 to 8,740,000 due to incorrect implementations by other network entities. Since these anomalies stem from mistakes at the application level and not from the EIP-4844 specification itself, they are excluded from the analysis. Going forward, slots from 8,570,000 to 8,626,175, before the introduction of EIP-4844, will be referred to as pre-4844, and slots from 8,626,176 and 8,720,000, and 8,740,000 to 8,839,999 after its implementation will be alternately referred to as post-4844.

The following data fields were extracted from the debug mode log file in the Prysm client:

\begin{itemize}
    \item \textbf{\texttt{receive\_time}}: The time when the consensus client received the slot, measured from the start of the slot
    \item \textbf{\texttt{chain\_service\_provide\_time}}: The time taken by the consensus client to execute the slot, update the consensus state and execution layer state, and the fork choice.
    \item \textbf{\texttt{data\_availability\_time}}: The time the consensus client waits for any blobs that have not yet arrived after executing the slot.
    \item \textbf{\texttt{sync\_time}}: The time when the consensus client synced the slot, measured from the start of the slot
\end{itemize}

\begin{figure}[ht]
    \centering
    \includegraphics[width=0.5\textwidth]{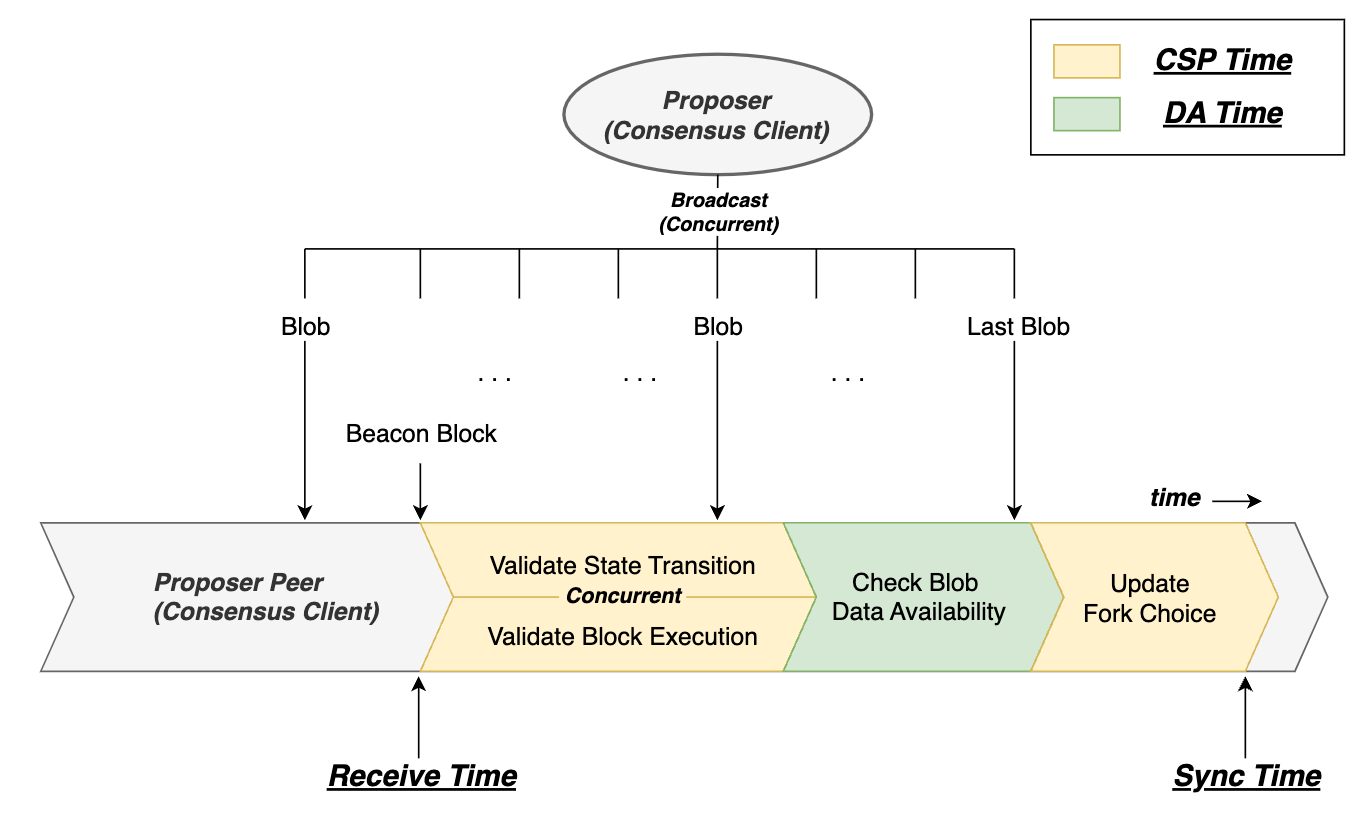}
    \caption{Operational flow of the consensus client}
    \label{fig:client_workflow}
    \Description{}
\end{figure}

To better understand the significance of these fields, it is essential to comprehend the operational flow of the consensus client. Figure \ref{fig:client_workflow} illustrates the client workflow, which details how data is processed and logged within the system.
   
When the consensus client receives the slot, it is recorded as \texttt{receive\_time}. Upon receiving a slot, the Prysm client simultaneously updates the consensus state and triggers the execution of the execution payload(block) within the execution layer. Since blobs are propagated separately from the block itself, it is necessary to verify the arrival of all blobs before selecting a fork. The duration required to ensure all blobs have arrived is referred to as the \texttt{data\_availability\_time}. Once all blobs are arrived, the fork choice process begins, which includes the verification of attestations and the handling of potential reorganizations. The moment when the slot processing is completed is recorded as the \texttt{sync\_time}. The time difference between the reception of the slot and its synchronization, excluding the \texttt{data\_availability\_time}, is recorded as the \texttt{chain\_service\_process\_time}. 



\begin{figure*}[ht]
  \centering
  \includegraphics[width=\linewidth]{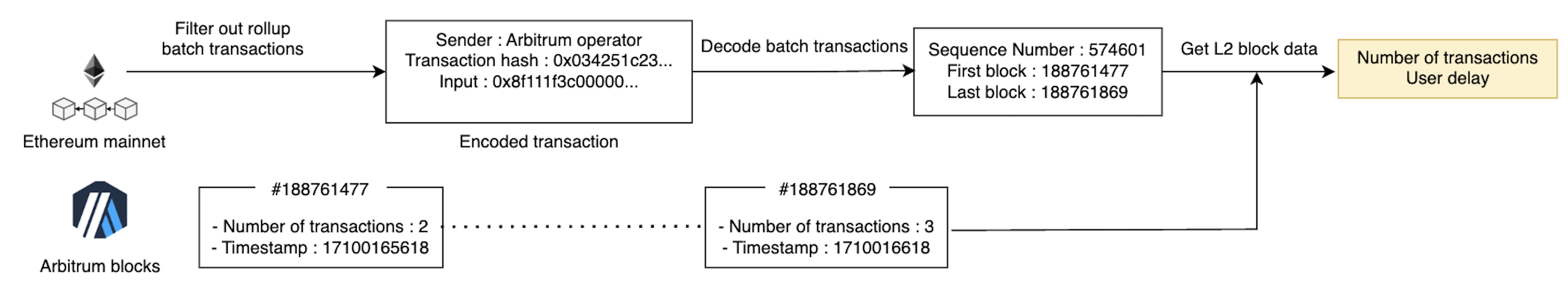}
  \caption{Illustration of the data collection and preprocessing for Arbitrum blocks}
  \label{fig:rollup_data_collection_example}
  \Description{}
\end{figure*}

\subsection{Ethereum usage data}
To assess the impact of EIP-4844 on Ethereum usage effectively, we focused on the top 10 rollups by Total Value Locked (TVL) as reported by L2BEAT\cite{l2beat2024}. As of April 28, 2024, these rollups—Arbitrum One, Optimism, Base, Blast, Starknet, zkSync Era, Linea, dYdX v3, Mode Network, and Scroll—account for 98.5\% of the ecosystem's TVL. We gathered data from sender addresses associated with these rollups, which are pivotal in committing transaction batches and verifying state differences. Our study spans 180,000 blocks before and after the implementation of EIP-4844, covering the period from February 17 to April 7, 2024.

A significant challenge in this analysis is distinguishing the transactions that contribute directly to Ethereum's role as a DA layer. Not all transactions from rollups serve this function; for instance, zk rollups often utilize Ethereum more for execution than for DA. Importantly, execution transactions, which are not DA transactions, cannot be replaced with blob transactions. To navigate this, we categorized transactions into two groups based on their roles: transactions from optimistic rollups that commit batches are deemed DA transactions, while for zk rollups, only those committing batches and posting state updates are classified as DA transactions. The specific sender addresses and their categorization are detailed in Appendix Table \ref{tab:transaction_classification}.

Data extraction was performed using an Erigon\cite{erigon2024} archive node, from which we retrieved gas fees, blob gas fees, and the sizes of calldata and blobs used by the identified rollup sender addresses. This data helped us derive three key metrics to analyze Ethereum's usage dynamics: posted data size, total fees paid, and the cost of posting 1MiB on the Ethereum network.

\subsection{Rollup Transactions Data}
To assess the impact of EIP-4844 on rollup transaction dynamics, we conducted a comprehensive analysis focusing on the changes in rollup transaction volumes and delays between rollup and Ethereum blocks. The analysis period spans 100,000 blocks before and after the implementation of EIP-4844.

Our data were sourced from transactions sent by recognized rollup addresses on the Ethereum network. Among the various transaction types initiated by rollups, we specifically collected batch transactions, which compress all individual rollup transactions. These transactions are crucial for ensuring user security by mitigating operator risk and safeguarding user funds. Batch transactions typically precede other transaction types such as proving and finalizing transactions, reflecting their foundational role in securing user interactions on rollups. We regard the timestamp of batch transaction sent to Ethereum as the settlement of rollup transactions, and calculate the delay by getting the time difference from rollup block timestamp.

The following common process was employed to extract data on rollup transactions and user delays:
\begin{enumerate}
    \item Filter rollup transactions from the Ethereum mainnet using known rollup sender addresses.
    \item Decode the data from rollup batch transactions.
    \item Acquire rollup block data from external sources and integrate this data with (2) to analyze user delays and transaction metrics.
\end{enumerate}

Figure \ref{fig:rollup_data_collection_example} illustrates an example of the process of our data collection and preprocessing for Arbitrum blocks.  

Each rollup employs unique encoding mechanisms, often modified by updates such as span-batch mechanisms\cite{opstack_spanbatches}, which posed significant decoding challenges. Additionally, the rapid block times and large data volumes of rollups like Arbitrum (0.26 seconds) and Optimism (2 seconds) necessitated the use of specialized tools and methods for data collection and analysis, as maintaining full nodes for all monitored rollups was infeasible.

We utilized a variety of rollup explorers and batch decoding tools tailored to each rollup's specific needs. Details on the specific tools and data sources used are provided in Appendix Table \ref{tab:data_sources_rollup_transactions}. Our analysis concentrated on six rollups—Arbitrum One, Optimism, Base, Starknet, zkSync Era, and Linea—where we were able to obtain decoded batch transaction data.

\subsection{Blob gas fee data}
\label{subsection:blob_gas_fee_data}
To conduct a comprehensive analysis of the blob gas fee mechanism, we collected data from our Erigon archive node on the base fees for blob gas, as well as the gas and blob gas usage for each transaction within selected blocks. To explore the new blob gas market, we specifically analyzed data from blocks 19,518,097 to 19,587,588, during which the blob gas base fee exceeded 0.1 Gwei.

\textbf{Blob gas market period}
The blob gas base fee update rule adjusts the base fee upward when average usage surpasses three blobs per block. Given the gradual uptake of blobs by rollups and their limited use by DApps, the blob gas base fee typically hovered around 1 wei for a considerable duration.

Our analysis concentrates on the period during which base fees rose above 0.1 Gwei, corresponding with heightened blob activity. This period commenced at block 19,518,097, triggered by the activation of blob submission services that briefly elevated the blob base fee. Although demand receded and the base fee reverted to 1 wei by block 19,587,588, the fluctuations within this interval are crucial for comprehending potential reactions of the blob gas fee market to increased DApp engagement. Focusing on this period allows for a detailed examination of the blob gas fee market's behavior under conditions of active blob utilization.

\begin{figure}[ht]
  \centering
  \includegraphics[width=0.8\linewidth]{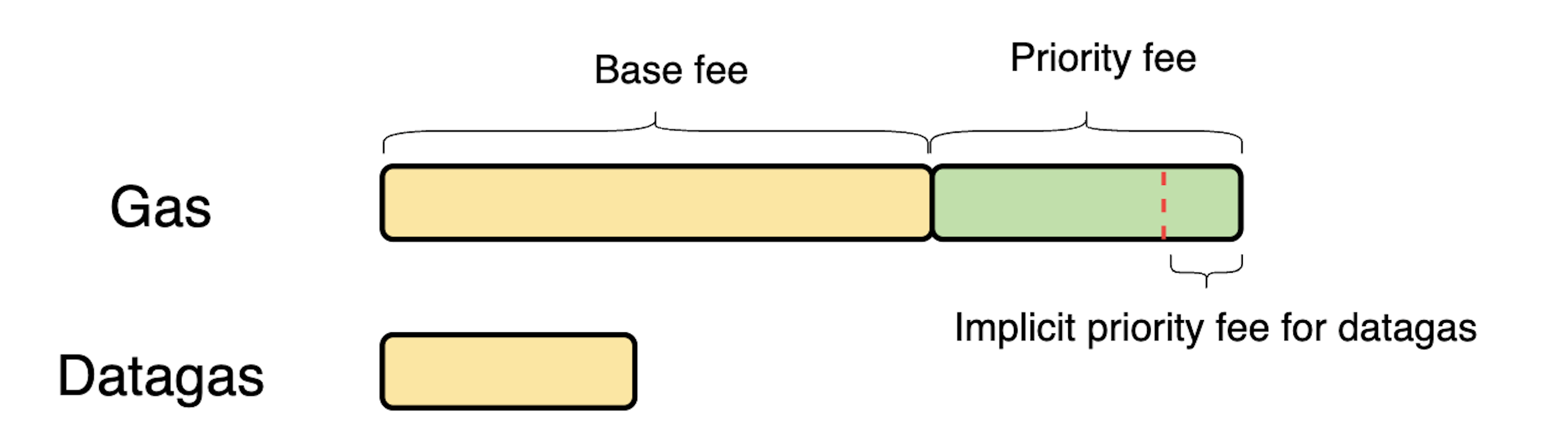}
  \caption{Implicit priority fee of blob gas}
  \Description{The figure describes that priority fee for blob gas is implicitly reflected in gas priority fee}
  \label{fig:blob_gas_priority_fee}
\end{figure}

\textbf{Blob gas priority fee.}
Unlike the gas fee update rule, where users can set a maximum priority fee per gas unit, the blob gas fee mechanism lacks this functionality. In the blob gas market, there is only a base fee, which is automatically adjusted based on network congestion. Users must implicitly set a blob gas priority fee, as illustrated in Figure \ref{fig:blob_gas_priority_fee}.

To effectively evaluate the blob gas base fee update rule, it is crucial to quantify the excess demand for blob gas. In the traditional gas market, the priority fee of a transaction serves as an indicator of how well the base fee reflects actual user demand. Consequently, we have developed a new metric to represent the blob gas priority fee using the following formula:

For each \( k \)-th block containing multiple transactions, we define the following parameters:
\begin{itemize}
    \item \( B_{\text{gas}, k} \) and \( B_{\text{blob gas}, k} \): Base fees for gas and blob gas in the \( k \)-th block.
    \item \( P_{\text{gas}, i, k} \) and \( P_{\text{blob gas}, i, k} \): Priority fees for gas and blob gas for the \( i \)-th transaction in the \( k \)-th block.
    \item \( E_{i,k} \) : Effective prices for \( i \)-th transaction in the \(k\)-th block.
    \item \( G_{i, k} \) and \( D_{i, k} \): Amounts of gas and blob gas used in the \( i \)-th transaction in the \(k\)-th block.
\end{itemize}

The blob gas priority for each transaction can be expressed as:

\[
\text{P}_{\text{blob gas}, i, k} := \left( \frac{(E_{i,k} - B_{\text{blob gas}, k} - \underset{\text{tx} \in k}{\mathrm{median}} (P_{\text{gas}, tx, k})) \times G_{i, k}} {D_{i, k}} \right)^+
\]

To find implicit priority fee for blob gas, we used median priority fee of other transactions in the same block as a proxy for gas priority fee, and subtracted it from the total fee paid.

\section{Empirical Results}
\subsection{Consensus security}
EIP-4844 introduces blobs, a new data type to be propagated and processed. Blobs add new burden for consensus layer\cite{buterin2024danksharding}, potentially slowing validator performance and prolonging the time required to achieve consensus. Such delays could threaten the security of the consensus network. Our thorough examination of the impact of EIP-4844 on Ethereum's consensus security has yielded the following key findings:

\begin{enumerate}
    \item \textbf{Fork Rate Increase}:     
    We present evidence that the fork rate has risen since the implementation of EIP-4844, from 3.097 to 6.707 slots per 2000 slots, suggesting a direct impact on network stability. This increase hints at possible challenges to network stability.
    
    \item \textbf{Slot Sync Time Increase}: We observed an increase in slot synchronization times, correlating with the quantity of blobs per slot. Given that delayed slot synchronization can heighten the fork rate, this extended sync time likely contributes to the observed increase in forks.

    \item \textbf{Analysis of Slot Sync Time Components}: Our analysis identifies receive time as the component most significantly affected by EIP-4844, which has led to increased sync times. Conversely, the DA time remains minimal and shows no direct correlation with the number of blobs, suggesting that blob propagation minimally impacts consensus delays.

\end{enumerate}

\subsubsection{Fork rate}
A fork occurs when different validators view distinct slots, which share the same parent slot, as valid. Frequent forks can compromise consensus security by invalidating transactions within orphaned slots and diminishing the network's ability to process transactions efficiently. Forks also potentially increase the vulnerability of forked slots to consensus attacks \cite{schwarz2022three, iqbal2021exploring} for malicious purposes, such as stealing MEV(Maximal Extractable Value) or double spending.

\begin{figure}
    \includegraphics[width=0.8\linewidth]{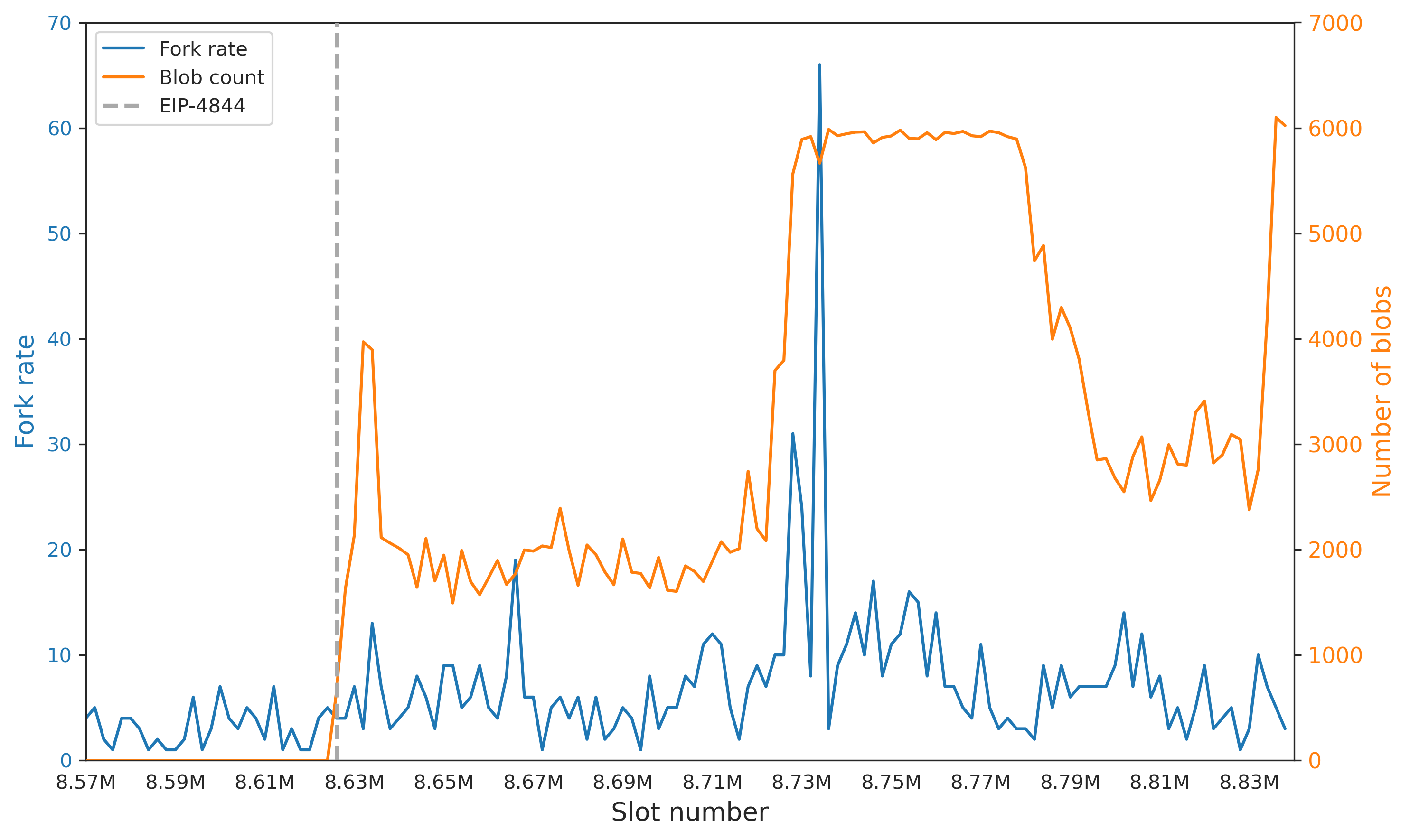}
    \caption{Change in fork rate pre-4844 and post-4844, with an anomaly between slots 8,720,000 and 8,740,000 due to an implementation bug}
    \label{fig:4_1_fork_rate_blob_line}
    \Description{}
\end{figure}

To determine if EIP-4844 has increased these risks, we analyzed changes in the fork rate, defined as the number of forked slots per 2,000 slots. Figure \ref{fig:4_1_fork_rate_blob_line} shows the fork rate in relation to the number of blobs. Notably, a significant increase is observed between slots 8,720,000 and 8,740,000, initially perceived as a threat to consensus security. This spike was later attributed to a bug in the implementation of EIP-4844 by some network participants, which went undetected until an increase in blob activity occurred \cite{hunter2024gist}.

Based on this analysis, we determined that the sharp increase in the fork rate during the spike was not a direct result of the EIP-4844 specification. Nevertheless, our further investigations reveal that, even after excluding this anomalous spike period, the average fork rate has still risen from 3.097 to 6.707 slots per 2,000 slots—an increase of 116.538\%. This notable elevation in the fork rate necessitates additional research to clarify its underlying causes.

\subsubsection{Sync time}
\begin{figure}
    \includegraphics[width=0.8\linewidth]{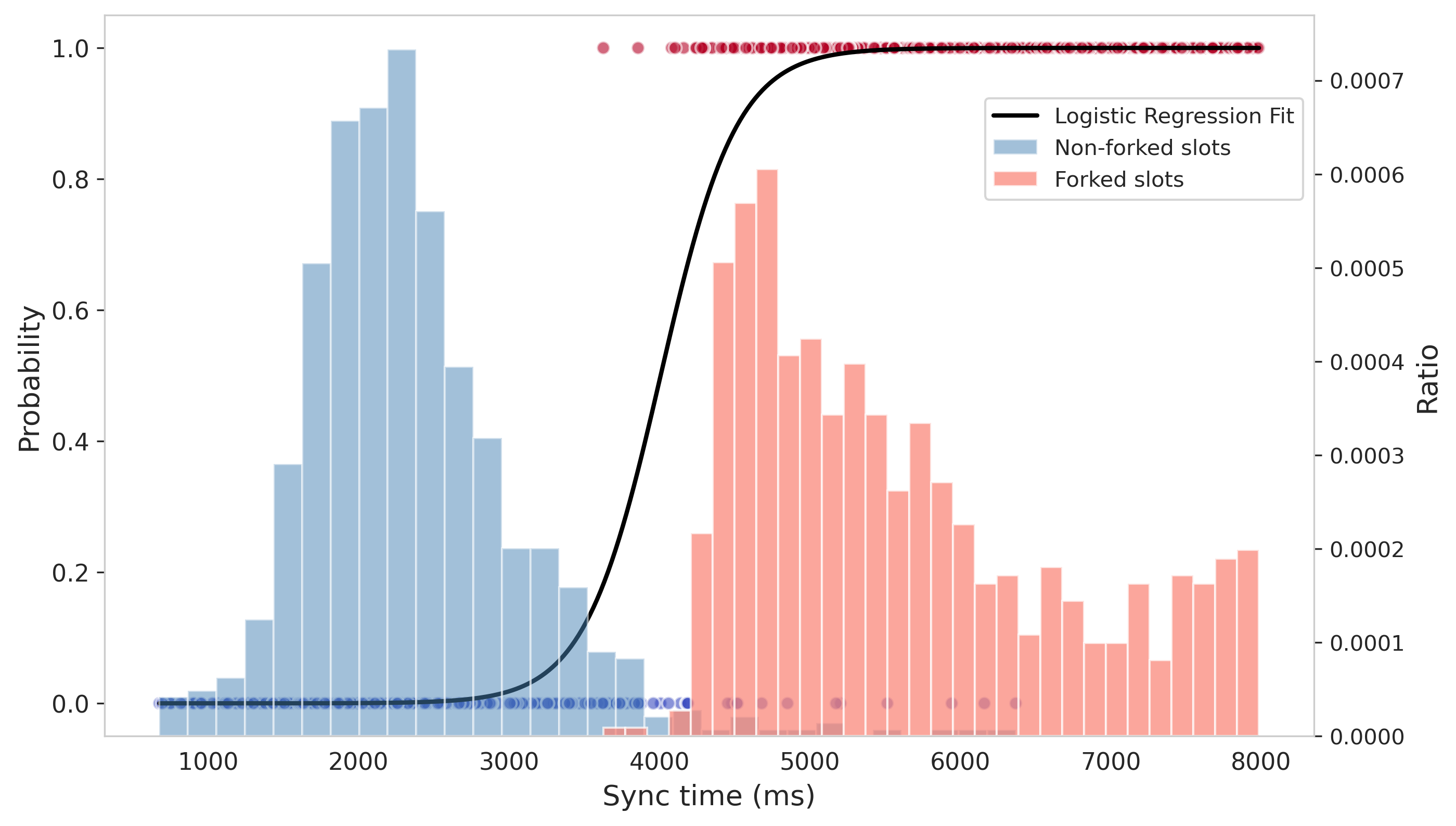}
    \caption{Distribution of sync time in forked blocks and non forked blocks with fitted logistic regression}
    \label{fig:4_1_logistic}
    \Description{}
\end{figure}

An increase in sync time can be a major contributor to a higher fork rate, as forks occur when a proposer fails to synchronize new slots with preceding ones. To confirm the relationship between fork rate and sync time, we applied a logistic regression model. The analysis showed that an increase in sync time is significantly associated with the likelihood of a slot being forked, with a coefficient of \(1.5 \times 10^{-3}\) (p-value < 0.001) for sync time, and a model intercept of -10.9497 (p-value < 0.001). These results suggest that slots with longer sync times have a higher probability of being forked, underlining the impact of sync time on fork rate. Figure \ref{fig:4_1_logistic} visualizes this relationship, clearly demonstrating that forked slots typically experience longer sync times than non-forked slots.

\begin{figure}
    \centering
    \hfill 
    \begin{subfigure}{0.45\linewidth}
        \centering
        \includegraphics[width=\linewidth]{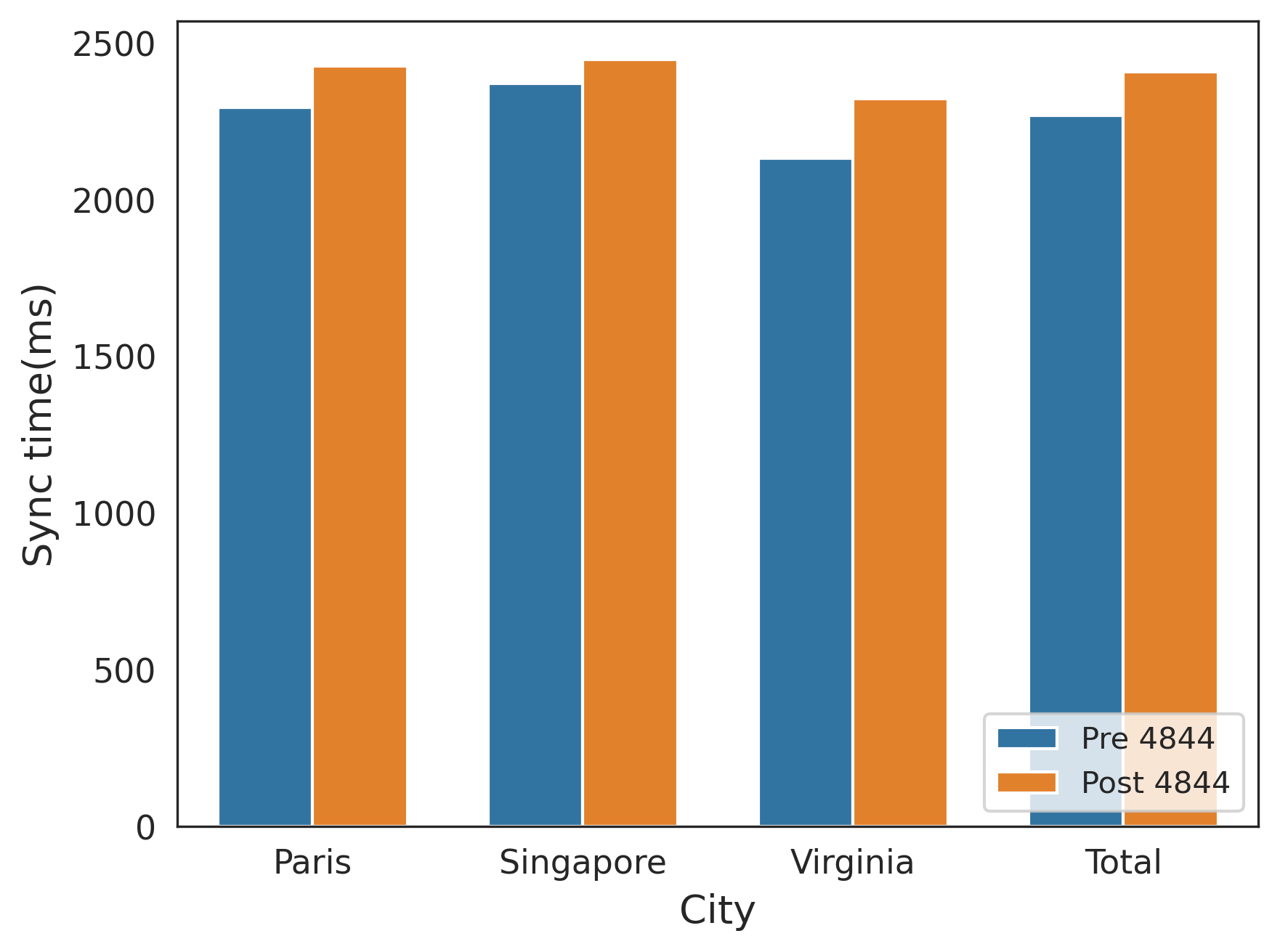}
        \caption{Average sync times across different cities and overall}
        \label{fig:4_1_sync_time_city}
    \end{subfigure}
    \hfill 
    \begin{subfigure}{0.45\linewidth}
        \centering
        \includegraphics[width=\linewidth]{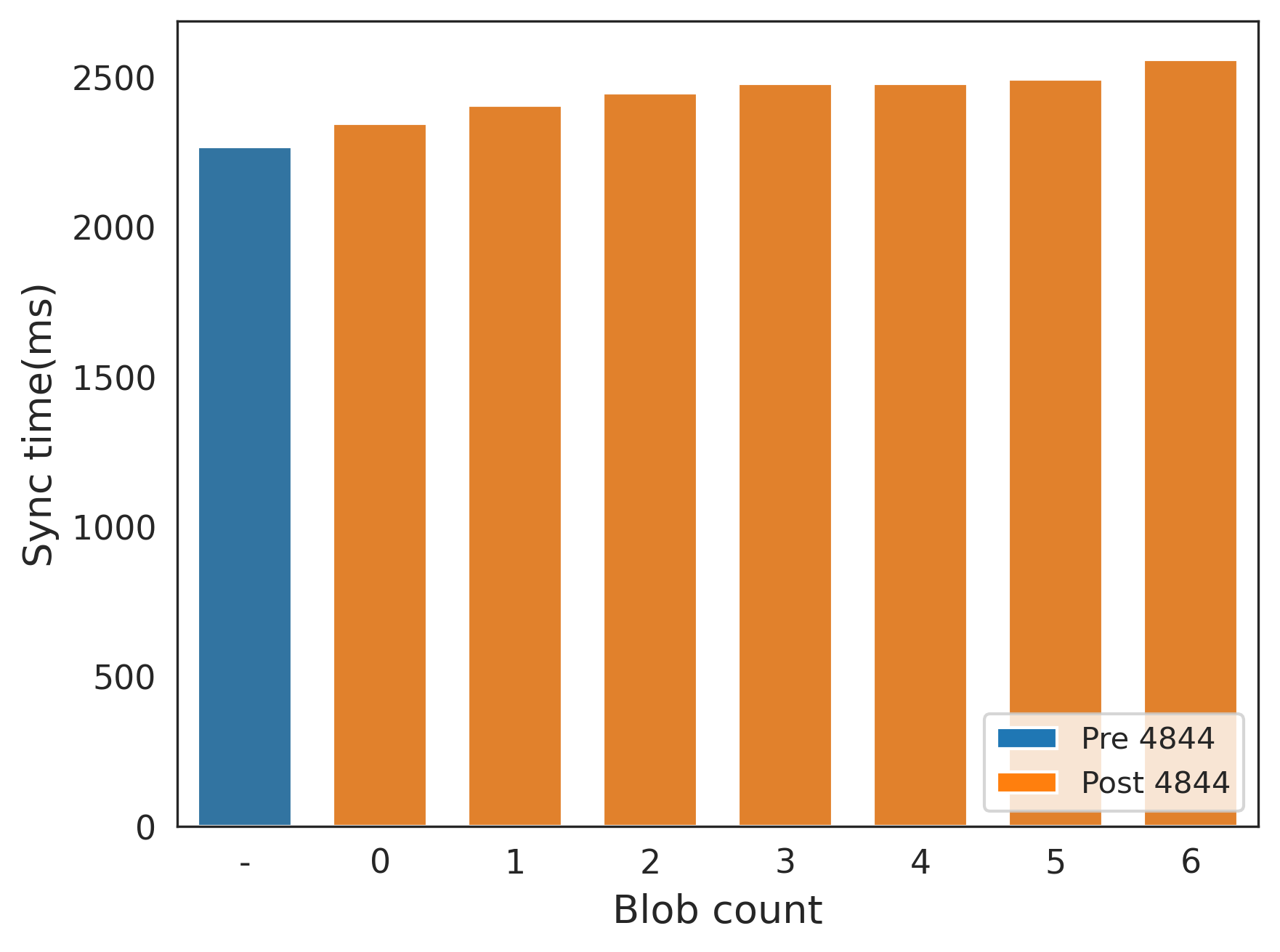}
        \caption{Sync time variation in relation to the number of blobs}
        \label{fig:4_1_sync_time_blob}
    \end{subfigure}
    \caption{Impact of EIP-4844 on sync times, illustrating an increase correlated with the number of blobs}
    \label{fig:sync_time}
    \Description{Impact of EIP-4844 on sync time. Sync time has increased after the implementation of EIP-4844 and rises with the increasing number of blobs.}
\end{figure}

The average sync time rose by approximately 140.065ms, from 2267.436ms to 2407.501ms. Figure \ref{fig:4_1_sync_time_blob} demonstrates that the increase in sync time correlates with the number of blobs per slot. Interestingly, slots without blobs also experienced an increase in sync time by about 77.967ms. Determining the precise cause of this increase is challenging; it could stem from operations which run fixed number of times to support EIP-4844, such as blob gas calculations, or possibly from other minor updates included in the Dencun upgrade. For our analysis, we have excluded the delays associated with slots without blobs, offering a conservative estimate of the impact of EIP-4844.

Therefore, the minimal increase attributable to EIP-4844 can be conservatively estimated at 62.098ms, accounting for only half of the observed increments in sync time and fork rates. However, it is vital to dissect which components of sync time EIP-4844 impacts. Such an analysis not only clarifies the direct impacts but also enriches future research on blob-related technologies like Danksharding\cite{dankrad2024sharding}. Comprehensive understanding of these dynamics is crucial for the refinement and development of forthcoming protocols. 

We have conducted detailed examinations of three specific components of sync time—receive time, CSP time, and DA time—as depicted in Figure \ref{fig:client_workflow}. Our findings reveal that the most significant impact of EIP-4844 was on receive time, contributing approximately 56.102ms to the overall increase in sync time. In contrast, DA time, newly added, did not show as significant an impact as expected, and CSP time appeared unaffected by EIP-4844, according to our conservative analysis.

\subsubsection{Receive time} 


Receive time is a critical component of sync time, particularly due to its influence on validator. According to Ethereum consensus specifications \cite{ethereum_consensus_specs}, validators are expected to attest to the current slot only if they receive a valid block within the first 4 seconds. Additionally, the proposer boost mechanism, introduced to mitigate balancing attacks \cite{neu2021ebb}, grants an extra 40\% of votes to slots arriving within this time frame in the fork choice rule \cite{proposer_boost}. Slots that fail to meet this attestation deadline do not receive the proposer boost, increasing their susceptibility to balancing attacks.

\begin{figure}
    \centering
    \begin{subfigure}{\linewidth}
        \centering
        \includegraphics[width=\linewidth]{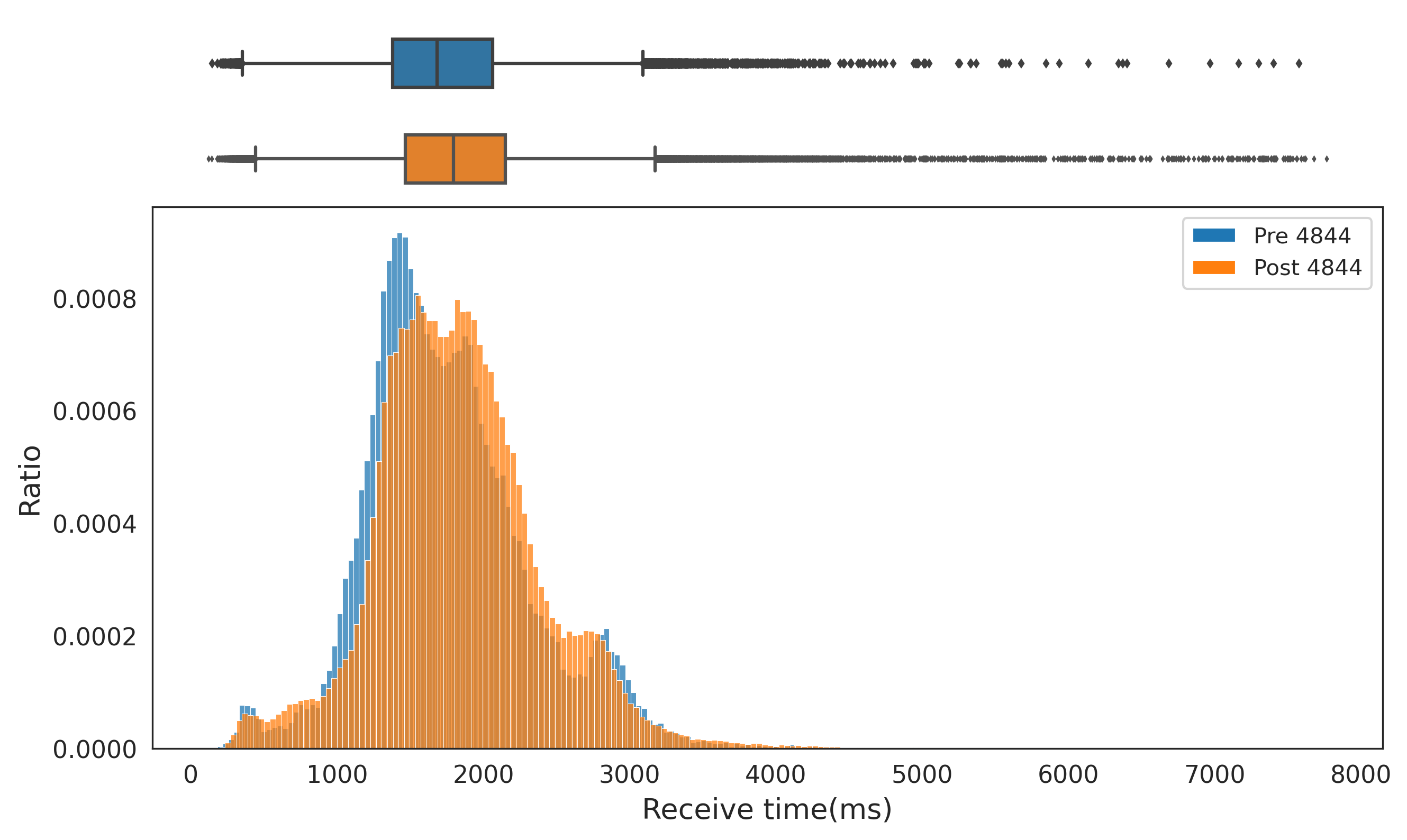}
        \caption{Distribution of receive time before and after EIP-4844}
        \label{fig:4_1_receive_time_hist}
    \end{subfigure}
    \hfill 
    \begin{subfigure}{0.45\linewidth}
        \centering
        \includegraphics[width=\linewidth]{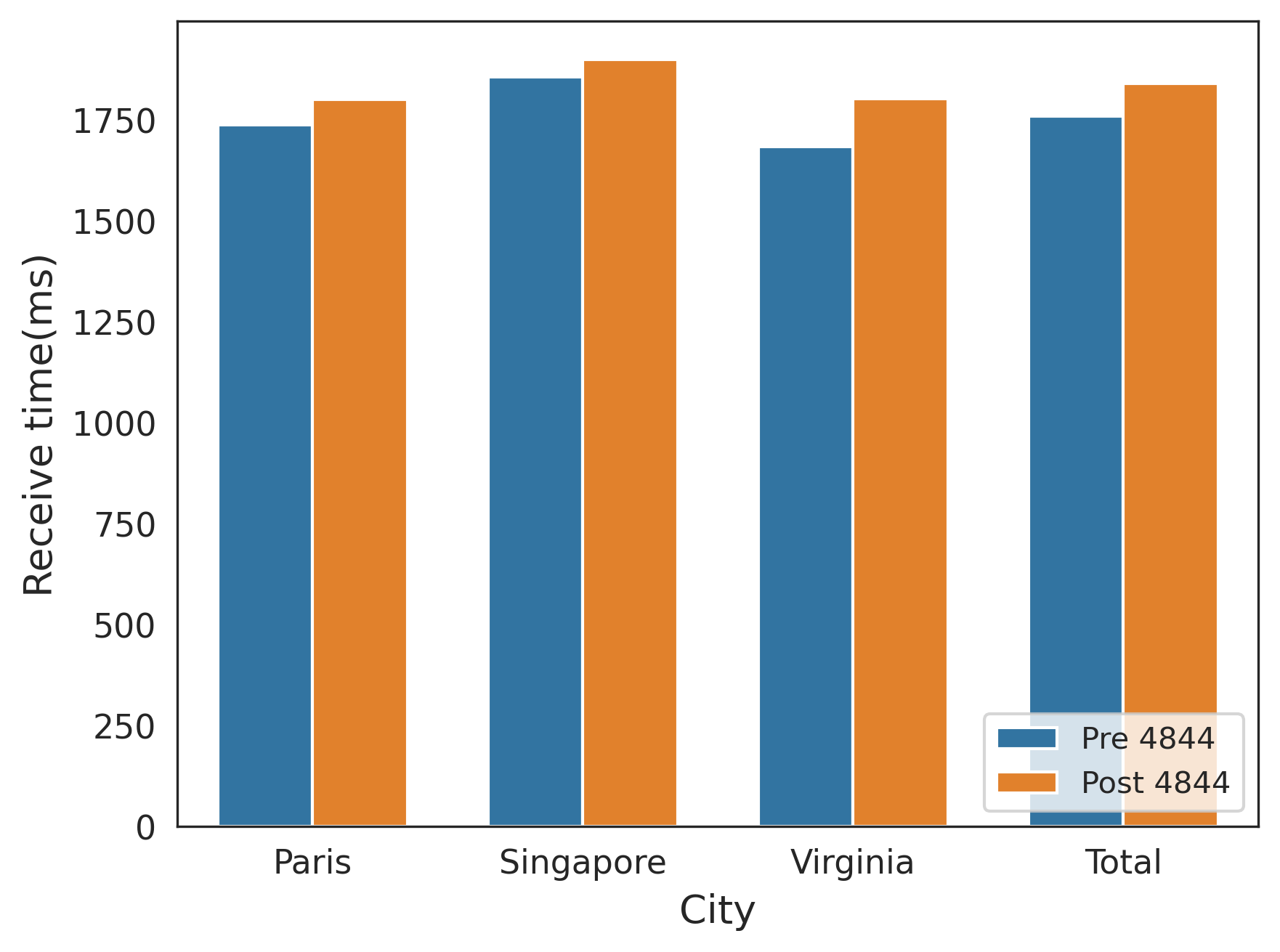}
        \caption{Average receive time by city and overall}
        \label{fig:4_1_receive_time_city}
    \end{subfigure}
    \hfill 
    \begin{subfigure}{0.45\linewidth}
        \centering
        \includegraphics[width=\linewidth]{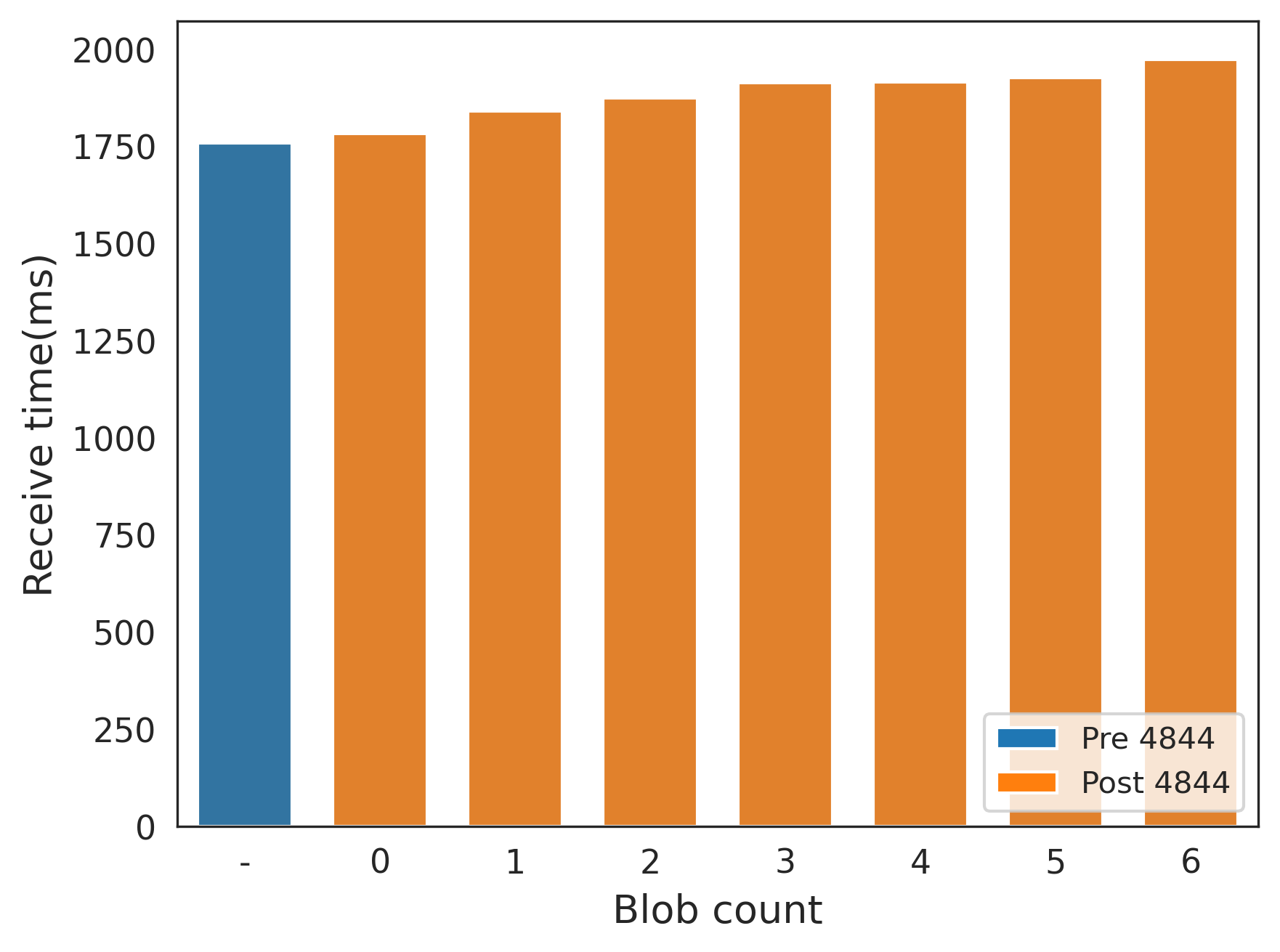}
        \caption{Average receive time by the number of blobs}
        \label{fig:4_1_receive_time_blob}
        \Description{}
    \end{subfigure}
    \caption{Impact of EIP-4844 on receive time. The increase in receive time, particularly in slots with higher blob counts, indicates heightened risk for reorganization attacks.}
    \label{fig:receive_time}
\end{figure}

Figure \ref{fig:4_1_receive_time_city} shows the average receive time by city and overall, with a noticeable increase from 1759.066 ms to 1840.032 ms—an elevation of approximately 80.966 ms. Figure \ref{fig:4_1_receive_time_blob} further demonstrates that receive time is correlated with the number of blobs; slots without blobs post-EIP-4844 have similar receive times to those before the implementation, while increases are proportional to blob counts.

After subtracting the 24.864 ms increase in receive time of slots without blobs, we estimate that the minimum impact of EIP-4844 is 56.102 ms. This increment comprises the majority of the observed rise in sync time, suggesting that EIP-4844 predominantly affect the receive time. This impact likely stems from the additional responsibilities of proposers to handle blob data from execution clients and generate KZG commitments for blobs.

Figure \ref{fig:4_1_receive_time_hist} illustrates the distribution of receive times before and after the implementation of EIP-4844, showing a slight overall shift towards longer times. Notably, the proportion of slots arriving after 4000ms increased from approximately 0.18\% to 0.415\% after EIP-4844—a more than two-fold increase. These data suggest that the likelihood of slots falling victim to reorg attacks has risen following the introduction of EIP-4844, as slots arriving later are more vulnerable to reorg attacks.

\subsubsection{Chain service process time}
Chain Service Process (CSP) time is a crucial component affecting the synchronization time of a slot. With the implementation of EIP-4844, new procedures were added, including the validation and storage of versioned hashes of KZG commitments within the execution engine. The increase in CSP time—from 482.565ms to 536.043ms, or about 53.478ms, while the average CSP time for blob-free slots was 52.779. This observation, specified in Appendix Figure \ref{fig:csp_time_blob}, suggests that CSP time does not correlate with the number of blobs. Therefore, we conclude that the overall rise in CSP time might not directly result from EIP-4844.

\subsubsection{Data availability time}
DA time, a new delay introduced by EIP-4844, represents the duration spent waiting for blobs to arrive after slot execution and beacon state updates are completed. Analyzing metric is particularly important as it is newly introduced by EIP-4844.

\begin{figure}
    \centering
    \begin{subfigure}{0.45\linewidth}
        \centering
      \includegraphics[width=\linewidth]{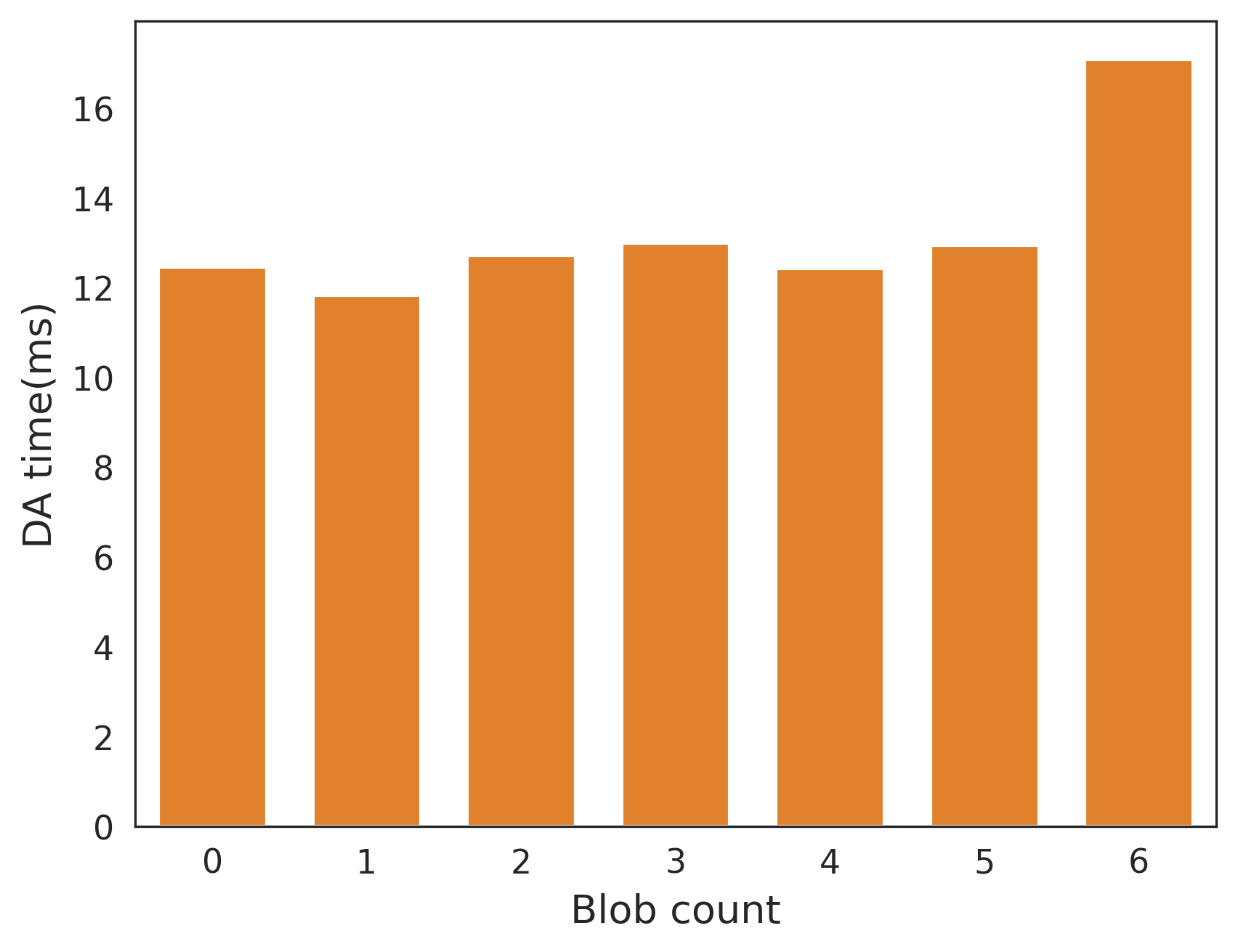}
      \caption{Average DA time by the number of blobs, showing minimal delay increases even with additional blobs}
      \label{fig:4_1_da_time_blob}
    \end{subfigure}
    \hfill 
    \begin{subfigure}{0.45\linewidth}
        \centering
        \includegraphics[width=\linewidth]{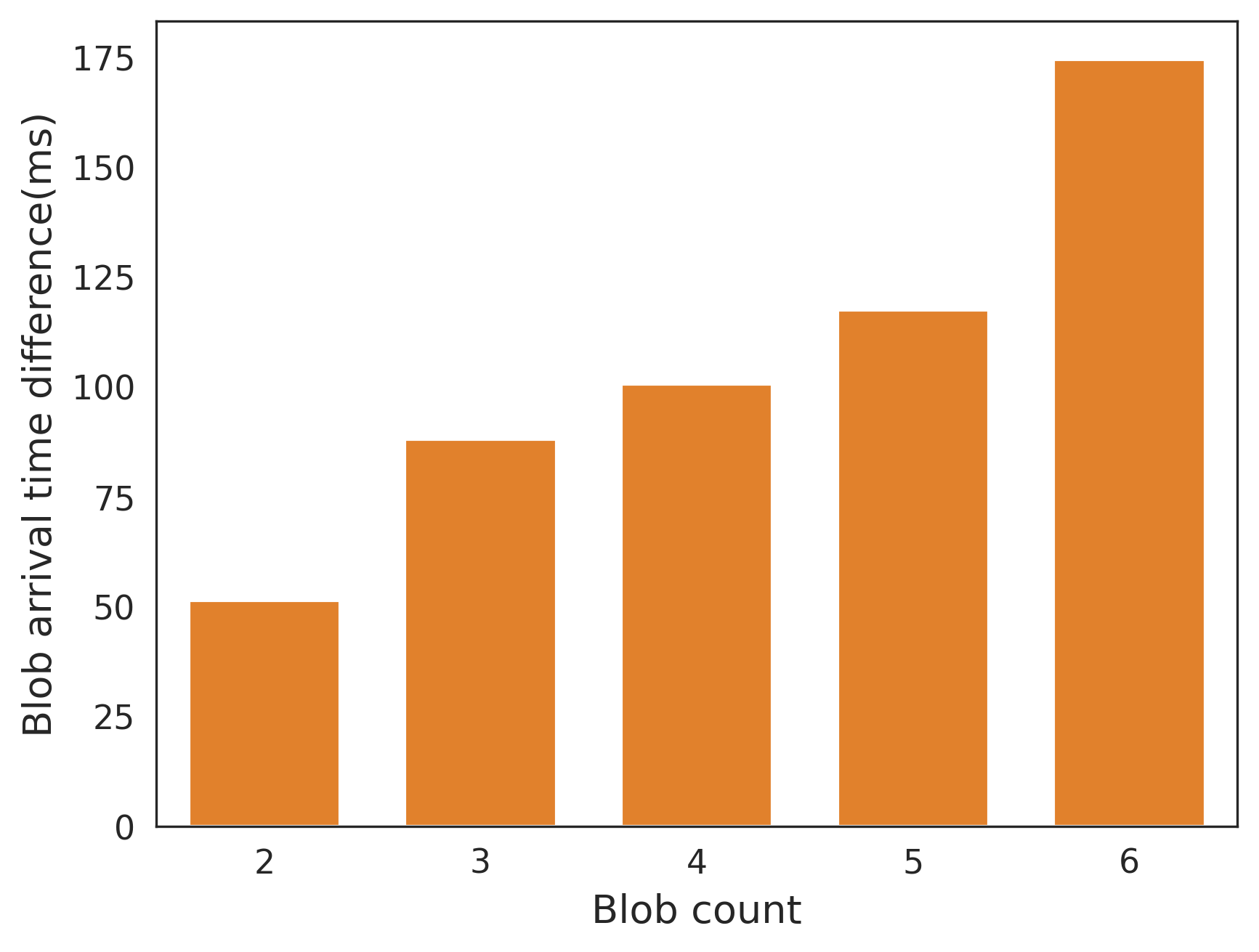}
        \caption{Difference in arrival times between the first and last blob, indicating increased delay with more blobs}
        \label{fig:4_1_first_last_blob}
    \end{subfigure}
    \caption{Impact of EIP-4844 on DA times}
    \label{fig:da_time}
    \Description{}
\end{figure}

Figure \ref{fig:4_1_first_last_blob} shows that as the number of blobs increases, the time difference between the arrival of the first and last blobs also increases, hinting at potential delays. However, Figure \ref{fig:4_1_da_time_blob} shows that the actual waiting time for blob arrival is minimal, averaging only about 13.417 ms overall, and just 0.956 ms bigger than slots without any blobs. Even in slots with six blobs, the delay remains below 17.5 ms, dropping to 4.619 ms when excluding slots without blobs, which is relatively minor.

This apparent contradiction is resolved by the independent propagation of blobs from the slot. It allows the consensus layer (CL) to proceed with tasks related to the slot, such as block execution and state transitions, while waiting for blobs, which may arrive more slowly. Remarkably, in 35.519\% of cases, the last blob arrived even before the slot itself. This strategy of separating blob propagation from slot propagation effectively minimizes any delays caused by blob transmission. Therefore, despite some impact on consensus security, our findings suggest that the effect of waiting times for blobs is not substantial.

\begin{figure}[ht]
  \centering
\end{figure}

\begin{table*}[ht]
\centering
\caption{Changes of Ethereum usage by rollups before and after EIP-4844}
\label{tab:4_2_rollup_metrics}
\begin{tabular}{@{}lcccccccccccc@{}}
\toprule
 & \multicolumn{3}{c}{Total Data Size (MiB)}  &  \multicolumn{3}{c}{Total Fees Paid (ETH)} & \multicolumn{3}{c}{Price For 1MiB (ETH)} & \multicolumn{3}{c}{Total Gas Used (Gas units)} \\ 
\cmidrule(lr){2-4} \cmidrule(lr){5-7} \cmidrule(lr){8-10} \cmidrule(lr){11-13}
Rollup Type & Before & After & Change & Before & After & Change & Before & After & Change & Before & After & Change \\
\midrule
All Rollups & 0.084 & 0.183 & +116.83\% & 0.075 & 0.021 & -71.38\% & 1.304 & 0.231 & -82.32\% & 1.725M & 0.784M & -54.53\% \\
Optimistic Rollups & 0.049 & 0. 111 & +127.4\% & 0.047 & 0.007 & -84.89\% & 0.905 & 0.239 & -73.55\% & 0.878M & 0.169M & -80.74\% \\
ZK Rollups         & 0.035 & 0.072  & +102.22\% & 0.028 & 0.014 &  -48.75\% & 1.516 & 0.280 & -81.53\% & 0.611M & 0.461M & -24.55\% \\
\bottomrule
\end{tabular}
\end{table*}

\subsection{Ethereum usage}
The effectiveness of EIP-4844 hinges on its primary objective: to enhance the efficiency of Ethereum as a DA layer for rollups. We performed a detailed examination of changes in Ethereum usage by the top 10 rollups following the implementation of EIP-4844. Our key findings include:

\begin{enumerate}
    \item The total data size posted by rollups on Ethereum as a DA layer has increased markedly, while the calldata size has decreased substantially. This shift is particularly pronounced in optimistic rollups compared to zk rollups.

    \item The total fees paid by rollups for DA have significantly decreased, alongside a considerable reduction in the cost per MiB of DA.

    \item There has been a significant reduction in the total gas used by rollups, primarily driven by decreased gas consumption in optimistic rollups.
\end{enumerate}

\subsubsection{Total amount of data posted.}
This section examines the total data size posted on Ethereum, focusing on its role as a DA layer. Our data include only transactions that commit batches or update states, specified in Appendix Table \ref{tab:transaction_classification}.

\begin{figure}
    \centering
    \begin{subfigure}{0.7\linewidth}
        \centering
        \includegraphics[width=\linewidth]{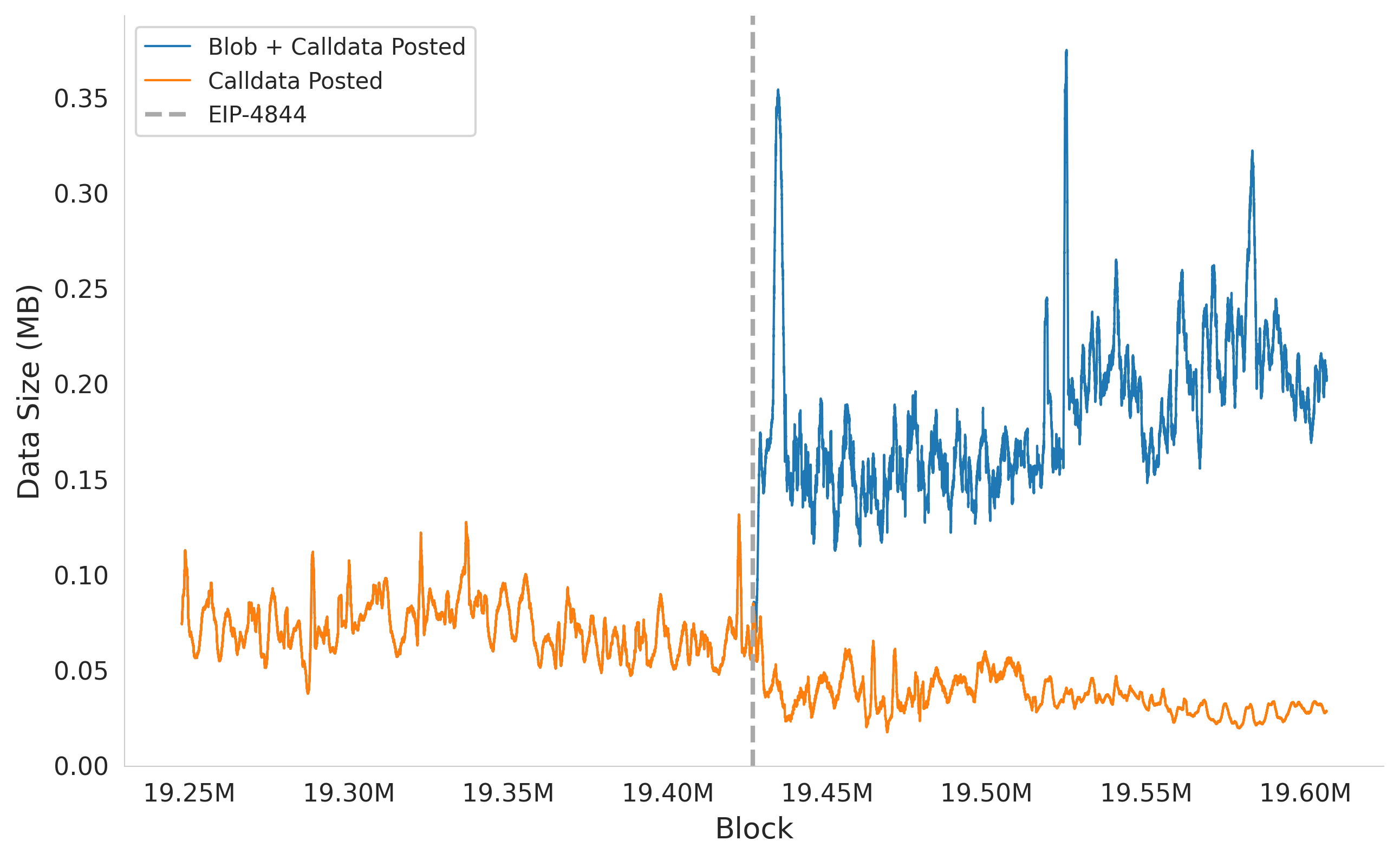}
        \caption{All rollups}  
        \label{fig:4_2_datasize_all}
    \end{subfigure}
    \\
    \begin{subfigure}{0.45\linewidth}
        \centering
        \includegraphics[width=\linewidth]{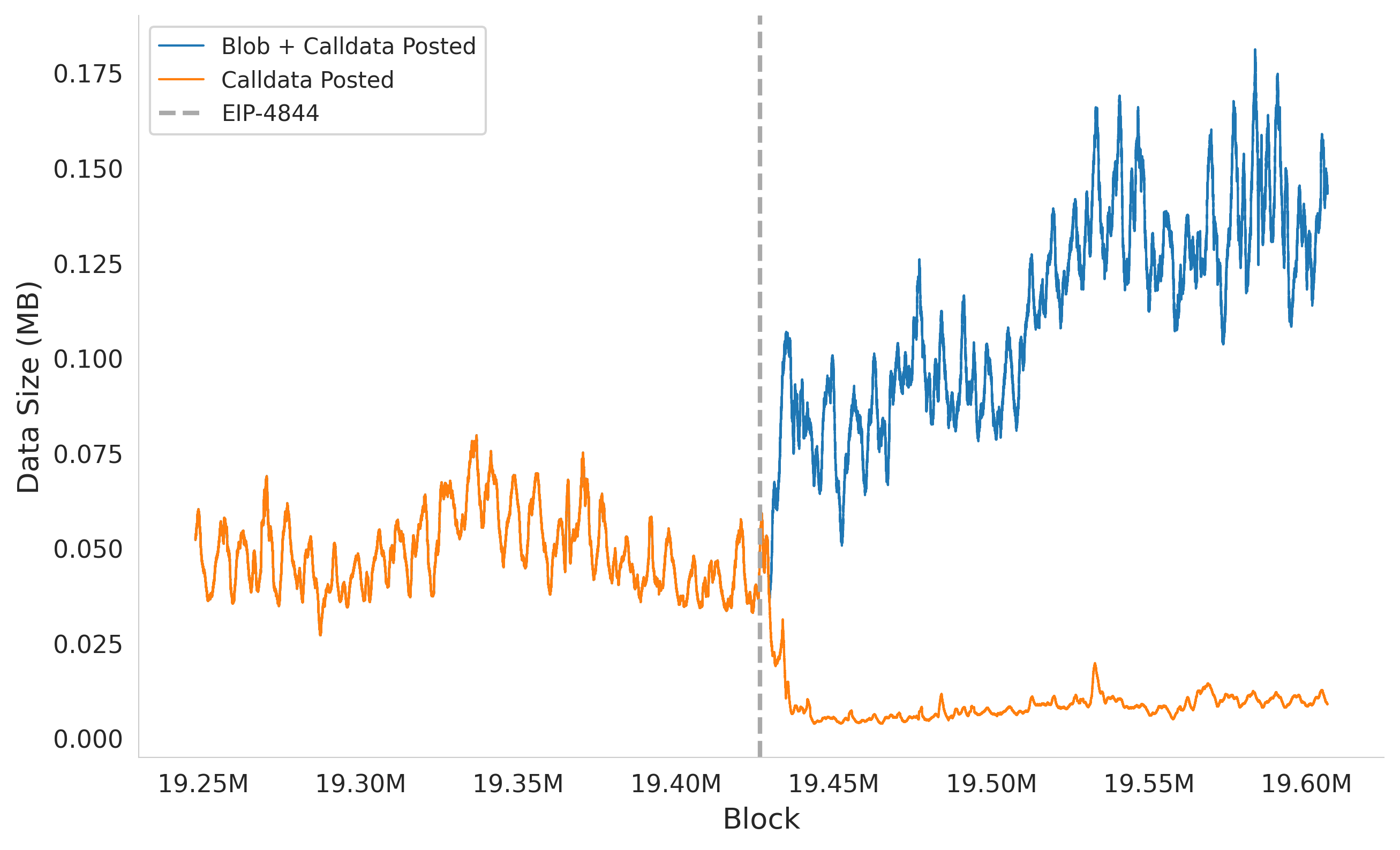}
        \caption{Optimistic rollups}  
        \label{fig:4_2_datasize_optimistic}
    \end{subfigure}
    \hfill 
    \begin{subfigure}{0.45\linewidth}
        \centering
        \includegraphics[width=\linewidth]{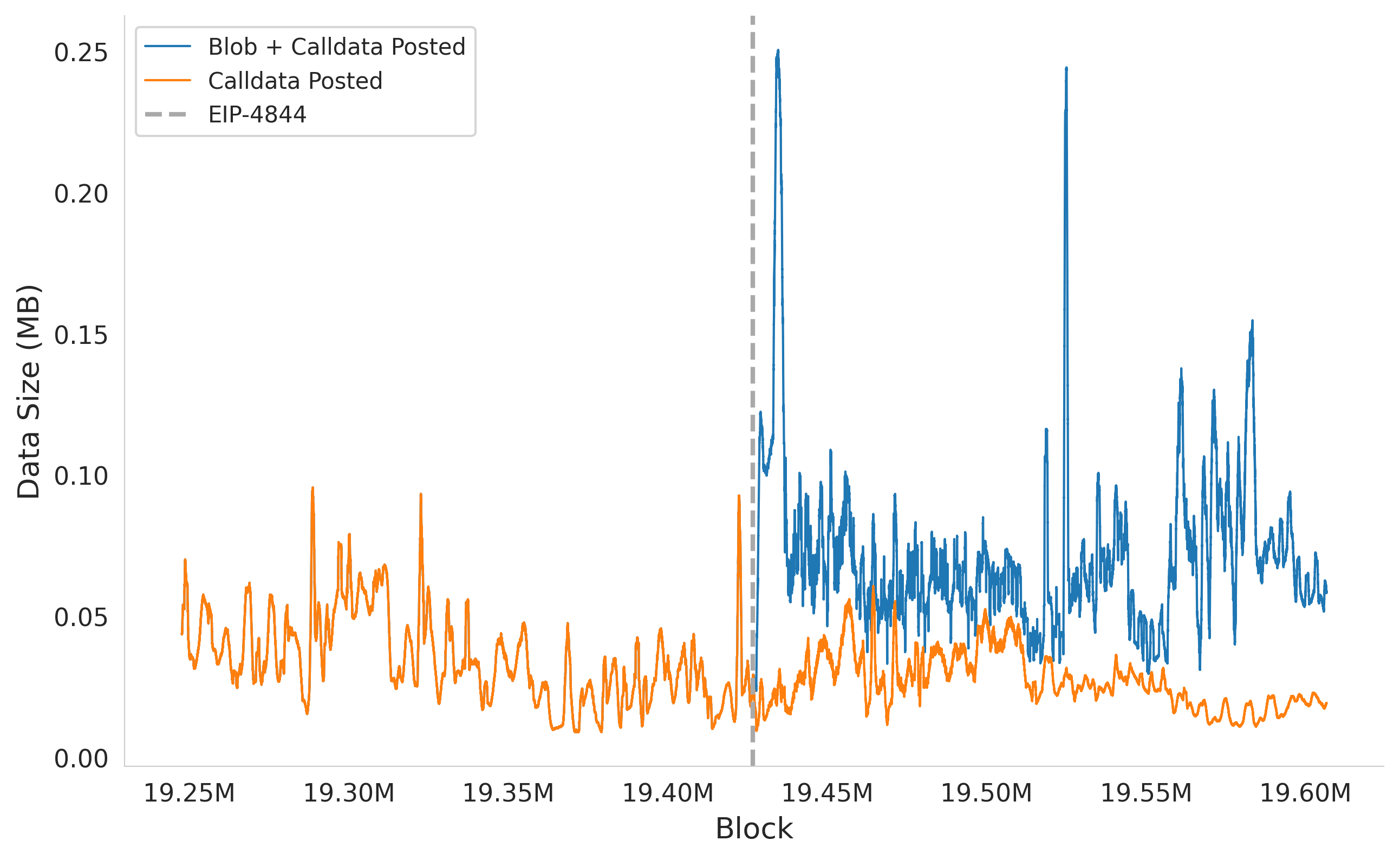}
        \caption{ZK rollups}  
        \label{fig:4_2_datasize_zk}
    \end{subfigure}
    \caption{Total data size and calldata size posted by top 10 rollups on Ethereum as a DA layer}
    \label{fig:4_2_datasizes}
\end{figure}

Figure \ref{fig:4_2_datasize_all} demonstrates a significant increase in the total data size posted, with the average data size per block rising from 0.084 MiB to 0.183 MiB, representing a 116.8\% increase. Conversely, the size of calldata posted decreased by 56.8\%, reducing to 0.036 MiB.

Figures \ref{fig:4_2_datasize_optimistic} and \ref{fig:4_2_datasize_zk} detail the changes in optimistic and zk rollups, respectively. Both types of rollups recorded substantial increases in total data size, each by over 100\%. Optimistic rollups, in particular, demonstrated a significant reduction in calldata size by 81\%, indicating a notable shift towards using blobs. Zk rollups showed a smaller decrease in calldata usage, from 0.035 MiB to 0.027 MiB. 

These findings evidence that EIP-4844 has effectively encouraged rollups to make greater use of Ethereum's DA layer capabilities. This is particularly evident in optimistic rollups, where the majority of transactions serve DA functions, possibly deriving more benefit from the protocol upgrade than zk rollups.

\subsubsection{Total amount of fee paid.}
Exploring the economic impact, we reviewed the total fees paid by rollups for utilizing Ethereum as a DA layer. As indicated in the Table \ref{tab:4_2_rollup_metrics}, the average fee paid by rollups per block experienced a substantial decrease following the implementation of EIP-4844. Prior to EIP-4844, rollups paid an average of 0.075 ETH per block. This fee reduced to 0.021 ETH per block after the policy change, marking a 72\% decrease.

The fee reduction was particularly notable among optimistic rollups. Before EIP-4844, these rollups paid approximately 0.047 ETH per block, constituting about 63\% of the total fee. Post EIP-4844, their contribution dropped dramatically to 0.007 ETH per block, which represents a reduction to 33\% of the total fees paid.

\subsubsection{The price of posting 1MiB}
\begin{figure}[ht]
  \centering
  \includegraphics[width=0.8\linewidth]{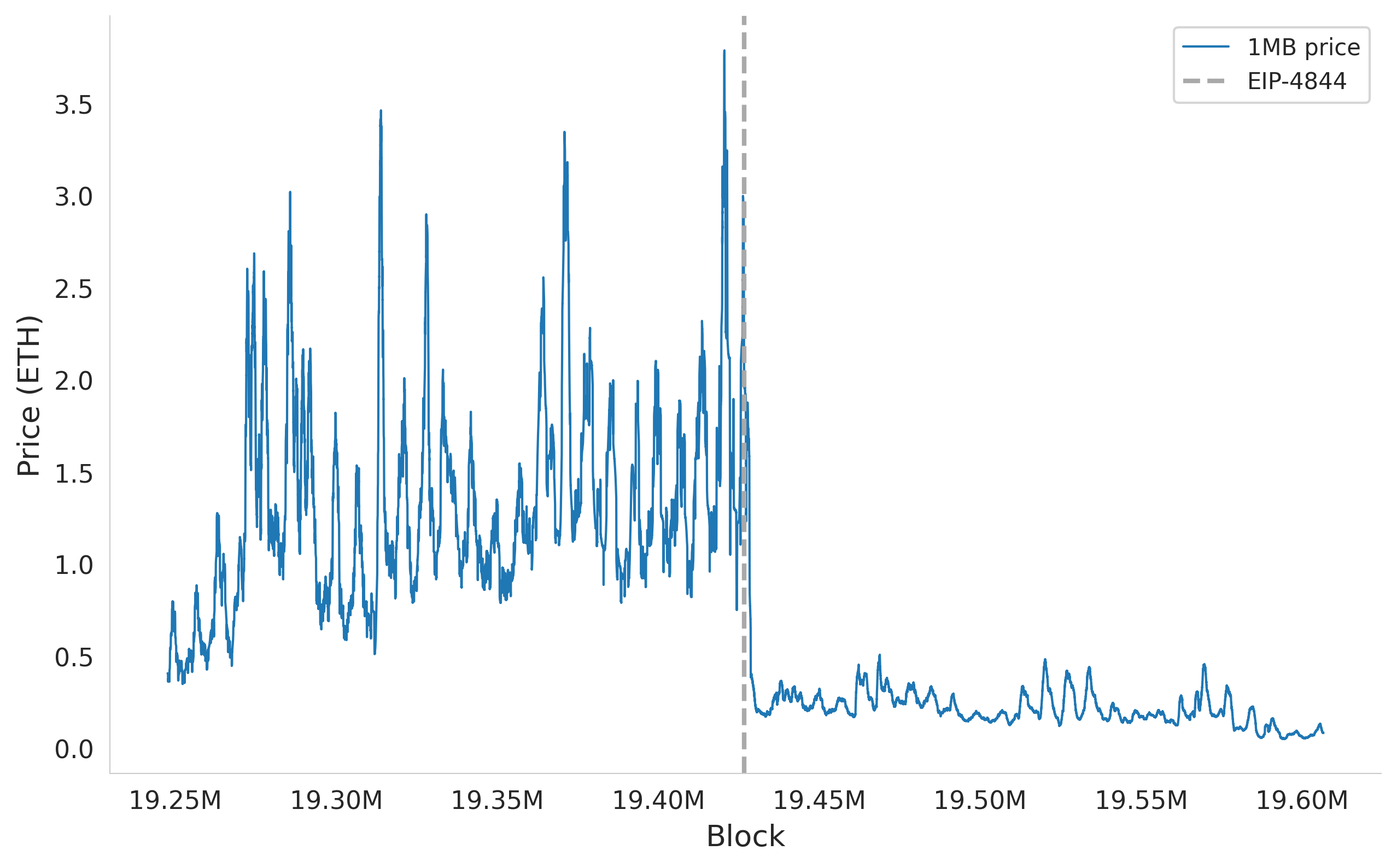}
  \caption{1MiB price of Ethereum as a DA layer}
  \Description{}
  \label{fig:4_2_1mb_price}
\end{figure}

In assessing the cost effectiveness of EIP-4844, we analyzed the price per MiB of data posted on Ethereum. As illustrated in Figure \ref{fig:4_2_1mb_price}, the price per 1MiB before the implementation of EIP-4844 was approximately 1.304 ETH per block. Following the changes introduced by EIP-4844, this price significantly decreased to 0.231 ETH per block, representing an 82\% reduction.

This substantial decrease underscores the effectiveness of EIP-4844 in reducing the costs associated with DA on Ethereum, thereby lessening the economic burden on rollups that rely on Ethereum security for data posting\cite{mandal2023investigating}. This reduction in cost is pivotal for enhancing the scalability and economic feasibility of using Ethereum as a DA layer.

\subsubsection{Gas used}
To gauge the overall impact of EIP-4844 on gas consumption, we looked into the total gas used by rollups, including transactions not related to DA. Table \ref{tab:4_2_rollup_metrics} illustrates the changes in gas usage. Specifically, the average gas used per block significantly decreased by approximately 0.95 million, from 1.73 million to 0.78 million. 

This reduction in gas usage was particularly pronounced in optimistic rollups, where gas consumption dropped by 0.71 million, from 0.88 million to 0.17 million. In contrast, ZK rollups exhibited a smaller decrease, with gas usage reducing by 0.23 million, from 0.85 million to 0.62 million. This modest decrease suggests that transactions involving zero-knowledge proofs, which cannot easily be converted to blob transactions, still use nearly the same amount of gas.

\subsection{Rollup transactions}
Understanding user interactions within the blockchain ecosystem is crucial, especially in response to changes introduced by EIP-4844. This section examines the impact of the upgrade on the number of rollup transactions that are posted on Ethereum, which reflects the level of rollup activity. Additionally, we consider user delay— the time it takes for transactions to be securely settled in Ethereum— as a crucial aspect of rollup security.

We analyzed six major rollups, comprising three optimistic rollups (Arbitrum One, Optimism, Base) and three ZK rollups (Starknet, zkSync Era, Linea), with a focus on transaction volume and user delays. Our principal observations are:

\begin{enumerate}
    \item All six rollups showed a marked increase in transaction volume, with Base experiencing a particularly significant rise, more than tripling its previous count.
    \item User delay, the time lag between the creation of rollup blocks and their posting on Ethereum, increased notably in four rollups. Conversely, Arbitrum and zkSync Era achieved significant improvements, witnessing reductions in their delay times. 
    \item The variability in user delay times also saw a notable increase, except in Arbitrum, where it remained more stable.
\end{enumerate}

\begin{table}[htbp]
\centering
\caption{Comparison of rollup transaction numbers and user delay(standard deviation) before and after EIP-4844}
\label{tab:4_3_rollup_results_summary}
\begin{tabular}{@{}lcccccc@{}}
\toprule
 & \multicolumn{3}{c}{\# of transactions} & \multicolumn{3}{c}{User Delay (s)} \\
\cmidrule(lr){2-4} \cmidrule(lr){5-7}
Rollup & Before & After & Change & Before & After & Change\\
\midrule
Arbitrum & 111.2 & 194.7 & \textbf{+75\%} & 519.9 & 197.5 & \textbf{-62\%} \\
& &  & & (197.5) & (100.8) & \\
Optimism & 53.8 & 91.8 & \textbf{+71\%} & 55.6 & 224.6 & \textbf{+304\%} \\
& &  & & (23) & (112.9) & \\
Base & 56.8 & 183.7 & \textbf{+224\%} & 59.4 & 161.1 & \textbf{+171\%} \\
&  &  & & (25.6) & (35.3) & \\
Starknet & 18 & 31 & \textbf{+72.5\%} & 17,156 & 27,675 & \textbf{+61.3\%} \\
& & & & (4467) & (9504) & \\
zkSync & 149.4 & 174.9 & \textbf{+17.1\%} & 261.4 & 154.9 & \textbf{-40.8\%} \\
& & & & (115.5) & (152.3) & \\
Linea & 127.5 & 187.5 & \textbf{+46.7\%} & 21,354 & 30,404 & \textbf{+42.4\%} \\
&  & & & (24489) & (25901) & \\
\bottomrule
\end{tabular}
\end{table}

\subsubsection{Number of rollup transactions}
\begin{figure}
    \centering
    \begin{subfigure}{0.45\linewidth}
        \centering
        \includegraphics[width=\linewidth]{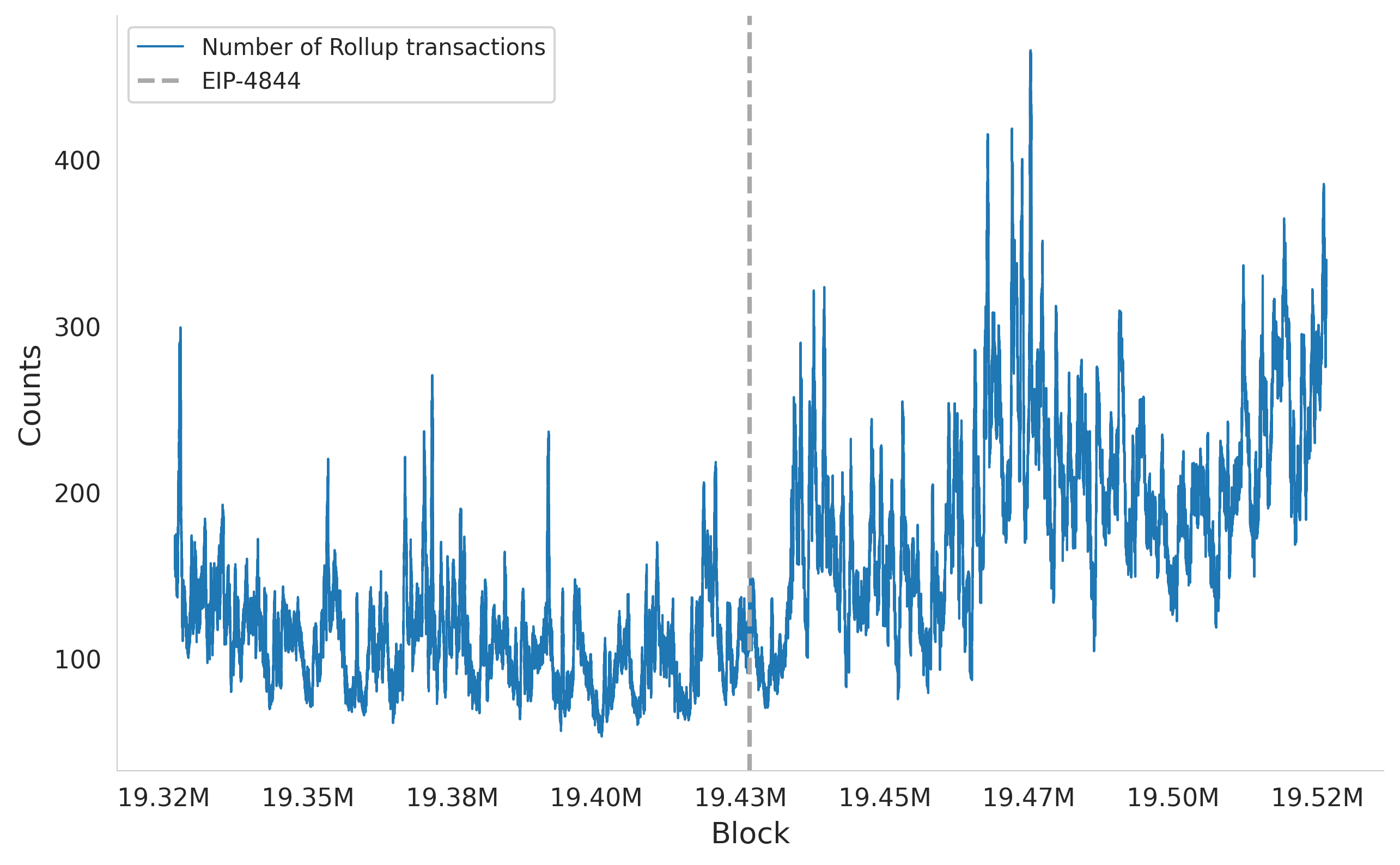}
        \caption{Arbitrum One}  
        \label{fig:4_3_arbitrum_tx}
    \end{subfigure}
    \hfill
    \begin{subfigure}{0.45\linewidth}
        \centering
        \includegraphics[width=\linewidth]{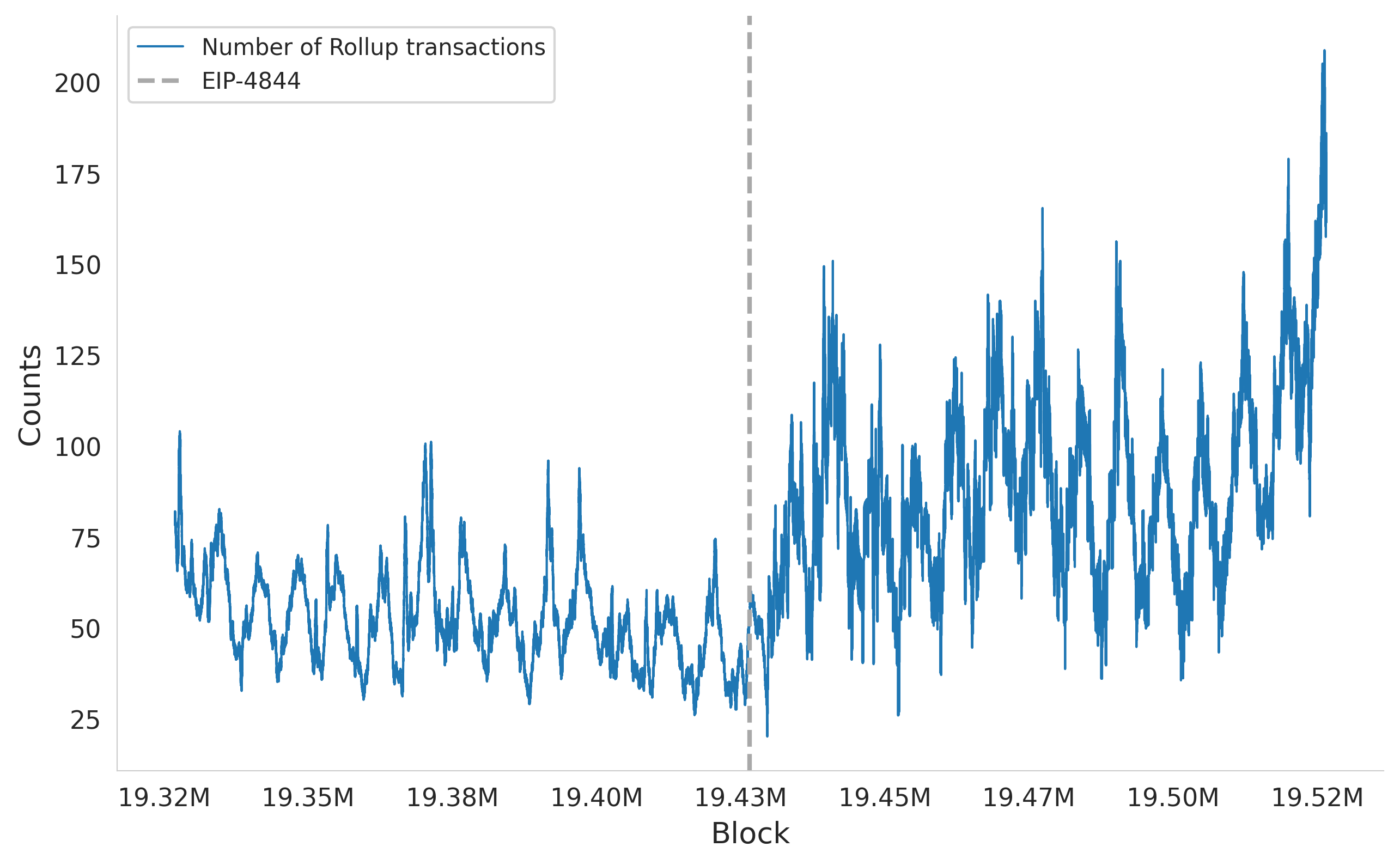}
        \caption{Optimism}  
        \label{fig:4_3_optimism_tx}
    \end{subfigure}
    \\
    \begin{subfigure}{0.45\linewidth}
        \centering
        \includegraphics[width=\linewidth]{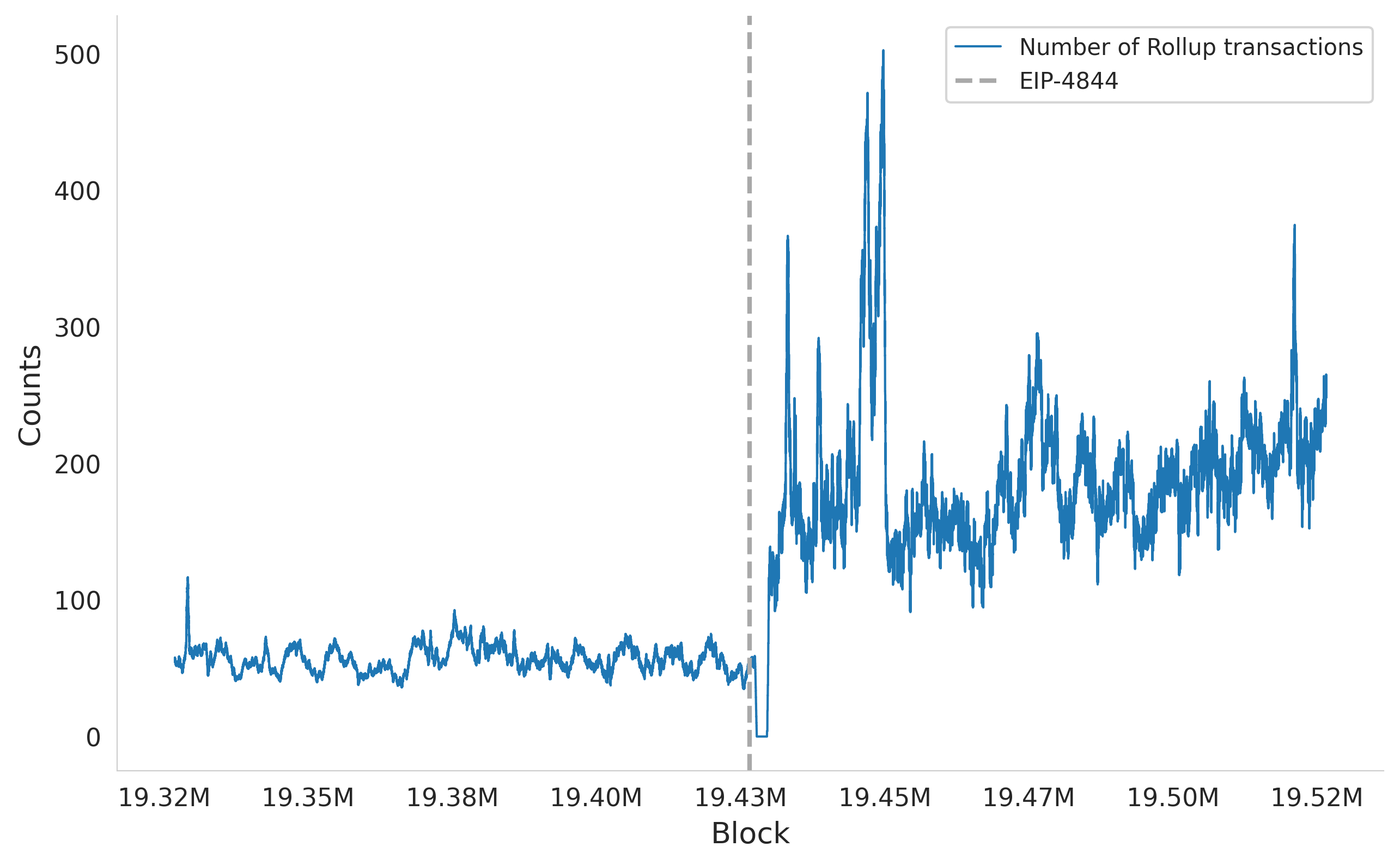}
        \caption{Base}  
        \label{fig:4_3_base_tx}
    \end{subfigure}
    \hfill
    \begin{subfigure}{0.45\linewidth}
        \centering
        \includegraphics[width=\linewidth]{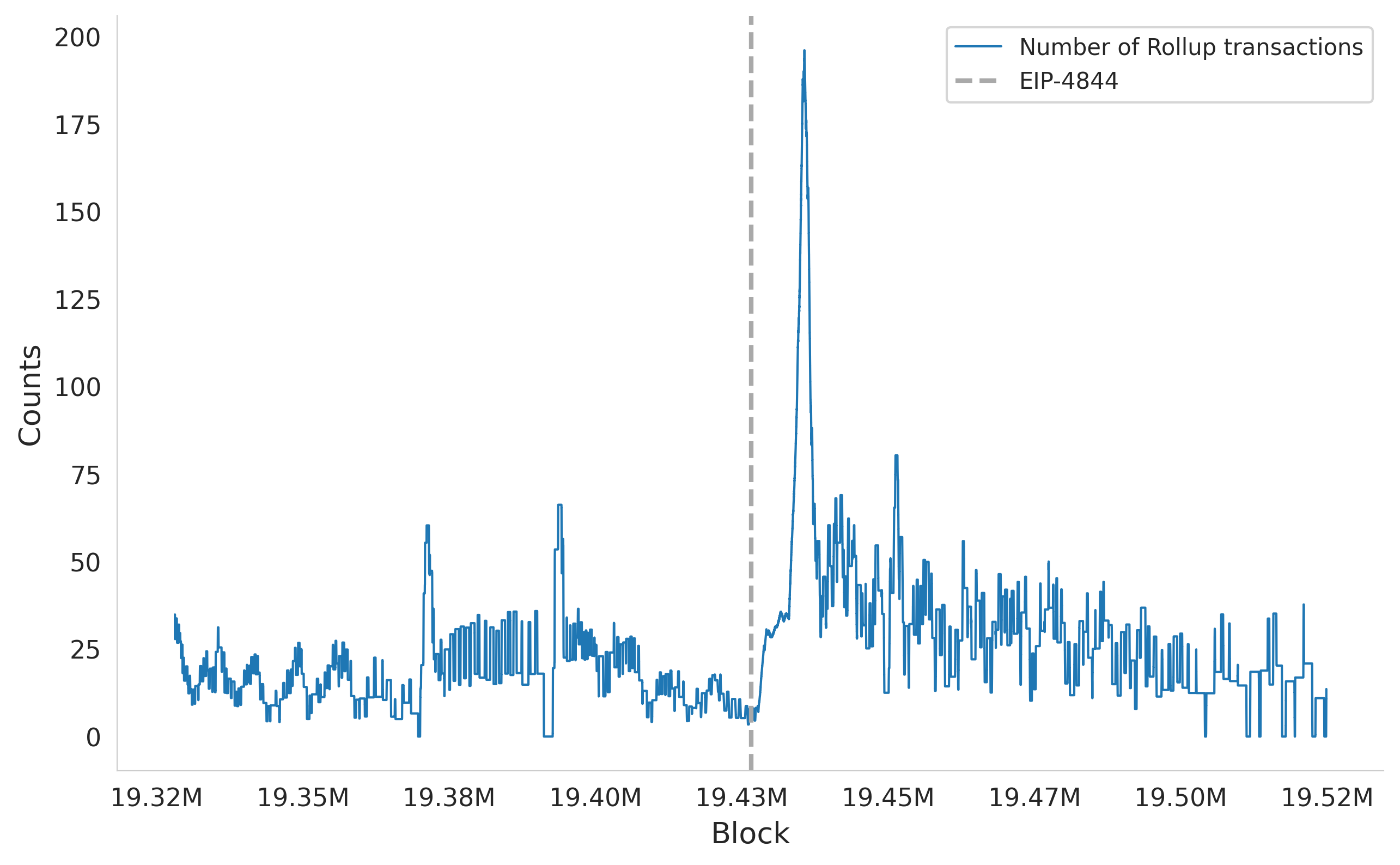}
        \caption{Starknet}  
        \label{fig:4_3_starknet_tx}
    \end{subfigure}
    \\
    \begin{subfigure}{0.45\linewidth}
        \centering
        \includegraphics[width=\linewidth]{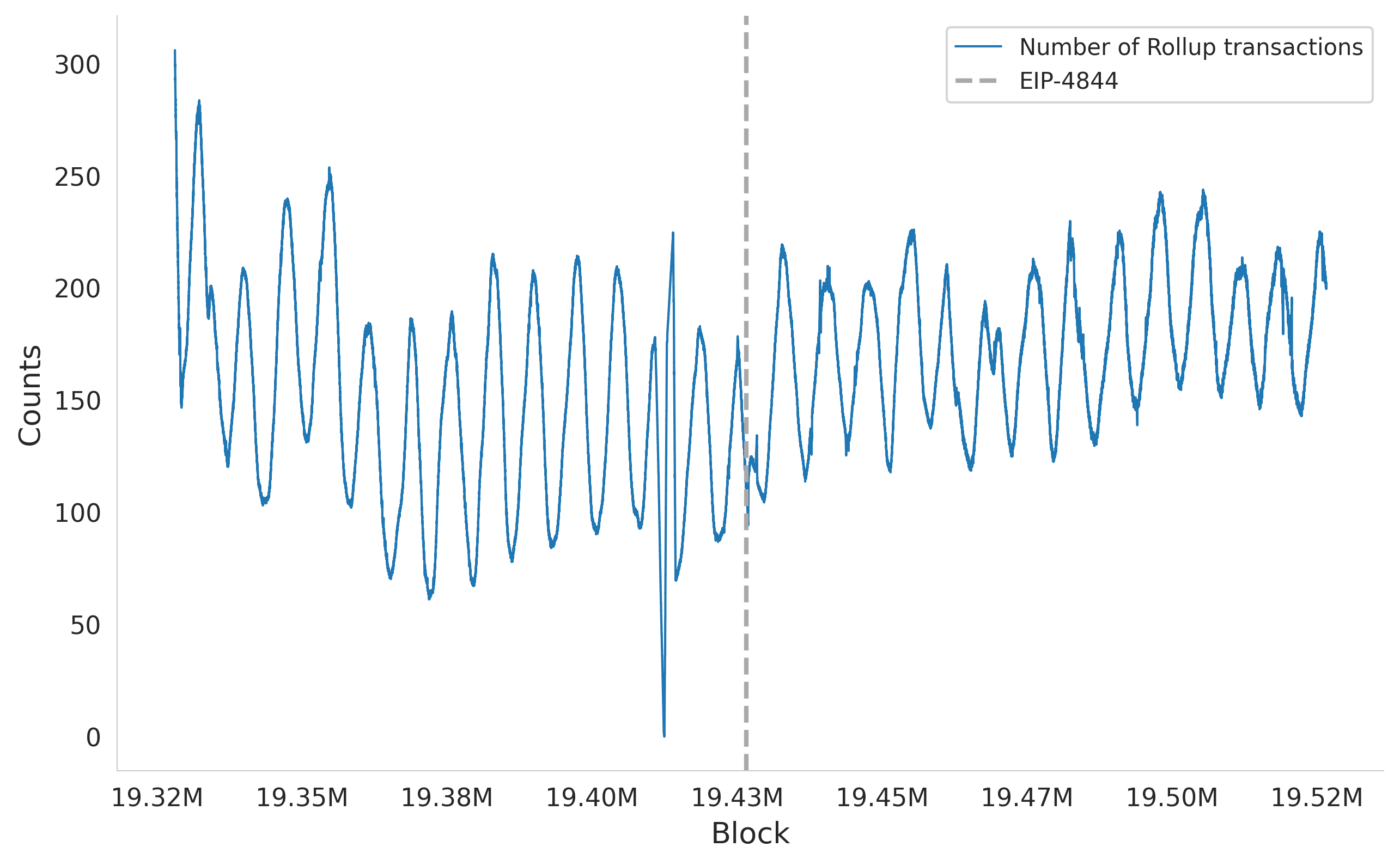}
        \caption{zkSync Era}  
        \label{fig:4_3_zksync_tx}
    \end{subfigure}
    \hfill
    \begin{subfigure}{0.45\linewidth}
        \centering
        \includegraphics[width=\linewidth]{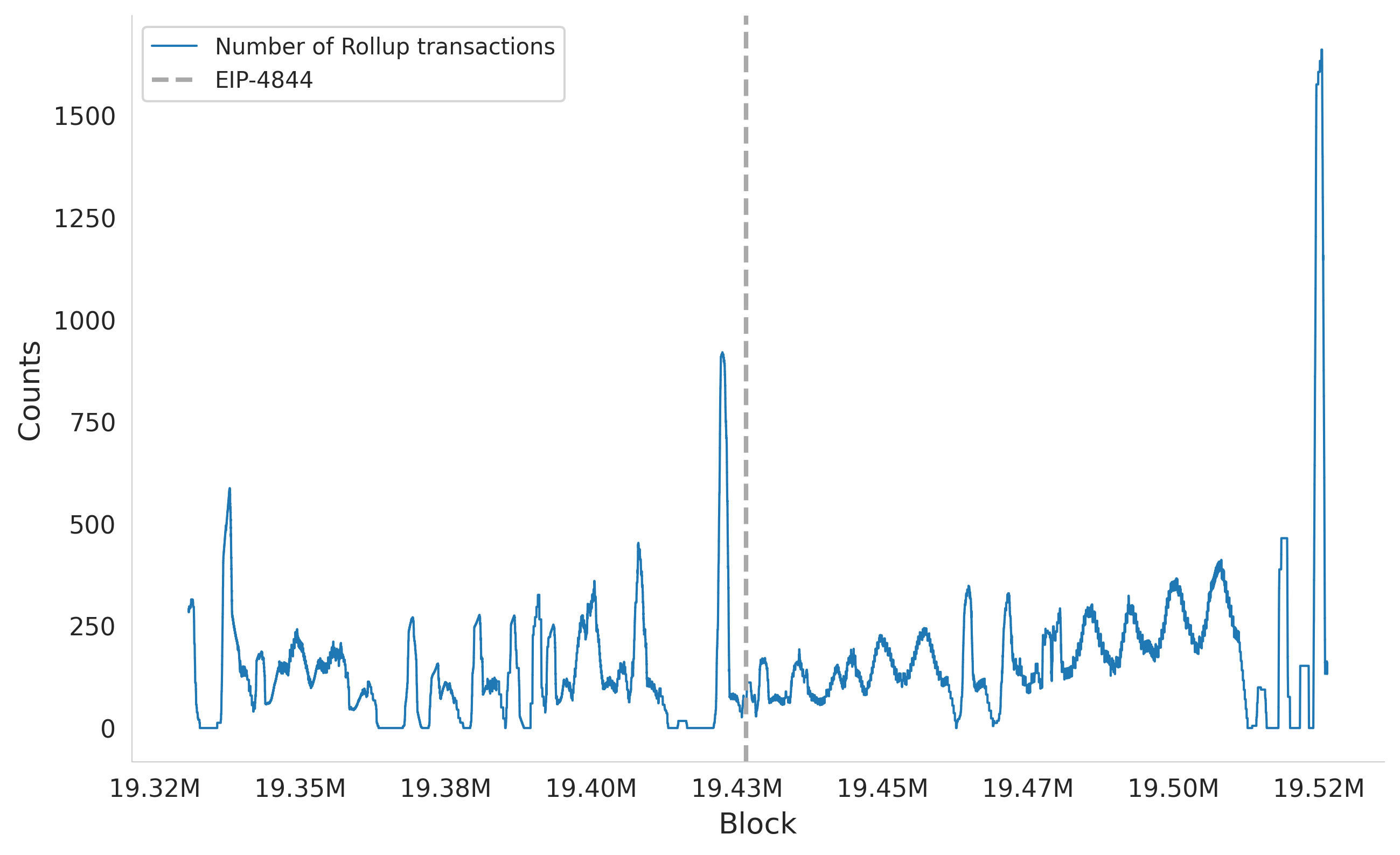}
        \caption{Linea}  
        \label{fig:4_3_linea_tx}
    \end{subfigure}
    \caption{Total number of rollup transactions that are posted on Ethereum}
    \label{fig:4_3_rollup_transactions_tx}
\end{figure}

To retrieve the number of transactions posted on Ethereum, we decoded the batch transactions on Ethereum sent by rollups. Table \ref{fig:4_3_rollup_transactions_tx} indicates a substantial increase across six rollups. Notably, Base experienced a significant rise of 224\%, while Arbitrum, Optimism, and Starknet each saw increases exceeding 70\%. This overall uptick suggests that the reduced fees resulting from EIP-4844 may have incentivized a greater volume of transactions on rollups.

However, attributing this increase solely to EIP-4844 would be premature without considering pre-existing growth trends. Rollups were already experiencing rapid expansion prior to the protocol's implementation, and the observed increases might partly reflect the ongoing market growth rather than the effects of EIP-4844 alone.

To rigorously assess the specific impact of EIP-4844, we employed a Regression Discontinuity Design (RDD) for each rollup. 

\[
\text{Number of Transactions}_i = \beta_0 + \beta_1 \text{Block Number}_i + \beta_2 \text{D}_i + \epsilon_i
\]

where \(\text{Number of Transactions}_i\) represents the number of transactions for rollup \(i\), \(\text{Block Number}_i\) is the block number, \(\text{D}_i\) is a dummy variable indicating whether the block number is after the implementation of EIP-4844 (1 if after, 0 otherwise), and \(\epsilon_i\) is the error term. The coefficient \(\beta_2\) on \(\text{D}_i\) captures the discontinuous jump at the threshold, providing an estimate of the impact of EIP-4844.

The results, presented in Table \ref{tab:4_4_rdd_results}, show statistically significant increases in transaction volumes for most rollups, affirming the effect of EIP-4844. However, Linea exhibited a negative coefficient for `D', suggesting that increases in Linea's transaction volumes may not be directly attributable to the effects of EIP-4844.

\begin{table}[ht]
\centering
\caption{RDD Analysis Results for Number of Rollup Transactions (Impact of `D')}
\label{tab:4_4_rdd_results}
\begin{tabular}{@{}lcccc@{}}
\toprule
Rollup      & Coefficient & Std. Error & t-Value & p-Value \\ \midrule
Arbitrum One & 43.91       & 6.74       & 6.51    & $<$0.001 \\
Optimism     & 24.87       & 3.47       & 7.17    & $<$0.001 \\
Base         & 100.34      & 5.94       & 16.89   & $<$0.001 \\
Starknet         & 36.28      & 1.08      & 33.69   & $<$0.001 \\
zkSync Era         & 41.38      & 5.5       & 7.45   & $<$0.001 \\
Linea         & -52.6      & 11.26       & -4.67   & $<$0.001 \\
\bottomrule
\end{tabular}
\end{table}

\subsubsection{Delay of L2 transactions}
To comprehensively assess the delays experienced by users of rollup transactions, we define the user delay metric as follows. This metric calculates the average timing difference between when rollup transactions are created and when they are committed on Ethereum.

\[
\text{User Delay}_b = \frac{\sum \limits_{tx \in b} \left( tx^{timing}_{eth} - tx^{timing}_{rollup} \right)}{N_b}
\]
\[
N_b = \sum \limits_{tx \in b} 1
\]

where \( b \) represents a specific Ethereum block, \( tx \) denotes a rollup transaction within block \( b \), \( tx^{timing}_{eth} \) is the timestamp when the transaction is committed on Ethereum, and \( tx^{timing}_{rollup} \) is the timestamp when the transaction is included in a rollup block. \( N_b \) is the total number of rollup transactions in block \( b \).

\begin{figure}
    \centering
    \begin{subfigure}{0.45\linewidth}
        \centering
        \includegraphics[width=\linewidth]{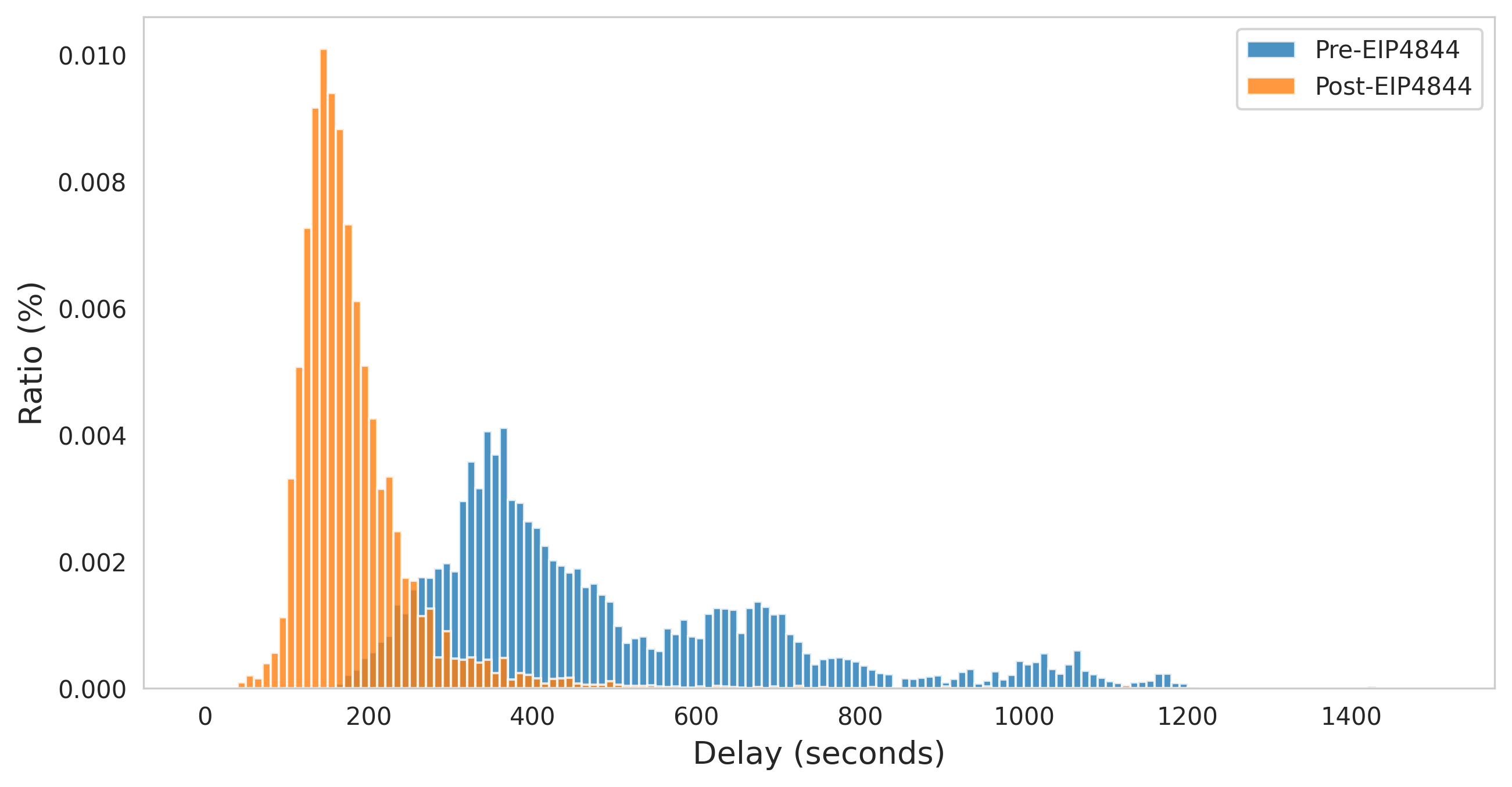}
        \caption{Arbitrum One}  
        \label{fig:4_3_arbitrum_delay_distribution}
    \end{subfigure}
    \hfill
    \begin{subfigure}{0.45\linewidth}
        \centering
        \includegraphics[width=\linewidth]{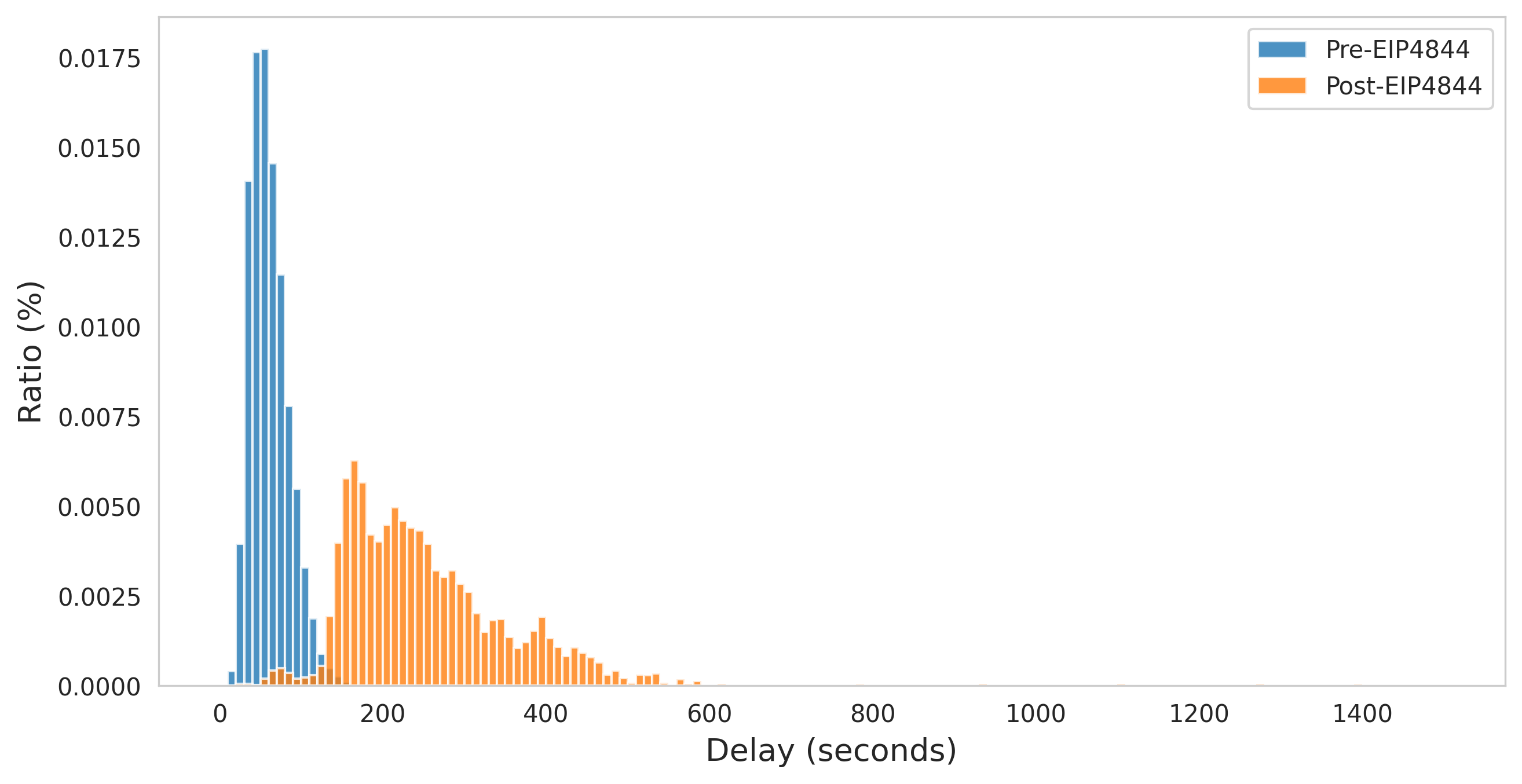}
        \caption{Optimism}  
        \label{fig:4_3_optimism_delay_distribution}
    \end{subfigure}
    \\
    \begin{subfigure}{0.45\linewidth}
        \centering
        \includegraphics[width=\linewidth]{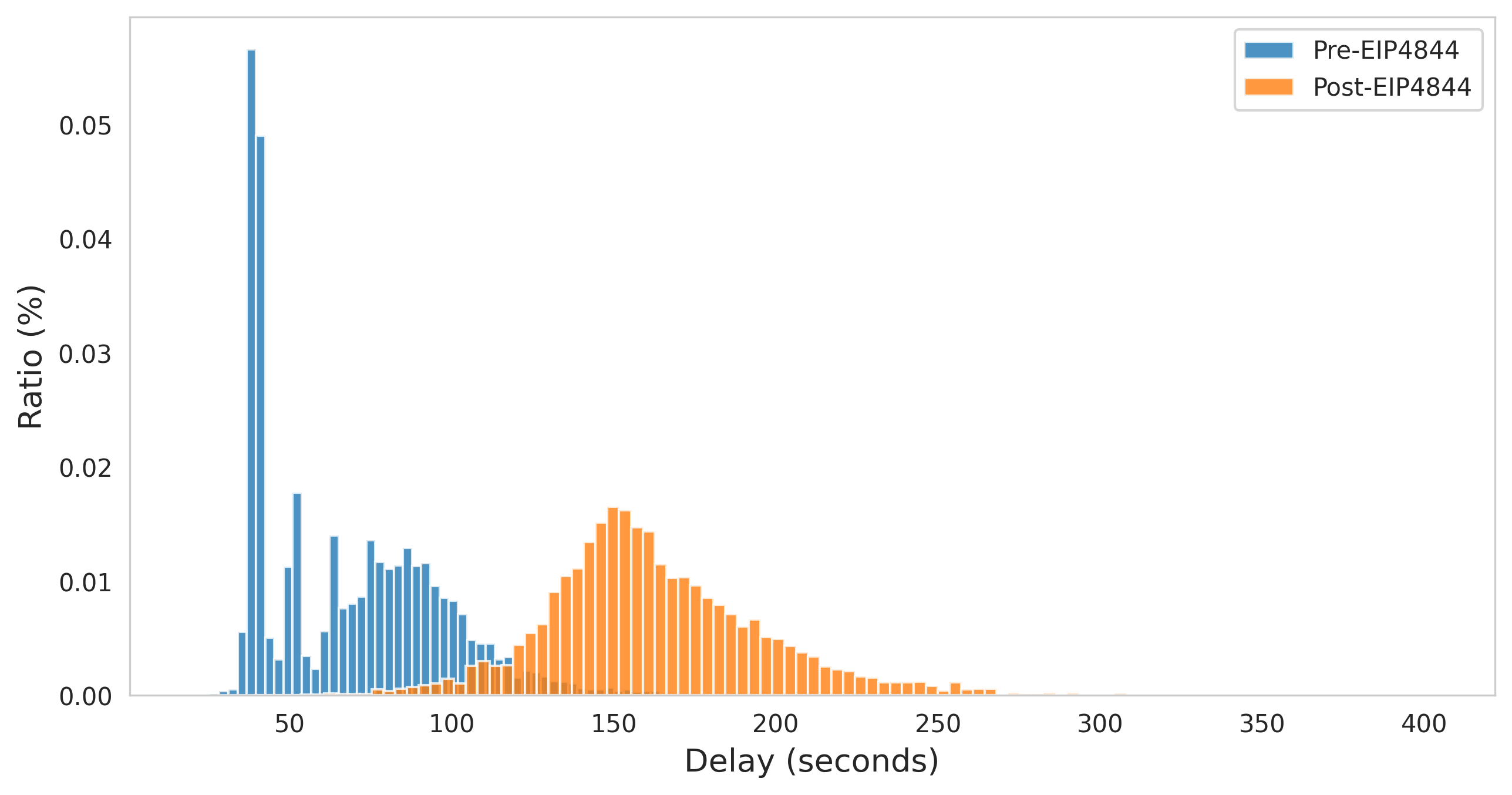}
        \caption{Base}  
        \label{fig:4_3_base_delay_distribution}
    \end{subfigure}
    \hfill
    \begin{subfigure}{0.45\linewidth}
        \centering
        \includegraphics[width=\linewidth]{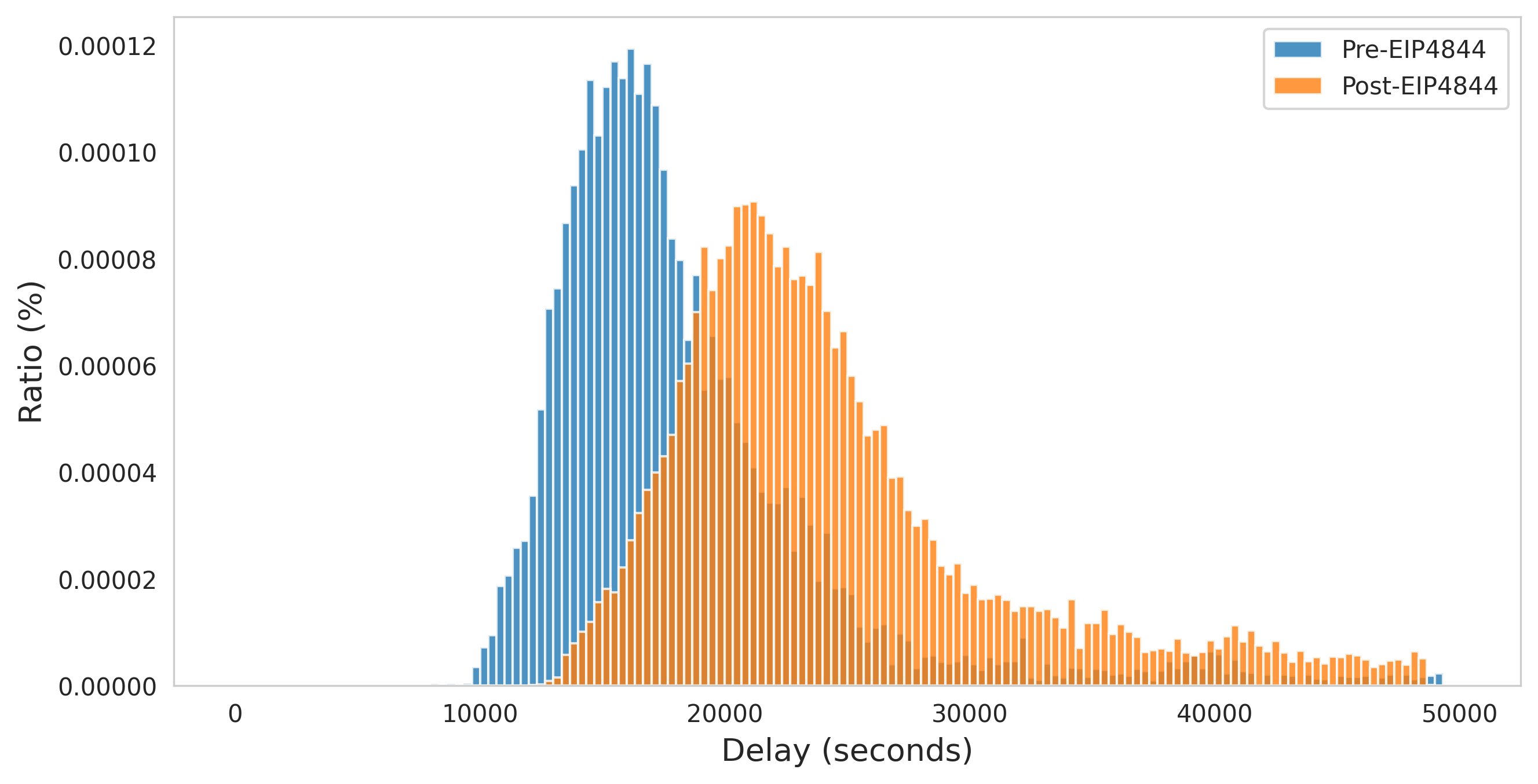}
        \caption{Starknet}  
        \label{fig:4_3_starknet_delay_distribution}
    \end{subfigure}
    \\
    \begin{subfigure}{0.45\linewidth}
        \centering
        \includegraphics[width=\linewidth]{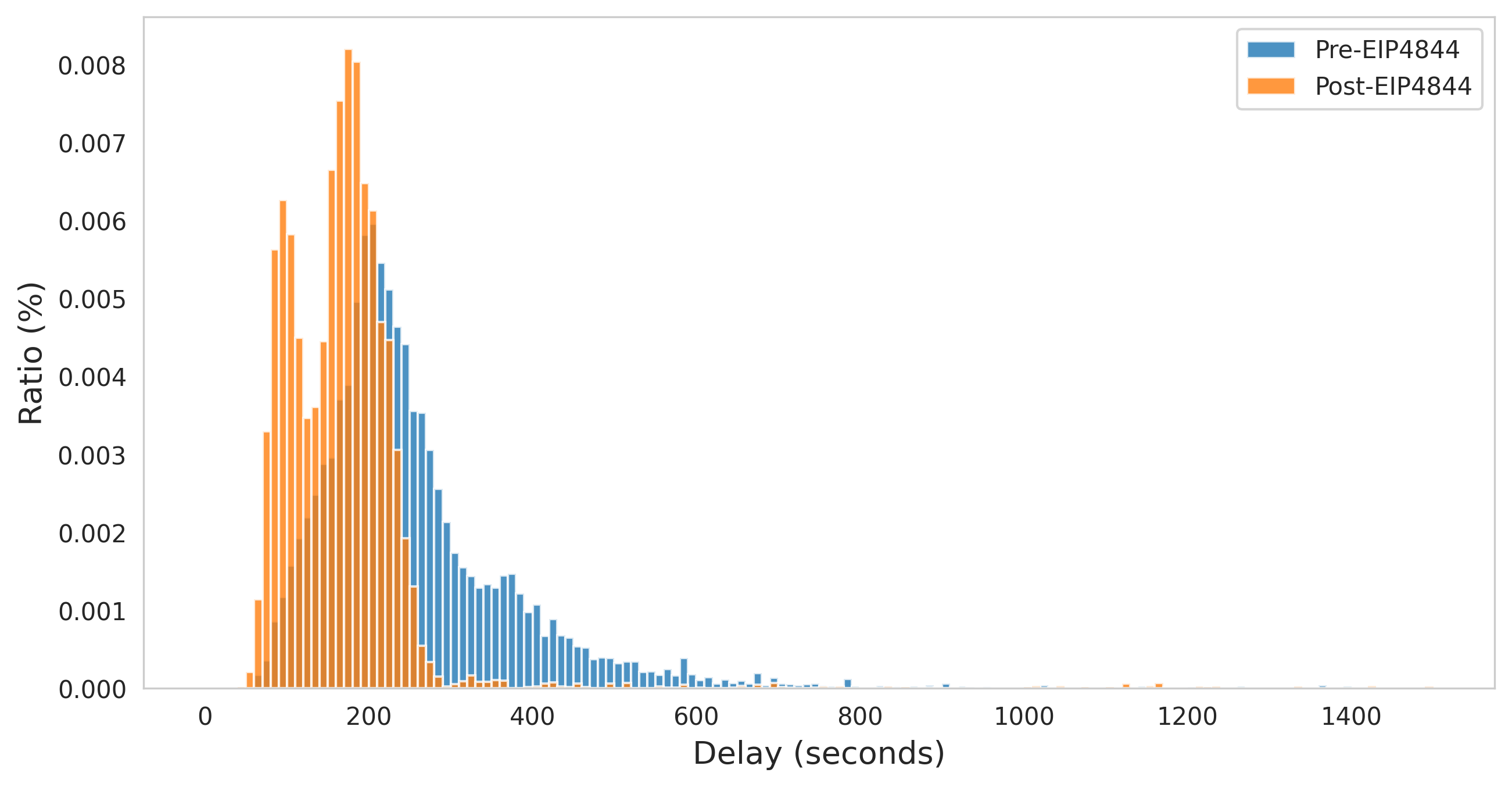}
        \caption{zkSync Era}  
        \label{fig:4_3_zksync_delay_distribution}
    \end{subfigure}
    \hfill
    \begin{subfigure}{0.45\linewidth}
        \centering
        \includegraphics[width=\linewidth]{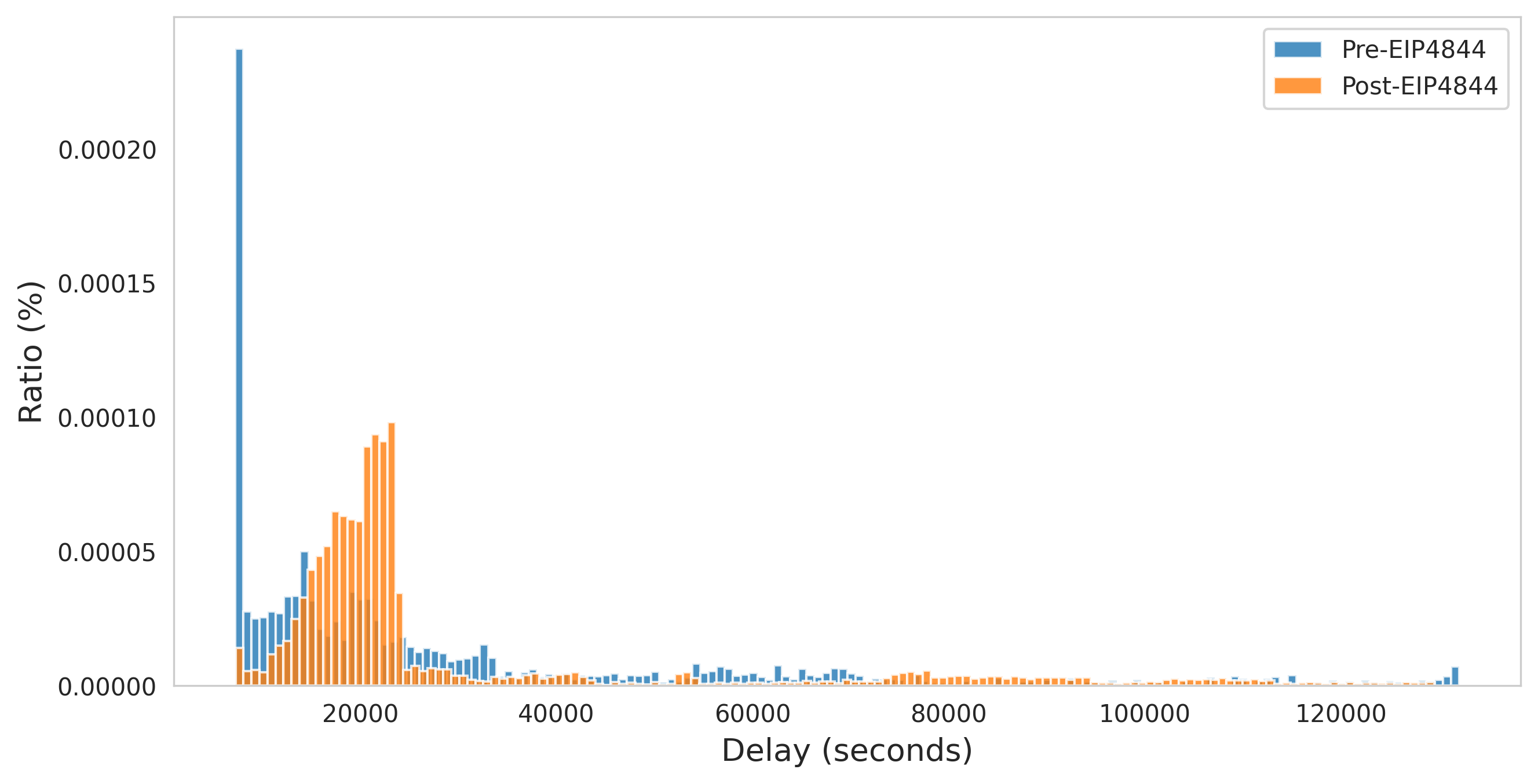}
        \caption{Linea}
        \label{fig:4_3_linea_delay_distribution}
    \end{subfigure}
    \caption{Distribution of user delay of rollup transactions before posted on Ethereum}
    \label{fig:4_3_rollup_delay_distribution}
    \Description{Distribution of user delay of rollup transactions before posted on Ethereum}
\end{figure}

Figure \ref{fig:4_3_rollup_delay_distribution} illustrates the distribution of user delays for each rollup before and after the implementation of EIP-4844. Despite an increase in the number of rollup transactions, four out of six rollups showed an increase in user delay. Notably, Arbitrum One saw a significant decrease in user delay by 62\%, and zkSync by 40\%, indicating faster transaction settlement times. Conversely, other rollups exhibited notable increases in delay.

Increased user delays indicate that users often experience longer wait times before their transactions are settled on Ethereum, requiring them to maintain trust in the rollup operators until their transactions are committed. A potential cause for increased delay could be rollups waiting to fill blobs to their capacity of 128KiB before committing them.

The standard deviation of user delay is also a crucial indicator of predictability in waiting time, essential for user assurance regarding transaction security. Table \ref{tab:4_3_rollup_results_summary} indicates that the standard deviation of user delays increased in all rollups except Arbitrum, suggesting that users frequently face longer waiting times for their transactions to be settled on Ethereum. 

While increasing transaction volumes that rapidly fill the 128KiB capacity may help reduce these delays, rollups with smaller or infrequent transactions might still face extended waiting times. This issue highlights the need for blob sharing protocols among different rollups to hasten transaction commit times and enhance user experiences by shortening wait periods.

\subsection{Blob gas fee market}
Understanding the blob gas fee market dynamics is essential for enhancing market predictability, enabling DApps and rollups to optimize data posting and fees. This predictability reduces variability in Ethereum transaction settlement times and minimizes costs. Previous research, such as \cite{diamandis2023designing,crapis2023optimal}, has investigated optimal strategies in multidimensional markets, providing practical insights that complement theoretical models.

Additionally, insights from this analysis could contribute to improvements in the blob gas fee market. Extensive studies on gas fee markets \cite{azouvi2023base,leonardos2023optimality,reijsbergen2021transaction} have explored how fee update rules might better capture user demands and reduce variability. Extending these studies to multidimensional fee markets could deepen our understanding.

This section presents an analysis of the newly emerged blob gas fee market and outlines the following key findings:

\begin{enumerate}
    \item The VAR model indicates that the gas base fee has a small, yet statistically significant influence on the blob gas base fee. Initially, this impact is positive but tends to diminish over time.
    \item We introduced a metric for `blob gas priority fee' to represent the priority demand for blob gas. The validity of this proxy was established by demonstrating its utility in enhancing the explainability of blob gas base fees.
    \item The blob gas fee market exhibits higher volatility compared to the gas fee market, indicating potential challenges in predictability and stability. Despite its volatility, the lower ratio of priority fee to base fee in the blob gas market suggests it captures market demands more effectively than the gas market.
\end{enumerate}

\subsubsection{Inter relationships between gas and blob gas market}
To investigate the dynamic interactions between the gas and blob gas markets, we employed a VAR model analyzing the base fees for both types of gas. The significant effects detected in the model are outlined in Table \ref{tab:4_4_VAR_gas_blob_gas}. For the gas base fee, positive effects are noted at lags 1 and 4, with a negative effect at lag 3, indicating oscillating impacts that diminish over time. In contrast, the equation for the blob gas base fee showed no significant effects. The near-zero correlation of residuals (-0.027446) suggests minimal unforeseen shared variations between these markets.

\begin{table}[ht]
\centering
\caption{Significant Inter-Variable Effects in the VAR Model for Gas and Blob Gas Markets}
\label{tab:4_4_VAR_gas_blob_gas}
\begin{tabular}{lccc}
\toprule
Variable & Coefficient & T-Statistic & P-Value \\
\midrule
L1.gas\_base\_fee & 0.058937 & 3.733 & 0.000 \\
L3.gas\_base\_fee & -0.056381& -2.934 & 0.003 \\
L4.gas\_base\_fee & 0.053983& 2.825 & 0.005 \\
\bottomrule

\end{tabular}
\end{table}

\subsubsection{Blob gas priority fee}
\begin{figure}[ht]
  \centering
  \includegraphics[width=0.7\linewidth]{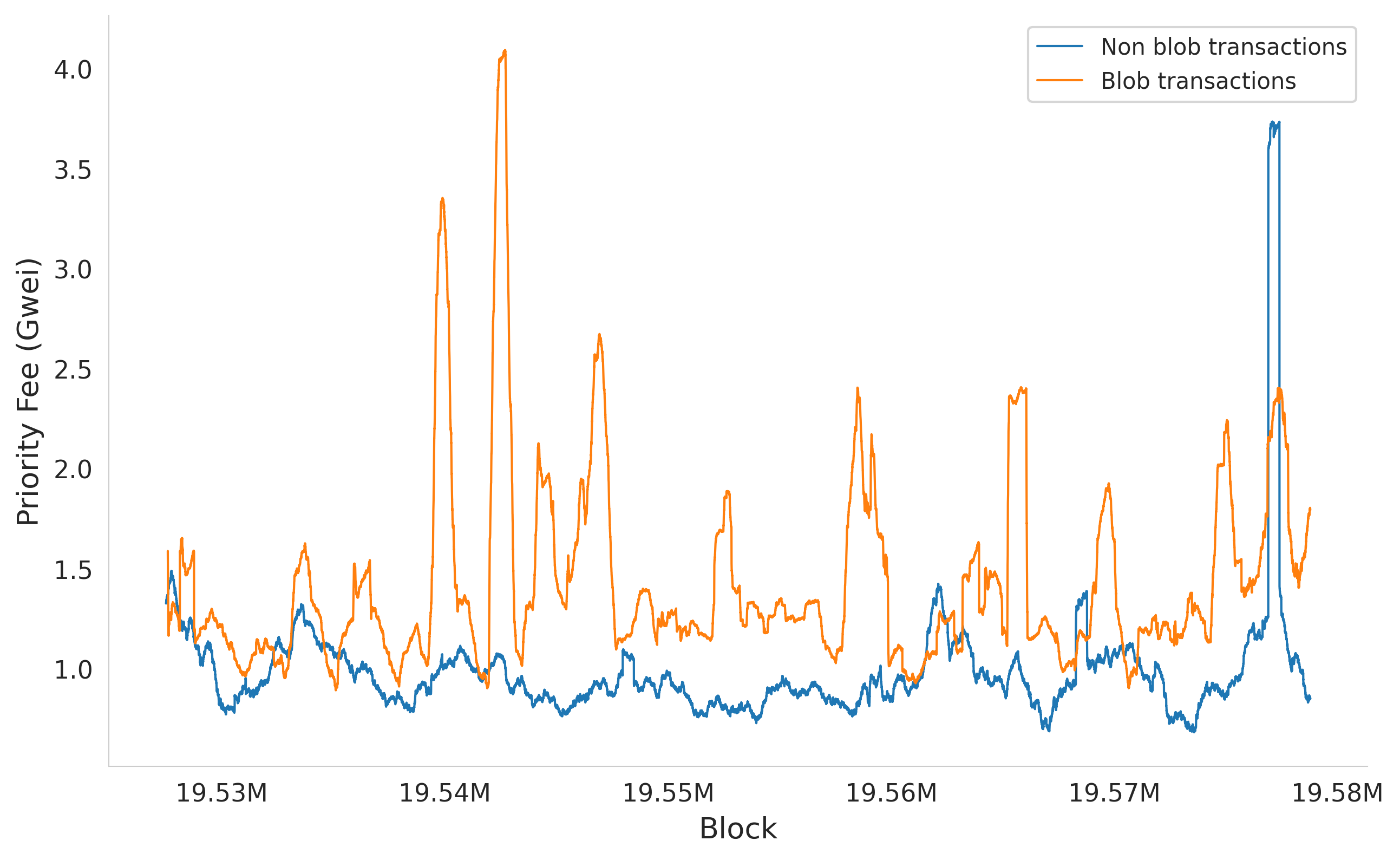}
  \caption{Comparison of median priority fees between blob and non-blob transactions}
  \Description{}
  \label{fig:4_4_priority_fee_comparison}
\end{figure}

We analyzed the economic implications of blob transactions by examining priority fees—calculated as the difference between the effective gas fee and the base fee. Median priority fees for blob and non-blob transactions were compared across various blocks. As shown in Figure \ref{fig:4_4_priority_fee_comparison}, blob transactions have a higher average median priority fee of 1.43 Gwei, which is 45.2\% greater than the 0.99 Gwei for non-blob transactions.

This notable difference implies that blob transactions are typically assigned a higher priority due to the need to handle additional blob data. Consequently, we have defined the priority fee for blobs, termed `blob gas priority fee,' using the following formula, as detailed in Section \ref{subsection:blob_gas_fee_data}:

\[
\text{P}_{\text{blob gas}, i, k} = \left( \frac{(E_{i,k} - B_{\text{blob gas}, k} - \underset{\text{tx} \in k}{\mathrm{median}} (P_{\text{gas}, tx, k})) \times G_{i, k}} {D_{i, k}} \right)^+
\]
The median gas priority fee from other transactions within the same block serves as the baseline. This baseline is subtracted from the total priority fee to isolate the component attributable to blob gas. This method underscores the additional costs imposed on blob transactions.

\textbf{Validating blob gas priority fee.}
The priority fee can act as a leading indicator of the base fee when the latter does not promptly reflect demand spikes. Conversely, a rising base fee, signaling increased demand, typically causes the priority fee to decrease. We validated the blob gas priority fee as a proxy for unmet demand using a Vector Autoregression (VAR) analysis, with findings detailed in Table \ref{tab:4_4_VAR_blob_gas_basefee_priority_fee} showing statistically significant interactions.

Panel A of the table illustrates the persistence of the blob gas priority fee across all time lags, indicating that these fees are applied consistently and reflect strategic adjustments within the network. Notably, a negative coefficient for the blob gas base fee at lag 1 suggests that a higher initial base fee might reduce subsequent priority fees, better aligning with market demands.

Panel B demonstrates a positive influence of the blob gas priority fee on the blob gas base fee, confirming that increases in the priority fee are promptly followed by rises in the base fee.

These results confirm the interrelationship between the priority and base fees, justifying the use of the blob gas priority fee metric. This metric provides essential insights into fee dynamics within the blob market, aiding developers and users in optimizing network interactions.

\begin{table}[ht]
\centering
\caption{Significant Effects in VAR Model}
\label{tab:4_4_VAR_blob_gas_basefee_priority_fee}
\begin{tabular}{lccc}
\multicolumn{4}{c}{Panel A: Results for Blob Gas Base Fee} \\
\toprule
Variable & Coefficient & T-Statistic & P-Value \\
\midrule
L1.blob\_gas\_priority\_fee & 0.545 & 9.927 & 0.000 \\
\bottomrule
\\
\multicolumn{4}{c}{Panel B: Results for Blob Gas Priority Fee} \\
\toprule
Variable & Coefficient & T-Statistic & P-Value \\
\midrule
L1.blob\_gas\_base\_fee & -0.0013 & -3.812 & 0.000 \\
L1.blob\_gas\_priority\_fee & 0.1151 & 26.271 & 0.000 \\
L2.blob\_gas\_priority\_fee & 0.1505 & 34.115 & 0.000 \\
L3.blob\_gas\_priority\_fee & 0.056 & 15.618 & 0.000 \\
L4.blob\_gas\_priority\_fee & 0.0967 & 21.78 & 0.000 \\
L5.blob\_gas\_priority\_fee & 0.0493 & 11.185 & 0.000 \\
\bottomrule
\end{tabular}
\end{table}

\subsubsection{Disscussion of Blob Gas Fee Mechanisms}
The blob gas base fee update rule's design is crucial for ensuring predictability for users. Previous studies often evaluate the gas base fee update rule based on two criteria: its volatility and how well it reflects actual demand, inferred from indirect metrics such as the effective gas fee \cite{ferreira2021dynamic,reijsbergen2021transaction,crapis2023optimal}.

\begin{figure}[ht]
  \centering
  \includegraphics[width=0.8\linewidth]{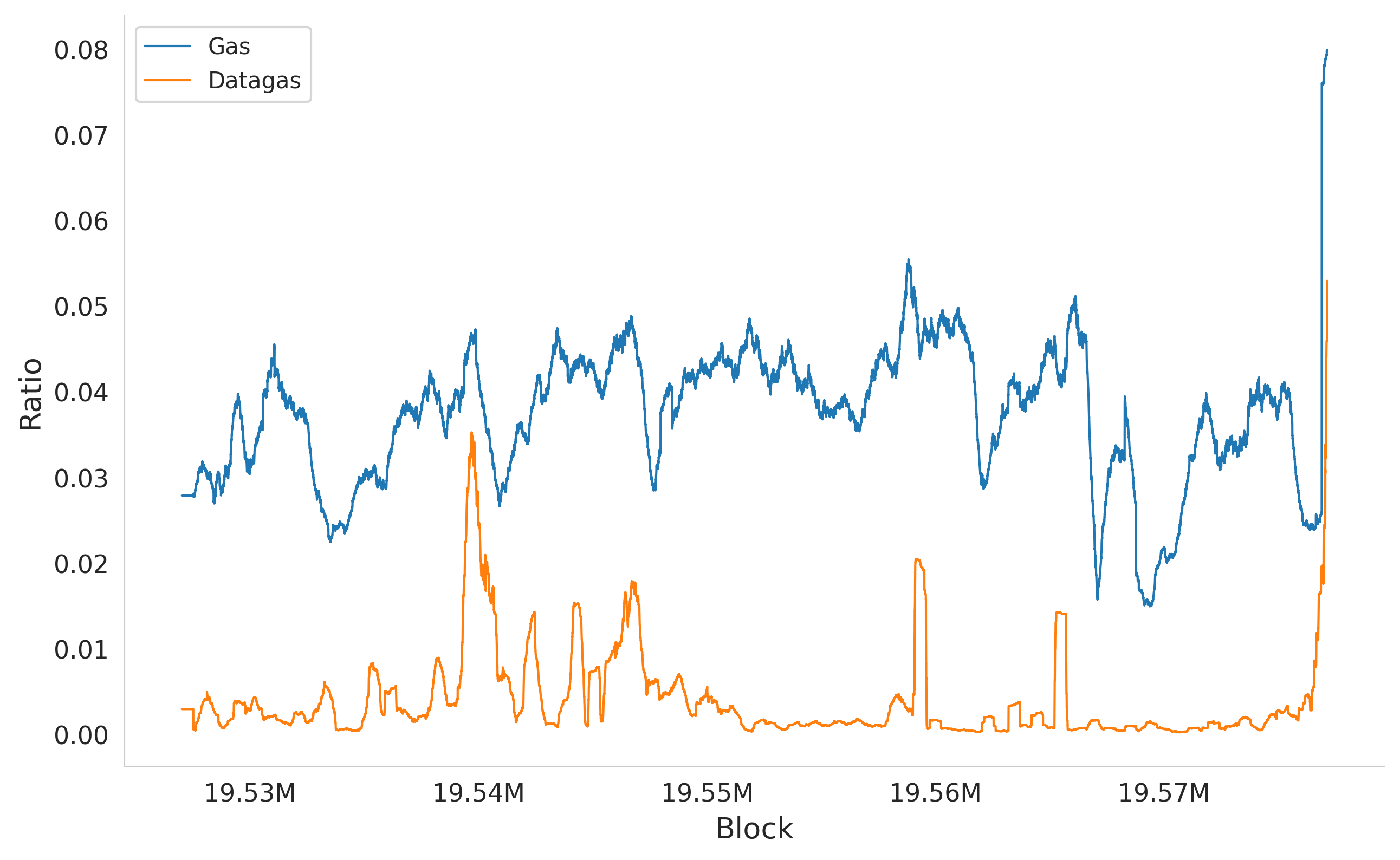}
  \caption{Ratio of priority fee to base fee for gas and blob gas}
  \Description{}
  \label{fig:4_4_priority_fee_ratio}
\end{figure}

In our analysis, the blob gas priority fee served as a proxy to gauge the deviation from actual demand within the blob gas market. We employed the ratio of the base fee to the priority fee as a critical metric, reflecting the proportion of unmatched user demands. Figure \ref{fig:4_4_priority_fee_ratio} illustrates significant differences between the gas and blob gas markets. The ratio in the gas market stands at 0.037, markedly higher than the blob gas market's 0.004. This disparity indicates that the blob gas market aligns more closely with user demand, despite its higher volatility, as detailed in Table \ref{tab:base_fees_summary}.

These observations indicate that the blob gas base fee generally aligns well with user demands, suggesting that the underlying mechanisms are effective. However, its heightened volatility poses challenges for the predictability and stability of transaction costs, which are critical for user strategies and overall market dynamics. This variability requires careful consideration; protocol designers must weigh trade-offs between market responsiveness and fee stability to improve the ecosystem’s operational efficiency.

\begin{table}[ht]
\centering
\caption{Summary Statistics of Blob Gas Base Fees in our analysis period}
\label{tab:base_fees_summary}
\begin{tabular}{lrrr}
\hline
\textbf{Statistic} & \textbf{Block} & \textbf{Gas Base Fee} & \textbf{Blob Gas Base Fee} \\ \hline
Count              & 51,778         & 51,778                & 51,778                    \\
Mean               & 19,552,890     & 28.25 \(\times 10^9\)  & 39.60 \(\times 10^9\)     \\
Std                & 14,947         & 11.73 \(\times 10^9\)  & 33.62 \(\times 10^9\)     \\ \hline
\end{tabular}
\end{table}

\section{Conclusion}
We have conducted a comprehensive analysis of EIP-4844 across four key dimensions: consensus security, Ethereum usage, rollup transactions, and user delays, and the blob gas fee market. Our analysis of the impact of EIP-4844 on the increase in fork rates and block delays provides essential insights into the security implications of the protocol update, addressing concerns within the Ethereum community. Our empirical findings demonstrate the changes in Ethereum and rollup ecosystem dynamics, highlighting the effectiveness of the upgrade and introducing new considerations for user security due to increased posting delays. Furthermore, our exploration of the blob gas fee market unveils new possibilities for evaluating fee structures and optimizing strategies for decentralized applications.

\bibliographystyle{ACM-Reference-Format}
\bibliography{arxiv}

\appendix
\section{Consensus security data}
This appendix presents a detailed evaluation of the consensus security data after the implementation of EIP-4844.

\subsection{Overall change of consensus metrics}
Table \ref{tab:consensus_metrics} shows the comparative metrics before and after the implementation of EIP-4844. The values in parentheses represent the increase excluding the impact of slots with zero blobs. The data reflects an increase in the metrics post-4844, with fork rate witnessing a noticeable increase. The synchronization time (Sync time), although showing a significant increase in average, only about half of this increase can be confidently attributed to the impact of EIP-4844. It was observed that CSP time and DA time did not significantly affect sync time, whereas receive time had a major impact.

\begin{table}[ht]
\caption{Comparative metrics before and after the implementation of EIP-4844}
\label{tab:consensus_metrics}
    \begin{tabular}{lccc}
    \hline
    \textbf{Metrics} & \textbf{Pre-4844} & \textbf{Post-4844} & \textbf{Increase} \\
    \hline
    Fork rate & 3.097 & 6.707 & 3.61 \\
    Sync time (ms) & 2267.436 & 2407.50 & 140.065 (77.967) \\
    Receive time (ms) & 1759.066 & 1840.032 & 80.966 (56.102) \\
    CSP time (ms) & 482.565 & 536.043 & 53.478 (0.699) \\
    DA time (ms) & - & 13.417 & 13.417 (0.956) \\
    \hline
    \end{tabular}
\end{table}

\subsection{Detailed analysis of consensus metrics}

\begin{figure}[ht]
    \centering
    \includegraphics[width=\linewidth]{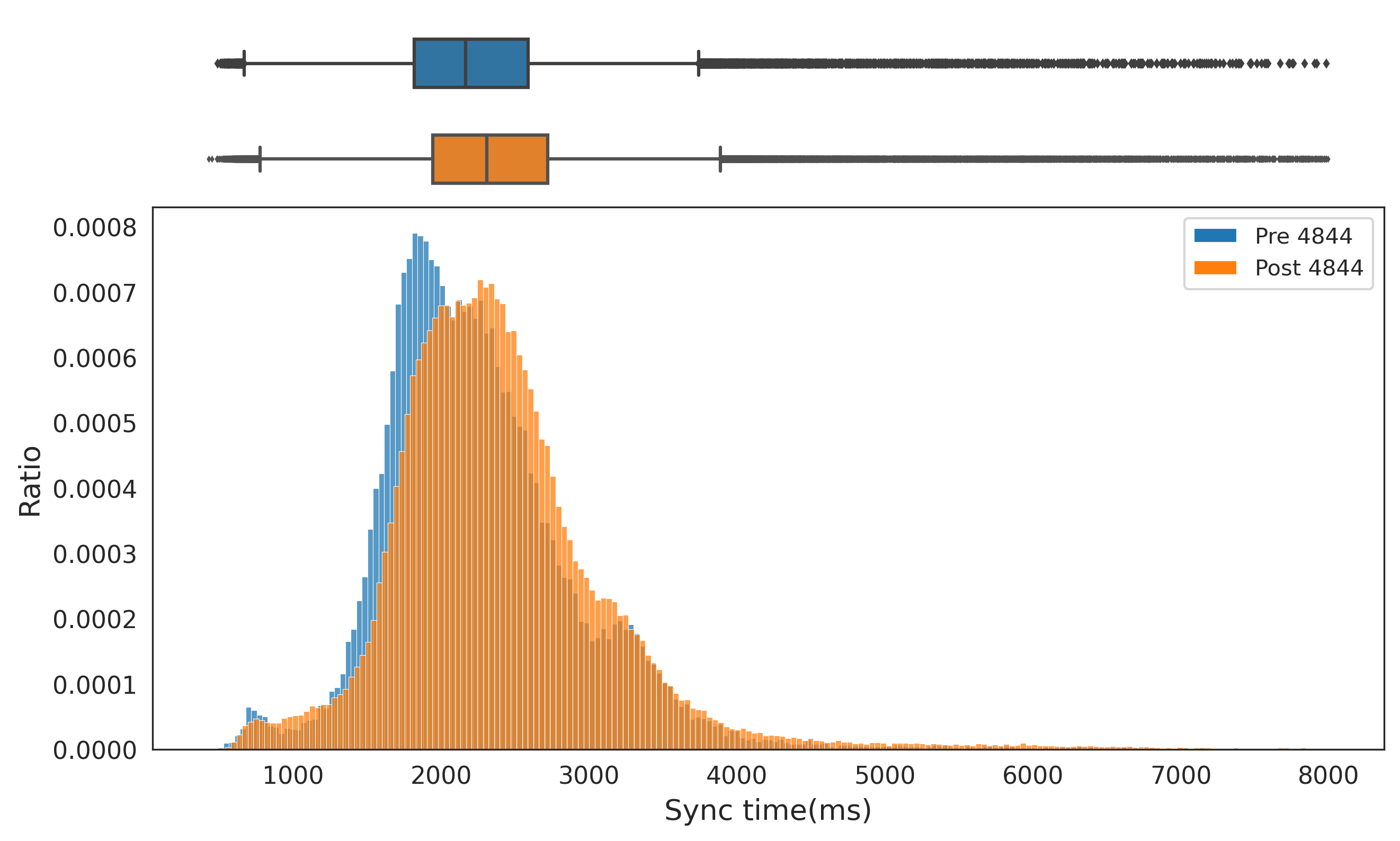}
    \caption{Distribution of sync time for slots before and after EIP-4844}
    \label{fig:sync_time_hist}
    \Description{}
\end{figure}

Figure \ref{fig:sync_time_hist} illustrates the distribution of sync time for slots before and after the enactment of EIP-4844. The box plots provide a visual comparison, while the histogram offers a distribution perspective, highlighting the shift towards slow synchronization times post-4844.

Table \ref{tab:logit_results} presents the results of a logistic regression analysis, which investigates the impact of synchronization time on the likelihood of a slot forking. The positive coefficient of 1.5 indicates that as synchronization time increases, the log odds of observing the fork also increase, suggesting a proportional relationship.

\begin{figure}[ht]
    \includegraphics[width=0.7\linewidth]{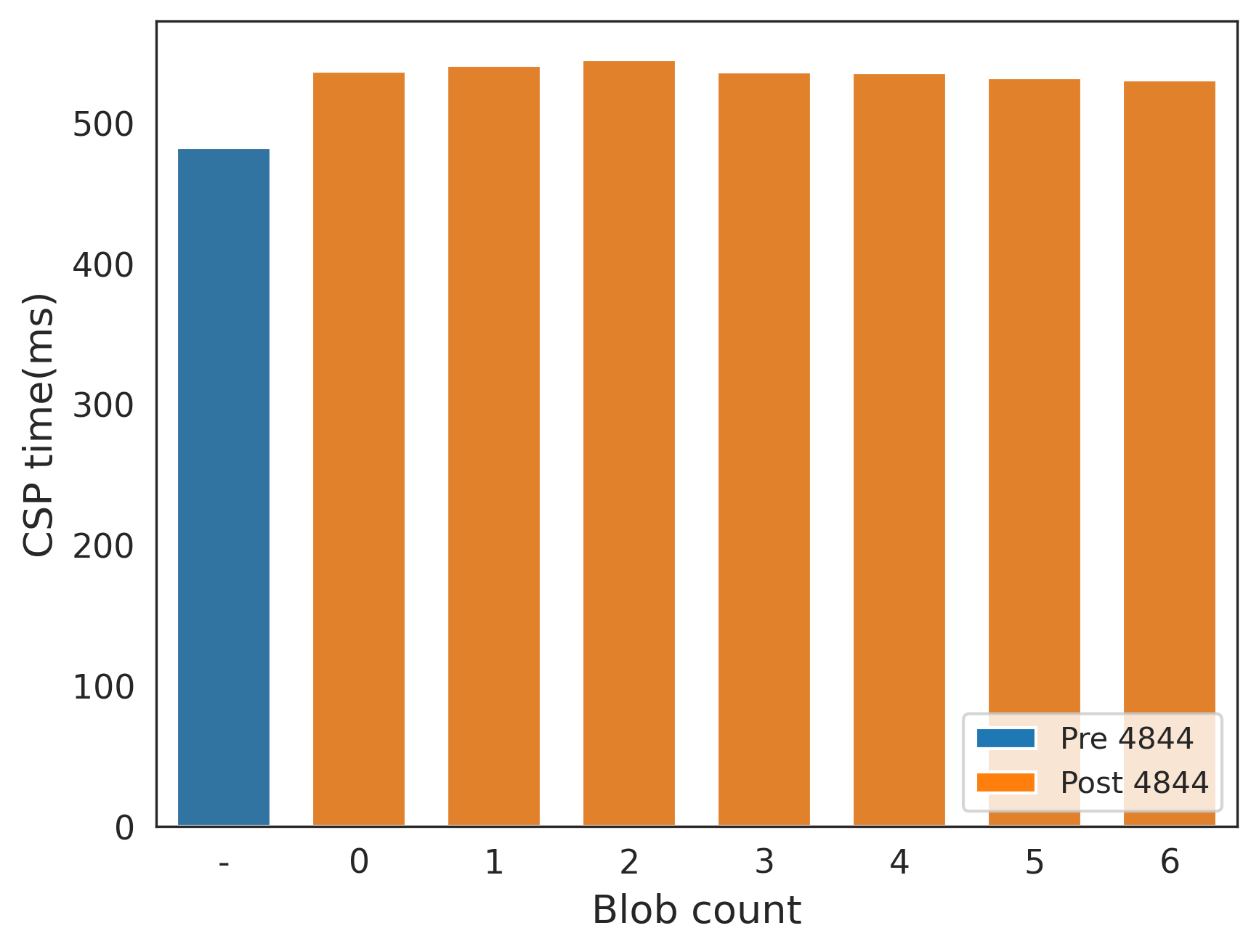}
    \caption{Average CSP time by the number of blobs}
    \label{fig:csp_time_blob}
    \Description{}
\end{figure}

Figure \ref{fig:csp_time_blob} analyzes the average CSP time by the number of blobs. The bar chart contrasts the average time taken for consensus proposals with varying blob counts before and after the implementation of EIP-4844. The comparison suggests that while there is an increase in CSP time post-4844, it is difficult to conclusively attribute this increase to the effects of EIP-4844.

\begin{table}[ht]
\centering
\caption{Summary of Logistic Regression Analysis}
\label{tab:logit_results}
    \begin{tabular}{lccc}
    \toprule
    \textbf{Parameter} & \textbf{Coefficient} & \textbf{Std. Error} & \textbf{P-Value} \\
    \midrule
    Constant & -10.9497 & 0.137 & $<$0.001 \\
    Sync Time ($\times 10^{-3}$) & 1.5 & 0.0248 & $<$0.001 \\
    \midrule
    \end{tabular}
\end{table}

\section{Rollup data collection}

\subsection{Rollup data sources}
Table \ref{tab:data_sources_rollup_transactions} represents data sources of decoded batch transactions and blocks of rollups. Combining this data with \ref{tab:data_sources_rollup_transactions}, we completed a detailed our dataset including transaction delay and L2 transaction data.

\begin{table}[ht]
\centering
\caption{Data sources of rollup transactions}
\label{tab:data_sources_rollup_transactions}
\begin{tabular}{@{}lll@{}}
\toprule
Rollup   & Decoded Batch Tx & Rollup Block \\ \midrule
Arbitrum One & ethernow.xyz           & arbiscan.io     \\
Optimism & ethernow.xyz           & optimistic.etherscan.io     \\
Base & ethernow.xyz           & basescan.org     \\
Starknet & voyager.online           & voyager.online     \\
zkSync Era & explorer.zksync.io           & explorer.zksync.io     \\
Linea & Ethereum event logs    & lineascan.build     \\
\bottomrule
\end{tabular}
\end{table}

\subsection{Rollup function classification}
Table \ref{tab:transaction_classification} represents the classification of rollup transactions based on their functionality and sender addresses. We investigated 10 rollups and their functions frequently used, and classified them to two parts: if they use Ethereum as DA layer or not. 

\begin{table*}[ht]
\centering
\caption{Classification of Rollup Transactions Based on Functionality and Sender Address}
\label{tab:transaction_classification}
\begin{tabular}{@{}l|llc@{}}
\toprule
Rollup & Sender Address & Function Called & Uses Ethereum as DA Layer \\ \toprule
Arbitrum & 0xC1b634853Cb333D3aD8663715b08f41A3Aec47cc & all & Yes \\
\cline{1-4}
Optimism & 0x6887246668a3b87F54DeB3b94Ba47a6f63F32985 & all & Yes \\
\cline{1-4}
Blast & 0x415c8893D514F9BC5211d36eEDA4183226b84AA7 & all & Yes \\
\cline{1-4}
Base & 0x5050F69a9786F081509234F1a7F4684b5E5b76C9 & all & Yes \\
\cline{1-4}
\multirow{4}{*}{Starknet} & \multirow{2}{*}{0x2C169DFe5fBbA12957Bdd0Ba47d9CEDbFE260CA7} & updateStateKzgDA & Yes \\
& & updateState & Yes \\
\cline{2-4}
& \multirow{2}{*}{0x22A82147A80747CFb1562e0f72F6be39F18B5F76} & verifyFRI & No\\

& & registerContinuousMemoryPage & No\\
\cline{1-4}
\multirow{3}{*}{zkSync Era} & 0x0D3250c3D5FAcb74Ac15834096397a3Ef790ec99 & commitBatches & Yes \\
\cline{2-4}
 & \multirow{2}{*}{0x3527439923a63F8C13CF72b8Fe80a77f6e572092} & proveBatches & No \\
 &  & commitBatches & Yes \\
 \cline{1-4}
\multirow{3}{*}{dYdX V3} & \multirow{3}{*}{0x8129b737912e17212C8693B781928f5D0303390a} & updateState & Yes \\
& & verifyFRI & No \\
& & verifyProofAndRegister & No \\
\cline{1-4}
\multirow{2}{*}{Scroll} & 0xcF2898225ED05Be911D3709d9417e86E0b4Cfc8f & commitBatches & Yes \\
\cline{2-4}
 & 0x356483dC32B004f32Ea0Ce58F7F88879886e9074 & finalizeBatches & No \\
 \cline{1-4}
Mode & 0x99199a22125034c808ff20f377d91187E8050F2E & all & Yes \\
\cline{1-4}
\multirow{2}{*}{Linea} & 0xa9268341831eFa4937537bc3e9EB36DbecE83C7e & submitBlob & Yes \\
\cline{2-4}
 & 0x9228624C3185FCBcf24c1c9dB76D8Bef5f5DAd64 & finalizeCompressedBlocksWithProof & No \\
\bottomrule
\end{tabular}
\end{table*}

\section{Detailed VAR model results for blob gas base fee and gas base fee}
This section presents the detailed results from the VAR model analysis conducted on the blob gas base fee and the gas base fee. The analysis includes various statistical tests and model estimations to assess the dynamics and relation between these two key metrics within the network's pricing mechanism.

\subsection{ADF test results}
Table \ref{tab:adf_results} displays the results of the ADF test, used to check the stationarity of the time series data for both the gas base fee and the blob gas base fee. Result confirmed that the Base Fee and Blob Gas Base Fee time series are stationary. The test statistics of -6.3719 and -10.5237 respectively, along with very low p-values. This indicates that the data are suitable for further econometric modeling, as they do not depend on time.

\begin{table}[htbp]
\centering
\caption{ADF test results}
\label{tab:adf_results}
\begin{tabular}{@{}lcc@{}}
\toprule
\textbf{Metric} & \textbf{Base Fee} & \textbf{Blob Gas Base Fee} \\ 
\midrule
Test Statistic & -6.3719 & -10.5237 \\
p-value & \(2.33 \times 10^{-8}\) & \(9.54 \times 10^{-19}\) \\
Number of Lags Used & 62 & 62 \\
Number of Observations & 69,429 & 69,429 \\
\midrule
\textbf{Critical Values} & & \\
\quad 1\% & -3.4304 & -3.4304 \\
\quad 5\% & -2.8616 & -2.8616 \\
\quad 10\% & -2.5668 & -2.5668 \\
\midrule
Conclusion & Stationary & Stationary \\
\bottomrule
\end{tabular}
\end{table}

\subsection{VAR model estimation output}
Detailed results for the VAR model estimation are provided below, showing the full regression output for both the gas base fee and blob gas base fee equations. 

Detailed results for the VAR model estimation are provided below, showing the full regression output for both the gas base fee and blob gas base fee equations. Table \ref{tab:var_model_summary_1} summarizes the overall model diagnostics including the number of equations, observations, log likelihood, and several information criteria that help in assessing the model fit. Table \ref{tab:appendix_VAR_model_1} presents the estimated coefficients and associated statistics for each variable within the equations, detailing the individual impacts in the model.

Table \ref{tab:var_model_summary_1} summarizes key metrics from the VAR model regression results. The table captures essential information such as the number of equations modeled, total observations considered, and various statistical measures including the Akaike Information Criterion (AIC), Bayesian Information Criterion (BIC), Hannan-Quinn Information Criterion (HQIC), and Final Prediction Error (FPE). These metrics provide insights into the model's performance and its predictive accuracy.

\begin{table}[htbp]
\centering
\caption{Summary of VAR Model Regression Results}
\label{tab:var_model_summary_1}
\begin{tabular}{lccc}
\toprule
\textbf{Metric}             & \textbf{Value} \\
\midrule
Number of Equations         & 2 \\
Number of Observations      & 51,773 \\
Log Likelihood              & -2.40029 $\times 10^{6}$ \\
AIC                         & 87.0488 \\
BIC                         & 87.0526 \\
HQIC                        & 87.0500 \\
FPE                         & 6.38013 $\times 10^{37}$ \\
\bottomrule
\end{tabular}
\end{table}

\subsection{Correlation matrix of residuals}
Table \ref{tab:correlation_matrix_1} presents the correlation matrix of residuals for the gas base fee and blob gas base fee. The near-zero correlation coefficients between the residuals of different equations suggest that the residuals are uncorrelated.

Table \ref{tab:appendix_VAR_model_1} provides detailed VAR model estimates for the gas base fee and blob gas base fee. For the gas base fee, all lagged values are significant predictors, with particularly strong influence from L1, evidenced by a high t-statistic of 155.252 and a p-value less than 0.001. 


\begin{table}[htbp]
\centering
\caption{Correlation Matrix of Residuals}
\label{tab:correlation_matrix_1}
\small
\begin{tabular}{@{}lcc@{}}
\toprule & \textbf{Base Fee} & \textbf{Blob Gas Base Fee} \\ \midrule
\textbf{Base Fee} & 1.000 & -0.027446 \\
\textbf{Blob Gas Base Fee} & -0.027446 & 1.000 \\
\bottomrule
\end{tabular}
\end{table}

\begin{table}[ht]
\centering
\caption{Summary of VAR Model Regression Results}
\label{tab:var_model_summary_2}
\begin{tabular}{lc}
\toprule
\textbf{Metric}             & \textbf{Value} \\
\midrule
Number of Equations         & 2 \\
Number of Observations      & 51,772 \\
Log Likelihood              & -2,335,520 \\
AIC                         & 84.5486 \\
BIC                         & 84.5530 \\
HQIC                        & 84.5500 \\
FPE                         & 5.23571 $\times 10^{36}$ \\
\bottomrule
\end{tabular}
\end{table}

\section{Detailed VAR model results for blob gas base Fee and blob gas priority fee}

In this subsection, we present the results of the Vector Autoregression (VAR) model analysis that examines the interactions between the blob gas base fee and the blob gas priority fee. Table 14 provides a summary of the regression results, offering an overview of the model's fit and diagnostic statistics. Following this, Table \ref{tab:appendix_VAR_model_2} details the estimated coefficients and statistics, allowing us to understand the influence of past values on current values for each variable.

\subsection{VAR model estimation output}
Table \ref{tab:var_model_summary_2} displays the VAR model's key statistics. The log likelihood value is notably large at -2,335,520, suggesting the model’s fit to the data under analysis. The model's complexity and goodness-of-fit are further quantified by AIC, BIC, and HQIC, all closely valued around 84.5. These criteria help in model selection, with lower values generally indicating a better model relative to the number of parameters used.

Table \ref{tab:appendix_VAR_model_2} outlines the VAR model results for datagas\_base\_fee and datagas\_priority\_fee\_per\_datagas. The model shows strong persistence in datagas\_base\_fee, as indicated by the significant coefficient of 0.9588 for its first lag, and an effect of prior priority fees on current base fees, though the impact diminishes over time, as seen in the insignificant coefficient for the sixth lag. In contrast, the datagas\_priority\_fee\_per\_datagas equation indicates a slight decrease in priority fees with an increase in base fees at the previous lag. This suggests an autocorrelation that indicates the complex, dynamic interplay between these fees within the network's pricing mechanism.

\begin{table}[ht]
\centering
\caption{Correlation Matrix of Residuals}
\label{tab:correlation_matrix_2}
\small
\begin{tabular}{@{}lcc@{}}
\toprule
 & \textbf{datagas\_base\_fee} & \textbf{datagas\_priority\_fee} \\
\midrule
datagas\_base\_fee & 1.000000 & -0.017866 \\
datagas\_priority\_fee & -0.017866 & 1.000000 \\
\bottomrule
\end{tabular}
\end{table}

\begin{table*}
\centering
\caption{Estimated coefficients and statistics for gas base fee and blob gas base fee}
\label{tab:appendix_VAR_model_1}
\begin{tabular}{>{\raggedright\arraybackslash}p{3cm}cccc}
\toprule
\textbf{Equation for gas\_base\_fee} & Coefficient & Standard Error & T-Statistic & P-Value \\
\midrule
Constant & 1.33977e+08 & 1.85917e+07 & 7.206 & <0.001 \\
L1.gas\_base\_fee & 0.681658 & 0.004391 & 155.252 & <0.001 \\
L2.gas\_base\_fee & 0.145536 & 0.005313 & 27.390 & <0.001 \\
L3.gas\_base\_fee & 0.066853 & 0.005344 & 12.510 & <0.001 \\
L4.gas\_base\_fee & 0.046873 & 0.005314 & 8.821 & <0.001 \\
L5.gas\_base\_fee & 0.054052 & 0.004391 & 12.310 & <0.001 \\
\addlinespace
\midrule
\textbf{Equation for blob\_gas\_base\_fee} & Coefficient & Standard Error & T-Statistic & P-Value \\
\midrule
Constant & 4.10045e+08 & 6.68550e+07 & 6.133 & <0.001 \\
L1.gas\_base\_fee & 0.058937 & 0.015789 & 3.733 & <0.001 \\
L2.gas\_base\_fee & -0.018849 & 0.019107 & -0.986 & 0.324 \\
L3.gas\_base\_fee & -0.056381 & 0.019216 & -2.934 & 0.003 \\
L4.gas\_base\_fee & 0.053983 & 0.019108 & 2.825 & 0.005 \\
L5.gas\_base\_fee & -0.035820 & 0.015790 & -2.269 & 0.023 \\
\bottomrule
\end{tabular}
\end{table*}

\begin{table*}
\centering
\caption{Estimated coefficients and statistics for datagas\_base\_fee and datagas\_priority\_fee\_per\_datagas}
\label{tab:appendix_VAR_model_2}
\begin{tabular}{>{\raggedright\arraybackslash}p{8cm}cccc}
\toprule
\textbf{Equation for datagas\_base\_fee} & Coefficient & Standard Error & T-Statistic & P-Value \\
\midrule
Constant & 410690637.474525 & 38312702.654541 & 10.719 & <0.001 \\
L1.datagas\_base\_fee & 0.958823 & 0.004394 & 218.233 & <0.001 \\
L1.datagas\_priority\_fee\_per\_datagas & 0.545003 & 0.054900 & 9.927 & <0.001 \\
L6.datagas\_priority\_fee\_per\_datagas & -0.068182 & 0.054934 & -1.241 & 0.215 \\
\addlinespace
\midrule
\textbf{Equation for datagas\_priority\_fee\_per\_datagas} & Coefficient & Standard Error & T-Statistic & P-Value \\
\midrule
Constant & 56044603.443251 & 3059021.581396 & 18.321 & <0.001 \\
L1.datagas\_base\_fee & -0.001337 & 0.000351 & -3.812 & <0.001 \\
L1.datagas\_priority\_fee\_per\_datagas & 0.115156 & 0.004383 & 26.271 & <0.001 \\
L6.datagas\_priority\_fee\_per\_datagas & 0.075670 & 0.004386 & 17.252 & <0.001 \\
\bottomrule
\end{tabular}
\end{table*}

\subsection{Correlation matrix of residuals}
Table \ref{tab:correlation_matrix_2} presents the correlation matrix of residuals for the VAR model involving datagas base fee and datagas priority fee per datagas. This matrix is critical for checking the assumption of no serial correlation among residuals, which is a fundamental requirement for the validity of model inferences in VAR analysis.

The values in the matrix show the correlation coefficients between the residuals of the two equations modeled. A coefficient close to zero between different variables' residuals, such as seen here between datagas base fee and datagas priority fee per datagas, suggests that there is no significant linear relationship between the residuals. This indicates that the model does not suffer from multicollinearity issues and that the residuals are behaving as expected—randomly and independently from each other, which supports the reliability of the model's forecasts and conclusions.

\section{Rollup transaction dynamics}
\subsection{Total fee paid by rollups}
\begin{figure}[ht]
  \centering
  \includegraphics[width=0.8\linewidth]{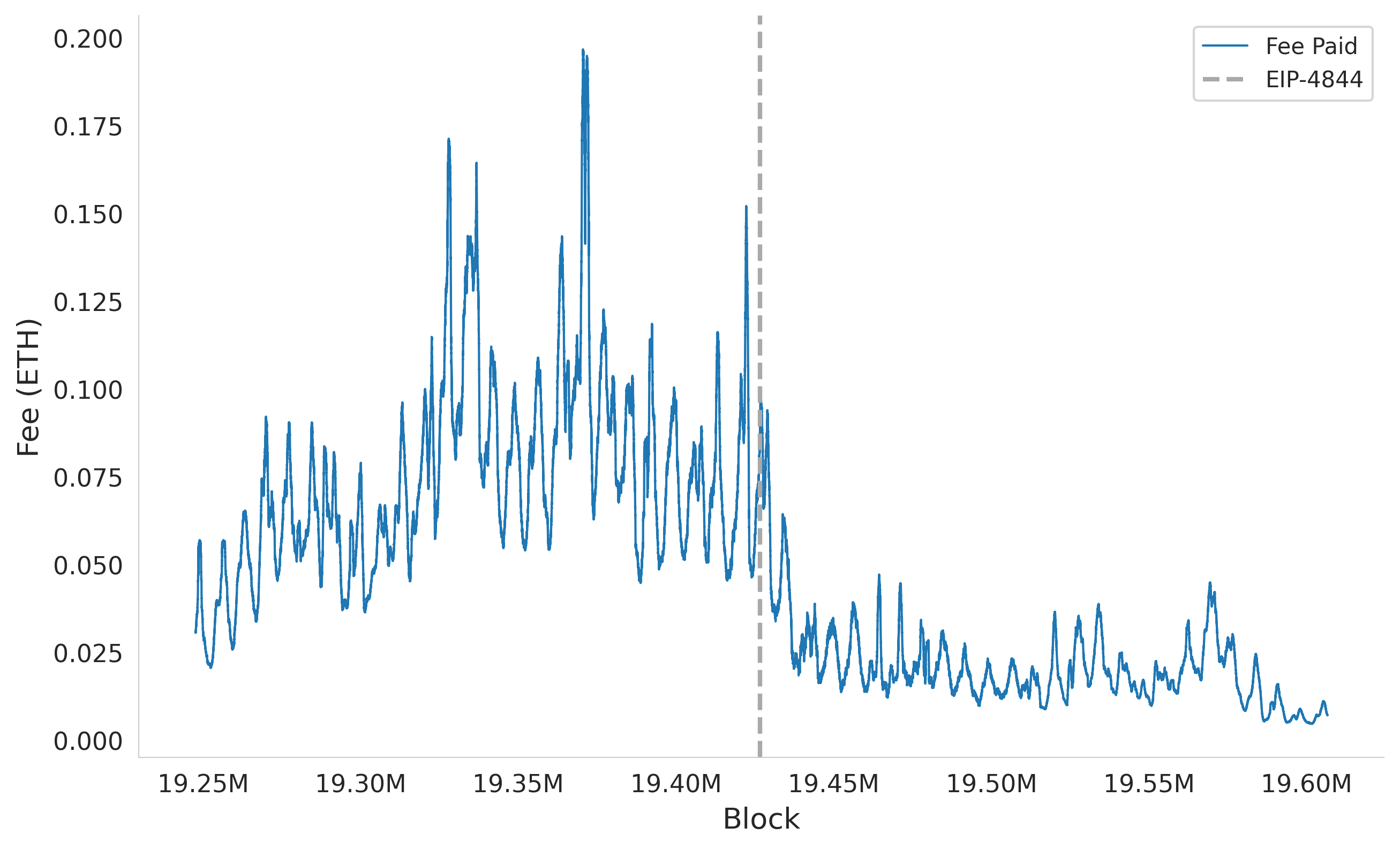}
  \caption{Total fee paid by top 10 rollups on Ethereum as a DA layer}
  \Description{}
  \label{fig:4_2_fee_paid}
\end{figure}

Figure \ref{fig:4_2_fee_paid} illustrates the total fee paid by the top 10 rollups on Ethereum to use Ethereum as a Data Availability (DA) layer. The vertical dashed line indicates the implementation of EIP-4844. The figure shows a noticeable decrease in fees after EIP-4844, reflecting the impact of protocol changes aimed at reducing transaction costs for rollup operations.

\begin{figure}[ht]
    \centering
    \begin{subfigure}{0.8\linewidth}
        \centering
        \label{fig:4_2_gas_used_all}
        \caption{All rollups}
        \includegraphics[width=\linewidth]{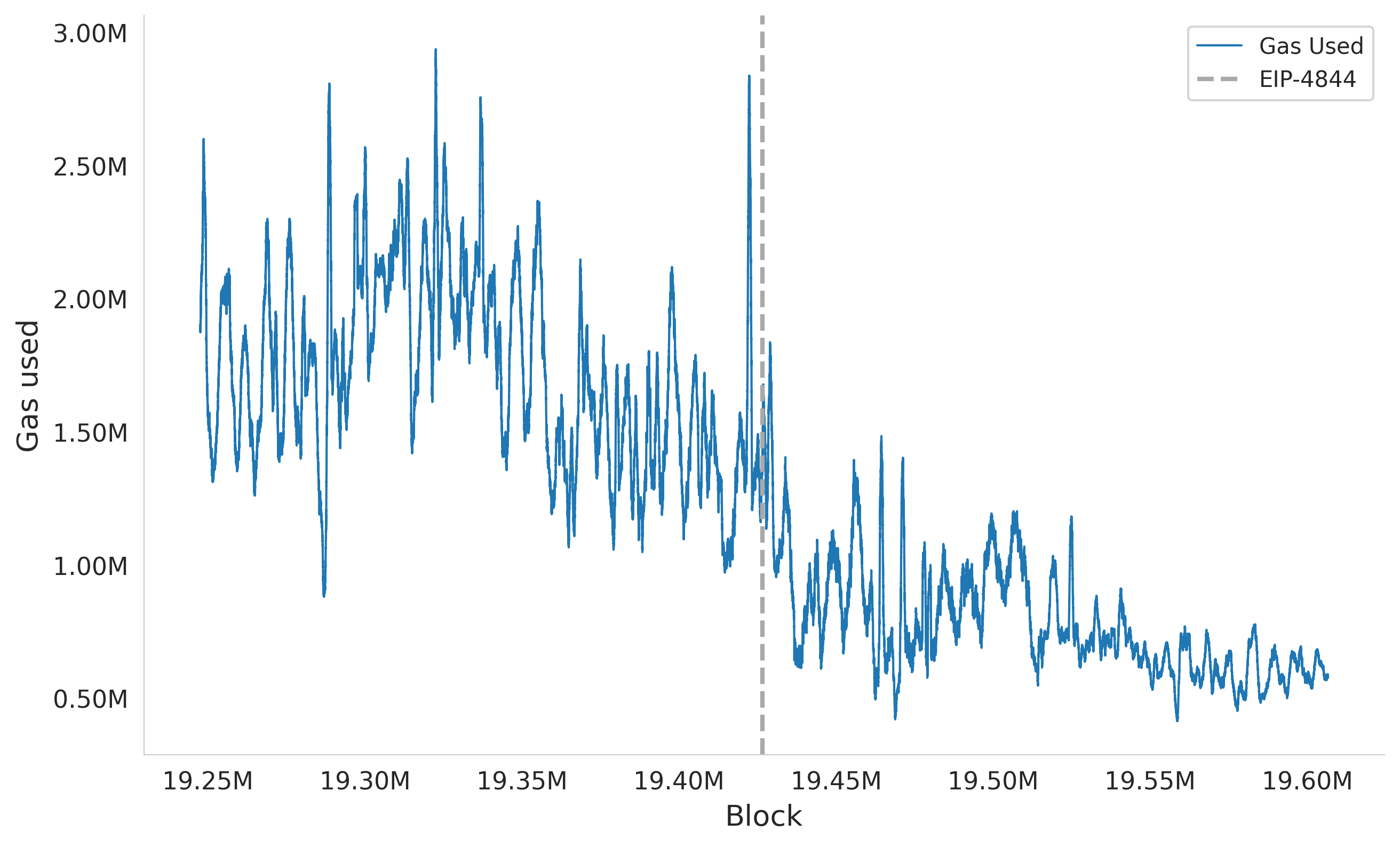}
    \end{subfigure}
    \\
    \begin{subfigure}{0.45\linewidth}
        \centering
        \label{fig:4_2_gas_used_optimistic}
        \caption{Optimistic rollups} 
        \includegraphics[width=\linewidth]{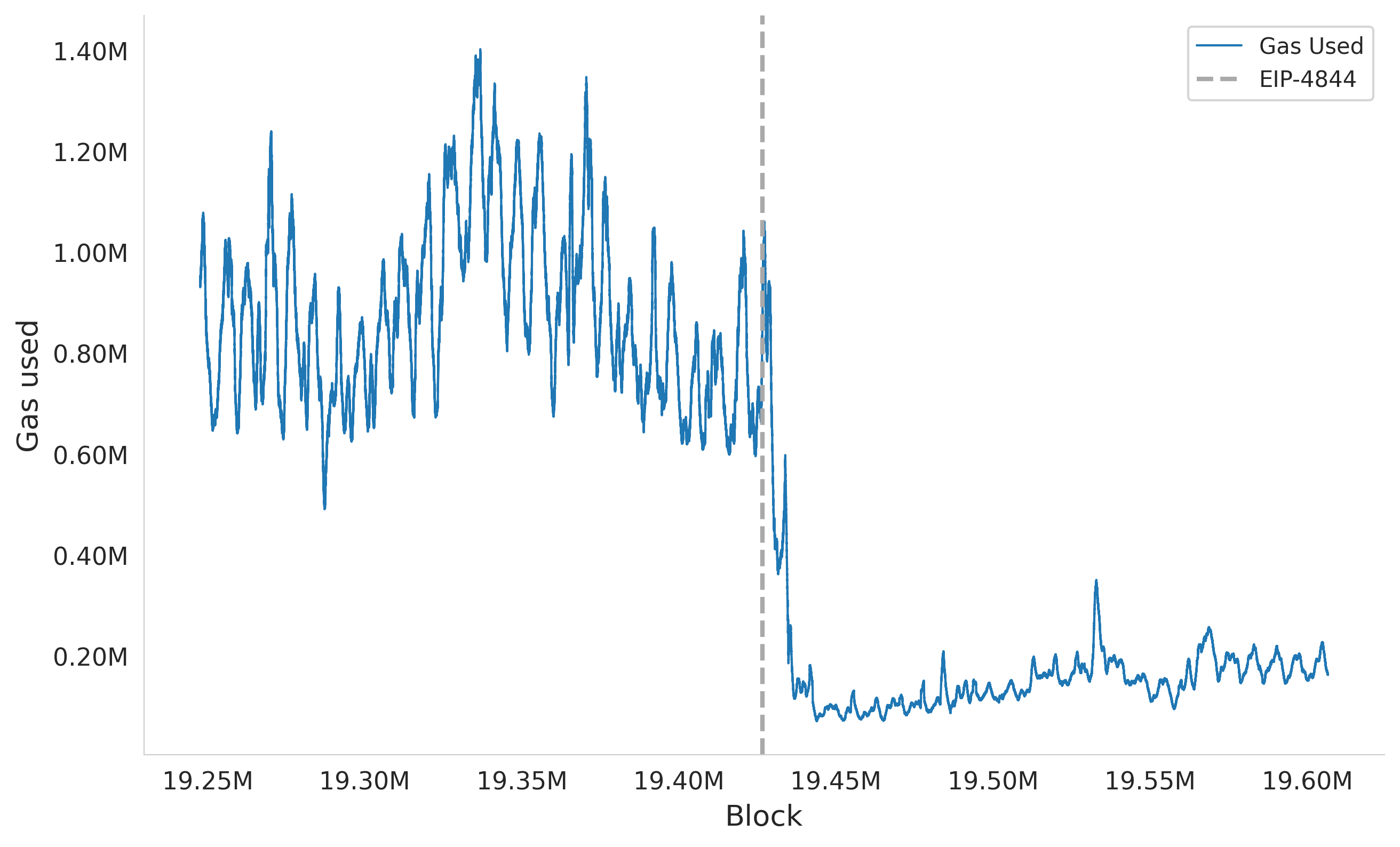}
    \end{subfigure}
    \hfill 
    \begin{subfigure}{0.45\linewidth}
        \centering
        \label{fig:4_2_gas_used_zk}
        \caption{ZK rollups} 
        \includegraphics[width=\linewidth]{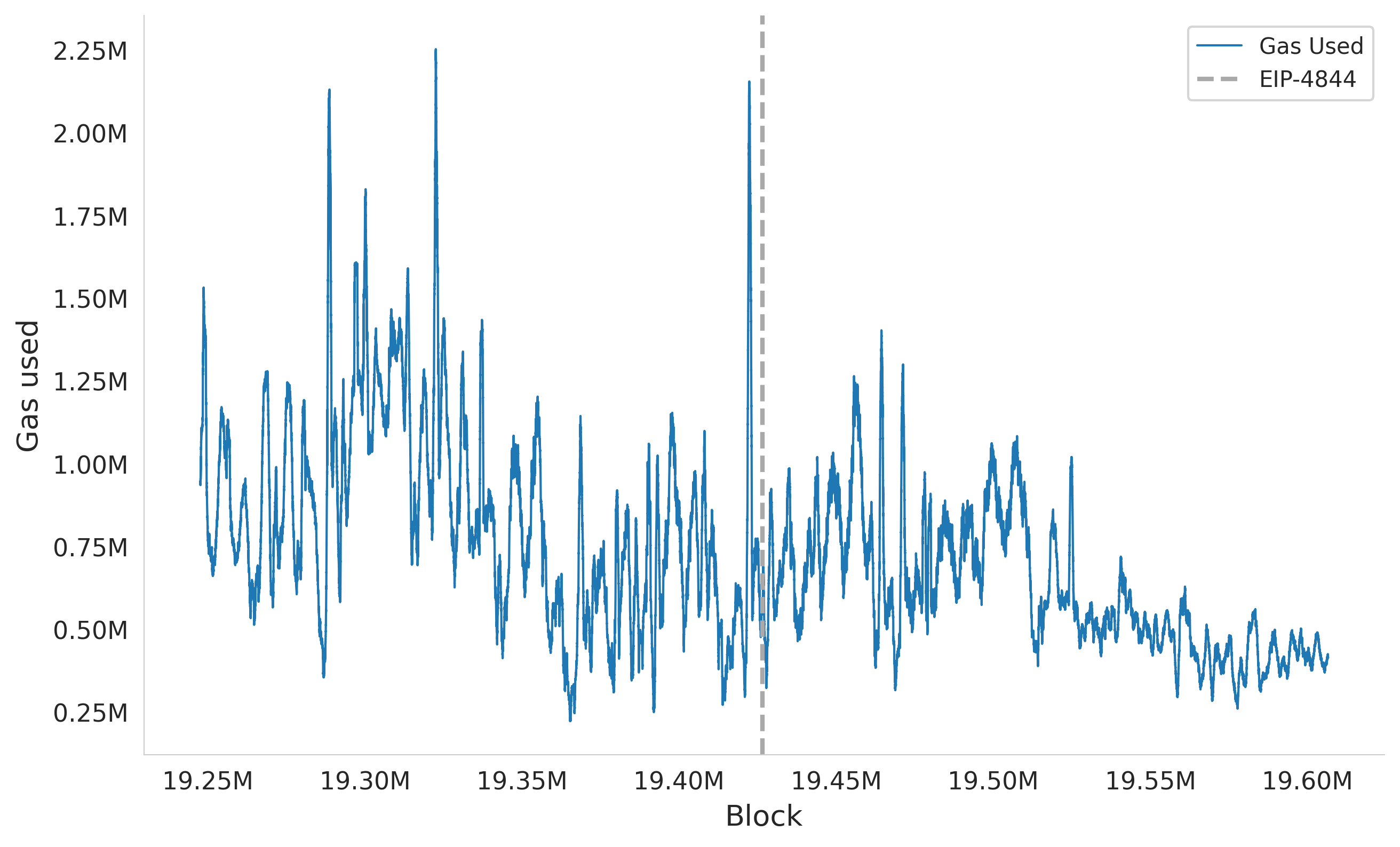}
    \end{subfigure}
    \caption{Total gas used by top 10 rollups on Ethereum}
    \label{fig:4_2_gas_used}
    \Description{Total gas used by top 10 rollups on Ethereum}
\end{figure}

\subsection{Gas used by rollups}
Figure \ref{fig:4_2_gas_used} shows the total gas usage by rollups in all rollups, optimistic rollups, and ZK rollups. The figure shows a noticeable decrease in fees after EIP-4844, especially for optimistic rollups.

\end{sloppypar}
\end{document}